\documentclass{aa}  

\usepackage{amsmath,amssymb}
\usepackage[breaklinks=true]{hyperref}
\usepackage{threeparttable}

\usepackage{graphicx}
\usepackage{xcolor}
\usepackage[normalem]{ulem}
\usepackage{listings}
\definecolor{dkgreen}{rgb}{0,0.6,0}
\definecolor{gray}{rgb}{0.5,0.5,0.5}
\definecolor{mauve}{rgb}{0.58,0,0.82}
\usepackage{txfonts}
\usepackage{tikz}
\usepackage{booktabs}

\providecommand{\teff}{\ensuremath{{T_{\rm eff}}}\xspace}

\providecommand{\radius}{\ensuremath{{R}}\xspace}
\providecommand{\Zabun}{\ensuremath{{Z}}\xspace}

\providecommand{\gmag}{\ensuremath{G}}

\providecommand{\kpc}{\ensuremath{\,\rm kpc}\xspace}

\providecommand{\Rsun}{\ensuremath{\,{\radius}_{\odot}}\xspace}
\providecommand{\Zsun}{\ensuremath{\,{\Zabun}_{\odot}}\xspace}

\newcommand{\galaxia}{{\sl Galaxia}}
\newcommand{\agama}{{\sl AGAMA}}

\newcommand\gaia{\textit{Gaia}}

\newcommand\gdrtwo{\gaia~DR2}
\newcommand\gedrthree{\gaia~EDR3}
\newcommand\gdrthree{\gaia~DR3}

\newcommand\kepler{\textit{Kepler}}
\newcommand\glh{GALAH}
\newcommand\apg{{APOGEE}}
\newcommand\rave{{RAVE}}
\newcommand\sdss{{SDSS}}

\newcommand\rc{\textsc{RC}}
\newcommand\rgbp{\textsc{RGB}}
\newcommand\gaiawise{\textit{gdr3wise}}
\newcommand\periodspacing{\textit{PS}}
\newcommand\deltanu{$\Delta \nu$}

\newcommand\lfrc{$LF_{RC}$}

\newcommand{\absg}{$M_{G}$}
\newcommand{\absw}{$M_{W1}$}
\newcommand{\mldist}{$D_{ml}$}

\newcommand{\gunlim}{{GaiaUnlimited}}
\newcommand{\absmag}{$M_{\lambda}$}
\newcommand{\appmag}{$m_{\lambda}$}
\newcommand{\appmaglim}{$m_{\lambda, lim}$}
\newcommand{\alambda}{$A_{\lambda}$}

\newcommand{\twomass}{{2MASS}}
\newcommand{\panstars}{{Pan-STARRS}}
\newcommand{\allwise}{AllWISE}
\newcommand{\wise}{{WISE}}
\newcommand{\bayestar}{{Bayestar}}
\newcommand{\gaussian}{{Gaussian}}

\newcommand{\pearson}{$\rho_{gw}$}

\newcommand{\alfe}{\ensuremath{[\mathrm{\alpha/Fe}]}}
\newcommand{\feh}{\ensuremath{[\mathrm{Fe/H}]}}

\newcommand{\logg}{\mbox{$\log g$}}
\newcommand{\kiel}{{\sl Kiel}}

\newcommand{\jkzero}{$(J-K)_{0}$}
\newcommand{\clr}{$(J-K)_{0}$}

\newcommand{\camd}{$CaMD$}

\newcommand{\hpix}{$HEALPix$}

\newcommand{\bprp}{{ $G_\mathrm{BP} - G_\mathrm{RP}$}}
\newcommand{\grp}{{ $G - G_\mathrm{RP}$}}

\newcommand{\errorovparallax}{$\sigma_{\varpi}/\varpi$}

\newcommand{\cbj}{{\sl CBJ21}}
\newcommand{\schlegel}{{\sl S98}}
\newcommand{\geo}{{$d_{\rm geo}$}}
\newcommand{\photogeo}{{$d_{\rm photgeo}$}}

\newcommand{\fsel}{{$F$}}
\newcommand{\fseli}{{$F_{\rm i}$}}
\newcommand{\fselitop}{{$S_{\rm top, i}$}}
\newcommand{\fselisub}{{$S_{\rm sub, i}$}}
\newcommand{\rcut}{{$R_{\rm cut}$}}

\newcommand{\rdtwo}{{$R_{\rm d2}$}}
\newcommand{\rd}{{$R_{\rm d}$}}
\newcommand{\hz}{{\sl $h_{z}$}}
\newcommand{\hztwo}{{\sl $h_{z2,\odot}$}}
\newcommand{\hzsun}{{\sl $h_{z,\odot}$}}

\newcommand{\rflare}{\ensuremath{R_{\rm fl}}}
\newcommand{\fdisc}{{\sl $f_{d1}$}}
\newcommand{\zwarp}{{$z_{\rm w}$}}
\newcommand{\awarp}{{$a_{\rm w}$}}
\newcommand{\rwarp}{{$R_{\rm w}$}}
\newcommand{\hwarp}{{$h_{\rm w0}$}}
\newcommand{\phiwarp}{{$\phi_{\rm w}$}}

\newcommand{\phiprime}{$\phi^{'}$}

\newcommand{\rgal}{{\sl $R$}}
\newcommand{\zgal}{{\sl $Z_{GC}$}}
\newcommand{\xgc}{{\sl $X_{GC}$}}
\newcommand{\ygc}{{\sl $Y_{GC}$}}
\newcommand{\zgc}{{\sl $Z_{GC}$}}

\newcommand{\mir}{{MIR}}

\newcommand{\nzproj}{{\sl $N(z|R)$}}

\newcommand{\nmin}{{$N_{\rm min, i}$}}

\newcommand{\lucey}{{L20}}
\newcommand{\jie}{{Yu18}}
\newcommand{\elsworth}{{Els19}}
\newcommand{\andrae}{{A23}}

\usepackage{graphicx}
\usepackage{txfonts}
\usepackage{hyperref}
\newcommand{\orcit}[1]{\protect\href{https://orcid.org/#1}{\protect\includegraphics[width=8pt]{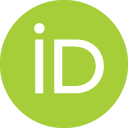}}}

\makeatletter
\renewcommand*\maketitle{%
  \thispagestyle{firstpage}
\begingroup
    \if@wideboxfn
    \setlength\bibindent{1.4\parindent}
    \else
    \setlength\bibindent{\parindent}
    \fi
    \renewcommand*\thefootnote{\@fnsymbol\c@footnote}%
    \renewcommand\@makefntext[1]{%
    \ifaa@longfn\hsize\textwidth\fi
    \noindent
    \hb@xt@\bibindent{\hss\@makefnmark\enspace}##1}
  \ifaa@twocolumn
  \begingroup
    \begin{aa@strip}
          \aa@maketitle
    \end{aa@strip}
    \@thanks            
  \endgroup
  \else
    \begingroup
      \let\thanks\footnote
      \aa@maketitle
    \endgroup
  \fi
\endgroup
  \setcounter{footnote}{0}%
}
\makeatother

\DeclareUnicodeCharacter{2212}{-}

\begin{document}

\title{GaiaUnlimited: The old stellar disc of the Milky Way \\as traced by the red clump}

\titlerunning{GaiaWISE RC catalogue}

\author{ 
Shourya Khanna\inst{1} \and
Jie Yu\inst{2,3,4} \and
Ronald Drimmel\inst{1} \and
Eloisa Poggio\inst{1} \and
Tristan Cantat-Gaudin\inst{5} \and
Alfred Castro-Ginard\inst{6} \and
Evgeny Kurbatov\inst{7} \and 
Vasily Belokurov\inst{7} \and
Anthony Brown\inst{6} \and
Morgan Fouesneau\inst{5} \and
Andrew Casey\inst{8,9} \and
Hans-Walter Rix\inst{5}
}

\institute{INAF - Osservatorio Astrofisico di Torino, via Osservatorio 20, 10025 Pino Torinese (TO), Italy\\    \email{shourya.khanna@inaf.it}  
 \and  
  School of Astronomy and Space Science, Nanjing University, Nanjing 210023, People's Republic of China.\\\email{jie.yu@nju.edu.cn} 
 \and  
 Key Laboratory of Modern Astronomy and Astrophysics, Ministry of Education, Nanjing 210023, People's Republic of China.   
 \and
  Research School of Astronomy \& Astrophysics, Australian National University, Cotter Rd., Weston, ACT 2611, Australia 
  \and
  Max Planck Institute for Astronomy, K\"{ o}nigstuhl 17, 69117 Heidelberg, Germany
  \and
  Leiden Observatory, Leiden University, Einsteinweg 55, 2333 CC Leiden, The Netherlands
  \and  
  Institute of Astronomy, University of Cambridge, Madingley Road, Cambridge, CB30HA, UK
  \and
  School of Physics and Astronomy, Monash University, Clayton VIC 3800, Australia
  \and
  Center for Computational Astrophysics, Flatiron Institute, 162 5th Avenue, New York City, New York
         }
\abstract{  
We present an exploration of the Milky Way's structural parameters using an all-sky sample of red clump (RC) giants to map the stellar density from the Galactic disc beyond 3 kpc. These evolved giants are considered to be standard candles due to their low intrinsic variance in their absolute luminosities, and this allows us to estimate their distances with reasonable confidence. We exploited all-sky photometry from the AllWISE mid-infrared survey and the \textit{Gaia} survey along with astrometry from \textit{Gaia} Data Release 3 and recent 3D extinction maps to develop a probabilistic scheme in order to select with high confidence \rc{}-like stars. 
Our curated catalogue contains about ten million sources, for which we estimated photometric distances based on the WISE $W1$ photometry. We derived the selection function for our sample, which is the combined selection function of sources with both \gaia{} and \allwise{} photometry.  Using the distances and accounting for the full selection function of our observables, we were able to fit a two-disc, multi-parameter model to constrain the scale height (\hz{}), scale length (\rd{}), flaring, and the relative mass ratios of the two-disc components. We illustrate and verify our methodology using mock catalogues of \rc{} stars. We find that the \rc{} population is best described by a flared disc with scale length \rd{}=$4.24\pm0.32$ kpc and scale height at the Sun of \hzsun{}=$0.18\pm0.01$ kpc, and a shorter and thicker disc with \rd{}=$2.66\pm0.11$ kpc, \hzsun{}=$0.48\pm0.11$ kpc, with no flare. The thicker disc constitutes 66\% of the \rc{} stellar mass beyond 3 kpc, while the flared disc shows evidence of being warped beyond 9 kpc from the Galactic centre. The residuals between the predicted number density of RC stars from our axisymmetric model and the measured counts show possible evidence of a two-armed spiral perturbation in the disc of the Milky Way. }  
\keywords{Galaxy: structure -- Galaxy: disc -- Galaxy: fundamental parameters --Stars: distances}
\maketitle

\section{Introduction}
\begin{figure*}
\includegraphics[width=1.\columnwidth]{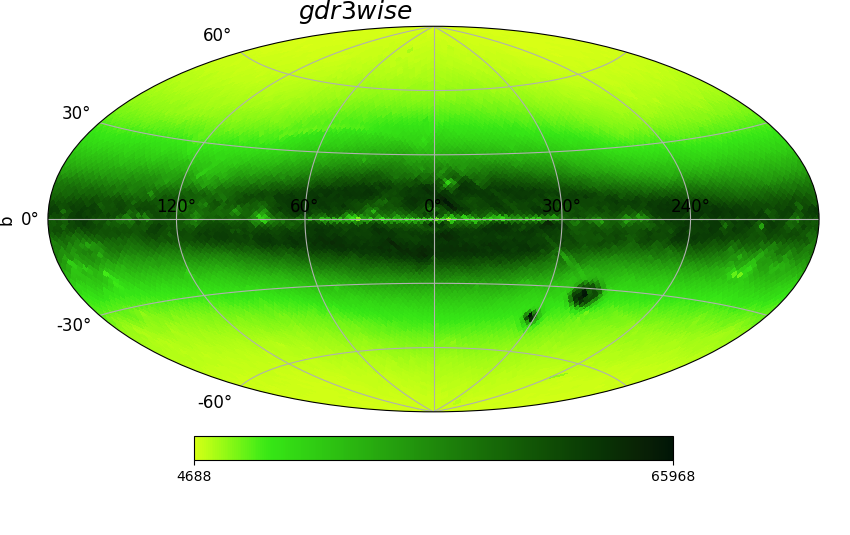} 
\includegraphics[width=1.\columnwidth]{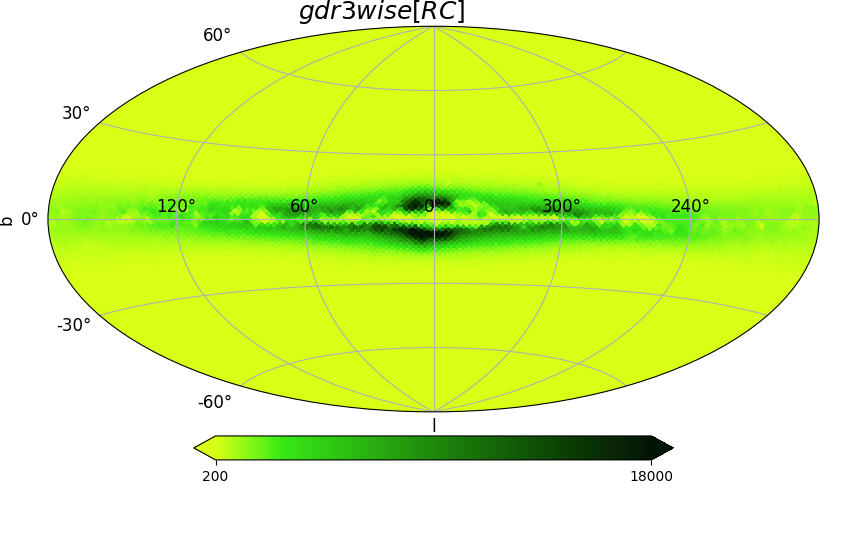} 
\caption{All sky spatial distribution of our dataset. Initial cross-match of \textit{Gaia} DR3 and \allwise{}, i.e. \gaiawise{} shown in the left panel, and the red clump candidates selected from this are shown in the right panel, i.e. \gaiawise{}[RC] for the region $|$\zgal{}$|<$2 kpc.  } \label{fig:allsky}
\end{figure*}

Our Galaxy presents a unique opportunity to study the properties and distribution of millions of individual stars in detail. Over the past few decades, this has been made possible in large part thanks to high-quality photometric surveys such as \twomass{} and \wise{} \citep{twomass, wise}; spectroscopic surveys such as \textit{Gaia}-ESO, \apg{}, \rave{}, LAMOST, and \glh{} \citep{Randich13, Majewski17, Steinmetz2020, Deng12, DeSilva15}; and more recently with additional synergy from the astrometry of the \gaia{} mission with unprecedented precision \citep{GaiaCollaboration:2016}. The wealth of data allows us to compare how various stellar populations trace the structure and shape of the Galaxy.

The generally accepted view is that the stellar content of the Milky Way is distributed along exponential discs in the galactocentric cylindrical coordinates (\rgal{}, \zgal{}). Early studies of stellar counts showed that the disc could be principally split into two components: a `thick' disc with scale height at the Sun of \hzsun{}{}$\sim 0.7-0.9$ kpc and a `thin' disc with a shorter scale height \hzsun{}{}$\sim 0.3$ kpc \citep{Yoshii:1982, gilmore_reid83, Juric:2008}. Conversely, the `thick' disc has a shorter scale length between the two, but typical estimates range between $2<$\rd{}$<4$ kpc \citep[see][for full review]{jbhreview2016}. Large-scale spectroscopic surveys have allowed further dissection of the disc using chemistry (\feh{},\alfe{}), showing that the scale parameters vary with age \citep{Hayden:2015}. For example, \cite{bovy16disc} analysed \apg{} data in bins of mono-abundance (approximately mono-age population) and found that the radial density of the high \alfe{} population can be described by a single exponential over a large range in galactocentric radius \rgal{}. The low \alfe{} population instead exhibits more complex radial density profiles, and it is understood as a series of sub-discs of similar ages, peaking in density at different radii from the inner to the outer disc.

In order to simultaneously model the vertical and radial structure of the disc, it would be advantageous to use a population of stars that is numerous, spans a large range in \rgal{}, and for which distances can be reliably derived. To this end, in this paper, we use stars in the \rc{} evolutionary phase. These are low-mass red giant stars burning helium in their convective core. During this stage, they have a relatively narrow distribution in their absolute magnitude, and their age distribution peaks between $1\sim2$ Gyr \citep{Cannon:1970,Girardi:2016}. Over the years, the \rc{} has been used to map out various features of the disc, such as the Galactic bulge \& bar \citep{Ness:2012,Wegg:2013}; to study the warp, flare, and scale parameters \citep{Lopez-Corredoira:2002,hWang_warp:2020,bovy16disc,Uppal:2024warp}; and to map the large-scale velocity field of the disc \citep{Bovy:2015,khanna18,anticentre_gaia21}. 

With the \gaia{} survey, we now have a dataset with homogenous all-sky photometry and astrometry, which allows for the selection of large samples of stellar tracers of similar type. In \cite{anticentre_gaia21},  \gedrthree{} and \twomass{} were combined to construct a catalogue of \rc{}-like stars to study the kinematics of the outer disc. This sample was limited to a narrow region about the Galactic anti-centre ($|l-180^\circ|<20^\circ$), where extinction is minimal, allowing for deep coverage far into the outer disc. In this contribution, we extend this approach to the entire sky. However, having a large sample of distance tracers is not enough by itself, as the goal is to compare the predictions from a sensible model to the data. Whereas the model lives in a perfect universe (i.e. it is unaffected by observational effects), the data are limited by the magnitude limits of the surveys used, incompleteness in data coverage due to inhomogeneity in data sampling, and any quality cuts that one imposes when compiling the data from one or more catalogues. Thus, to properly model the Galaxy (or indeed any system), one needs to bring the ideal model to the data space in order to take into account the selection effects and, importantly, the extinction. This is illustrated well by \cite{libinneyob22}, who recently modelled the youngest component of the disc using OB-type stars and explored the effects of incorrect dust models.

This study was made in the context of the \gunlim{} project,\footnote{\url{https://gaia-unlimited.org/}} which has provided tools to the community to build selection functions for various datasets based on \gaia{} \citep{Rix:2021, Cantat-Gaudin:2023, Castro-Ginard:2023, Cantat-Gaudin:2024rc}. Here, we apply some of these tools to model the density profile of \rc{} stars in the Galactic disc. 
The text is organised as follows: In Sect. \ref{sec:data} we describe the scheme to select \rc{} stars and derive distances; Sect. \ref{sec:method}
lays out the scheme for building the selection function and the galactic models we fit; in Sect. \ref{sec:results} we present the results of the density modelling on both mock and observed data; and we discuss these in detail in Sect. \ref{sec:discussion}.

\section{Data}
\label{sec:data}
Our primary dataset is the official cross-match between \gdrthree{} and \allwise{}, obtained from the \textit{allwise\_best\_neighbour} table provided on the \gaia{} data archive; the details of the cross-match algorithm are provided in \cite{dr2Marrese:2019}\footnote{\href{https://gea.esac.esa.int/archive/documentation/GDR3/Catalogue_consolidation/chap_crossmatch/}{Gaia crossmatch documentation}}. We retain only those sources with finite measurement for parallax ($\varpi$), as well as photometry in both \gaia{} ($G$) and \allwise{} ($W_{1}$ and $W_{2}$) bands. This gives a count of $N=303,183,850$ sources. The all-sky distribution of this dataset, which we call \gaiawise{}, is shown in Fig. \ref{fig:allsky} (left panel). Our algorithm to select \rc{} stars is agnostic to the data quality; hence, at this stage we do not filter out poor data. 

\subsection{Photo-astrometric selection of the red clump}
\label{sec:rc_method}

\begin{figure}
\centering
\includegraphics[width=1.\columnwidth]{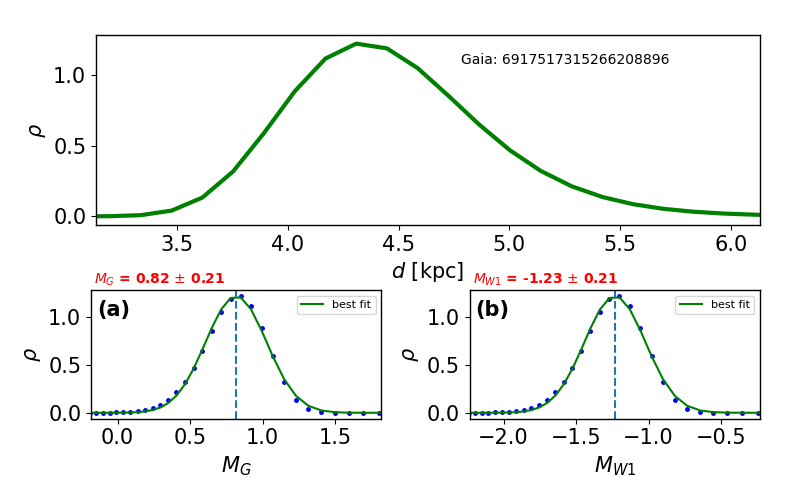} 
\caption{Procedure to obtain the distribution of the absolute magnitude starting from distance priors illustrated for an example source. The top inset shows the distance prior used, which is turned into a grid of absolute magnitude using 3D extinction maps. The bottom figures show a Gaussian fit to the resulting absolute magnitude distributions in the $G$ and $W1$ bands used to select the \rc{} candidates.} \label{fig:absmag_fit}
\end{figure}
For each star, one can in principle compute the absolute magnitude (\absmag) in both the \allwise{} ($\lambda=W1$) and \textit{Gaia} ($\lambda=G$) passbands as
\begin{equation}
\label{eqn:absmag_cal}
M_{\lambda} = m_{\lambda} - A_{\lambda} - {\rm \mu} ,   
\end{equation} where $\mu$ = 5 $\log_{10} (100 /\varpi^\prime [\rm mas])$ is the distance modulus, and $A_\lambda$ is the extinction in the passband $\lambda$, whose estimation is described in the following section. We adopt $\varpi^\prime = \varpi + 0.017$ assuming a global parallax offset from \cite{Lindegren:2021}. However, applying Eq. \ref{eqn:absmag_cal} would mean inverting the parallax to obtain an estimate for the distance modulus, which would yield problematic distances for sources with high parallax uncertainty, \errorovparallax{} $>0.2$ \citep{2015PASP..127..994B,Luri2018}. In this light, we make use of the Bayesian distances estimated by \cite{Bailer-Jones:2021}[hereafter \cbj]. Their catalogue provides geometric distances (\geo) for 1.47 billion sources, requiring only the \gedrthree{} parallaxes. Additionally, for a large majority, they also provide photo-geometric distances (\photogeo) for sources where \gmag{} magnitude and \bprp{} colour are also available. In general, the \textit{photogeo} distances are considered better for sources with high \errorovparallax. However, these have a dependence on the stellar population modelling in the \gedrthree{} mock catalogue \citep{Rybizki:2020} used as a prior, and could have complex behaviour at low latitudes, as noted by \cbj.  In any case, both distance estimators are dependent on the assumed 3D density distribution of stars in the Milky Way, and in the case of \cbj{} this is derived from the Besançon Galactic model \citep{besanconRobin:2003} used to construct the \gedrthree{} mock catalogue of \citet{Rybizki:2020}.
 Following \cbj, we construct the posterior probability distribution function (PDF hereafter) of distances ($d$) for every source in our sample. This can only be done for their geometric distances, and in any case our intention here is to use an informed prior with as few assumptions as possible, and only for the purpose of selecting \rc{} candidates. We use the distance priors and a likelihood function (see \autoref{app:bailerjones}), to compute the PDF on a grid of 500 points ($d_{1},d_{2},...d_{n}$), with $d_{1} =0$, and $d_{n} = 2 \times ($\photogeo$|$\geo $)$, i.e. setting the maximum grid point to be twice the distance prior for each source. In general, we use \photogeo{} distances where available, but for the minority of sources that don't have these provided, we use \geo{} distances instead. We use the median value of the distance PDF, along with sky position ($l,b$) to estimate the reddening $E(B-V)$, and using the reddening coefficients listed in Sect. \ref{sec:dustmodel} to convert to extinction $A_{\lambda}$. 
\begin{figure}
\centering
\includegraphics[width=.8\columnwidth]{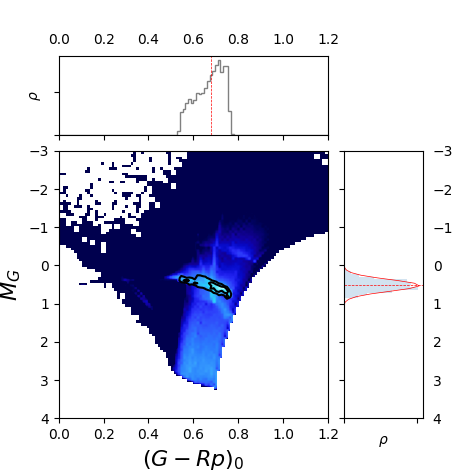} 
\includegraphics[width=.8\columnwidth]{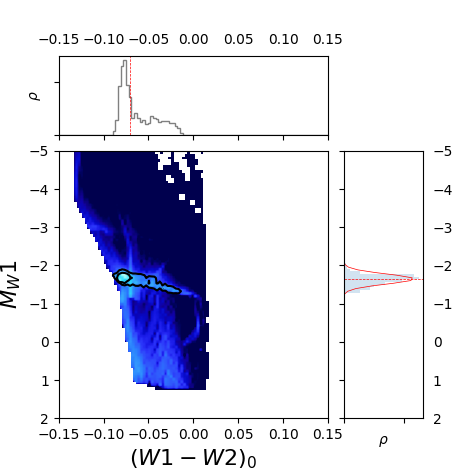} 
\caption{Distribution of giants on the \camd{} diagram in a mock realisation of the Milky Way using the \galaxia{} code (see Appendix \ref{app:rc_feh_cdf}). The top panel shows \absg{} vs \gaia{} colours, while the bottom panel shows \absw{} versus \allwise{} colours. The \rc{} is marked by the black contours. The marginalised histogram (normalised) for the x (top sub panel) and y (right sub panel) axes are also shown.  For the \rc{}, both \absg{} and \absw{} can be approximated by a quasi-Gaussian as is shown by the red curves on the right insets.} \label{fig:glx_rc_camd}
\end{figure}
 
We then converted the distance grid to that in absolute magnitude (i.e. $M_{\lambda,1},M_{\lambda,2},...M_{\lambda,n}$), thus obtaining the posterior PDF in $M_{\lambda}$, to which we fit a broad \gaussian{} ($\mathcal{N}$) profile in order to obtain the peak and spread ($\sigma_{ M_{\lambda}}$) in the inferred distribution. Since any given star has a single true absolute magnitude, $\sigma_{ M_{\lambda}}$ quantifies our uncertainty on our inferred $M_{\lambda}$ value, taken as the mean of the fitted Gaussian due to distance and extinction uncertainties. This procedure is illustrated in Fig. \ref{fig:absmag_fit}. 

\begin{table}
\caption{Absolute magnitude for RC stars.\label{tab:tab_extinct}}
\begin{tabular}{|l|l|l|}
\hline
Passband ($\lambda$) & $\Bar{M_{\lambda}}$ & $\sigma_{\Bar{M_{\lambda}}}$ \\
\hline
$W_{1}$    & $-1.68 \pm 0.01  $ & $0.10 \pm 0.02  $\\
$G$ & $+0.44 \pm 0.01 $ & $0.20 \pm 0.02 $ \\
\hline
\end{tabular}
\tablefoot{
Median absolute magnitude  $\Bar{M_{\lambda}}$, and dispersion in absolute magnitude $\sigma_{\Bar{M_{\lambda}}}$ for typical \rc{} stars, as listed in \citet{Hawkins:2017}
}
\end{table}

The absolute magnitude for the \rc{} has been calibrated in several passbands, for example by \cite{Hawkins:2017} and \cite{Ruiz-Dern:2018}, who also considered variations due to colour. \citet[][K19 hereafter]{khanna18} calibrated the absolute magnitude in the $K$ band as a function of metallicity $M_{K}$([FeH]) using the \galaxia{} code \citep{Sharma:2011}. They showed that the dependence of $M_{K}$ on metallicity is non-negligible for stars with \feh{}$<-0.4$ dex, while beyond this metallicity, the $M_{K}$ is roughly flat with \feh{} (see their Fig. A.4). How many \feh{}$<-0.4$ \rc{} stars we expect to observe depends on the age distribution, and the apparent magnitude limit of the sample. In Appendix \ref{app:rc_feh_cdf}, we select \rc{} stars from a Milky Way analogue generated using \galaxia{}. For these, Fig. \ref{fig:glx_rc_camd} shows the distribution of \absg{} versus ($G-Rp$)$_{0}$, and \absw{} versus $(W_{1} -W_{2})_{0}$, with the locus of the \rc{} marked by black contours.
Figure \ref{fig:rc_feh_cdf} shows the cumulative distribution function (CDF) of metallicity for the \rc{}. We show the CDF profile for three different age caps, that is assuming the maximum age of the \rc{} sample in the Galactic disc is $\tau=$(5,7,10) Gyr. We find that the expected fraction of \rc{} at \feh{}$<-0.4$ is about 12 \% for a cap of $\tau<7$ Gyr, and reaches about 20\% for $\tau<10$ Gyr. Since the age distribution of \rc{} generally peaks around $1\sim 2$ Gyr \citep{Girardi:2016}, we do not expect many \rc{} stars at \feh{}$<-0.4$ dex and thus neglected this small metallicity dependence in the absolute magnitude.  \autoref{tab:tab_extinct} lists the absolute magnitude ($\Bar{M_{\lambda}}$) and intrinsic dispersion $\sigma_{\Bar{M_{\lambda}}}$ for the \rc{} in the photometric bands used here, according to \cite{Hawkins:2017}. 

In order to select the \rc{}, we assumed that the distribution in absolute magnitude space is a 2D Gaussian with the centroid at ($\Bar{M_{G}}$,$\Bar{M}_{W1}$). In general for a given distribution, the distance between its centroid ($x_{0}$) and a point of interest ($x_{1}$) can be given in terms of the \citet[][\mldist{}]{mahalanobis1936} distance given as 
\begin{eqnarray}
D_{ml}^{2} = (x_{1} - x_{0})^{T} \Sigma^{-1} (x_{1} - x_{0}),  
\end{eqnarray} which respects the combined covariance ($\Sigma$) of $x_{0}$ and $x_{1}$. In our context, $x_{0}$ and $x_{1}$ are the two dimensional mean absolute magnitudes of the \rc{} and of our inferred absolute magnitudes ($M_{G}$,$M_{W1}$). The covariance matrix is then given as

\begin{equation}
\label{eqn:covariance_matrix}
  \Sigma = \bordermatrix{~  &  &  \cr
   &\varsigma_{g}^{2}  & \rho_{gw} \varsigma_{g}\varsigma_{w}  \cr
   & \rho_{gw}\varsigma_{g}\varsigma_{w} & \varsigma_{w}^{2} \cr
                    } \quad,  
\end{equation} 
where the diagonal terms combine the spread in the observed ($\sigma_{M_{\lambda}}$) and true dispersion of absolute magnitudes ($\sigma_{\Bar{M_{\lambda}}}$) such that $\varsigma_{g}^{2}
=\sigma_{M_{G}}^{2} +  \sigma_{\Bar{M_{G}}}^{2}$ and likewise for W1 (we used $\varsigma_{g}$ to avoid confusion with $\sigma_G$, the photometric uncertainty of $G$). The off-diagonal elements of $\Sigma$ depend on the correlation (\pearson) between the absolute magnitudes in the two passbands inferred by us as well as between the absolute magnitudes in the two passbands for genuine RC stars.  Since we have assumed the same distance prior and extinction map to derive the pdf of the absolute magnitudes in $G$ and $W1$, the correlation between the inferred absolute magnitudes is very close to 1. In order to determine the correlation in the intrinsic absolute magnitudes of RC stars, we selected from \gaiawise{}, stars with \errorovparallax{} $<0.05$, for which we could determine the inverse parallax distance ($d_{\varpi^\prime} = 1/\varpi^\prime$) and hence the absolute magnitudes using Eq. \ref{eqn:absmag_cal} in both the passbands. We further retained only those stars for which $|M_{\lambda} - \Bar{M_{\lambda}}| < \sigma_{\Bar{M_{G}}}$, i.e. where the absolute magnitudes were consistent with literature values of the absolute magnitudes of RC stars to within $1\sigma$. For this sample we could then compute the \textit{Pearson} correlation coefficient between the predicted (and observed) apparent magnitudes (see Appendix \ref{app:intrinsic_corr} which turns out to be close to 0.4. This would suggest that the overall correlation has to be between 0.4 \& 1.0, and we thus assumed a naive average of \pearson{}=0.7.

For a $N_{\rm dim}$ (=2 here) dataset, one can then write a likelihood function for every source in terms of the Mahalanobis distance as
\begin{equation}
\label{eqn:likelihood_mldist}
    ln (L) = -\frac{1}{2} ln (|\Sigma|) - \frac{1}{2}D_{ml}^{2} - \frac{N_{\rm dim}}{2}ln(2\pi),
\end{equation} where $|\Sigma|$ is the determinant of the covariance matrix. To select a source as an RC star, we require $ln (L) > ln (L)_{\rm th}$, where the threshold $(L)_{\rm th}$ is set by considering a hypothetical RC star placed exactly at \mldist{}=3 that is observationally `perfect' in the sense that $\varsigma_{g}^{2}
=  \sigma_{\Bar{M_{G}}}^{2}$ (and likewise for W1). Then using Eq. \ref{eqn:likelihood_mldist} we derive $ln (L)_{\rm th}$. 

Using the procedure above for each star in the \gaiawise{} dataset we find a total number of 9,761,294 \rc{} stars if we neglect the correlation (i.e. \pearson{}=0) in Eq. \ref{eqn:covariance_matrix}, and instead 10,269,109 \rc{} stars with the correlation taken into account (\pearson{}=0.7). Hereafter \gaiawise{}[RC]  refers to the latter case unless otherwise specified. We find that over 97\% of our sources have an Astrometric Fidelity (AF) $>$0.5, which is typically considered to be a threshold for good quality astrometry \citep{Rybizki:2020}.

\subsection{Coordinate system}
\label{sec:coord_trans}
\begin{figure}
\includegraphics[width=1.\columnwidth]{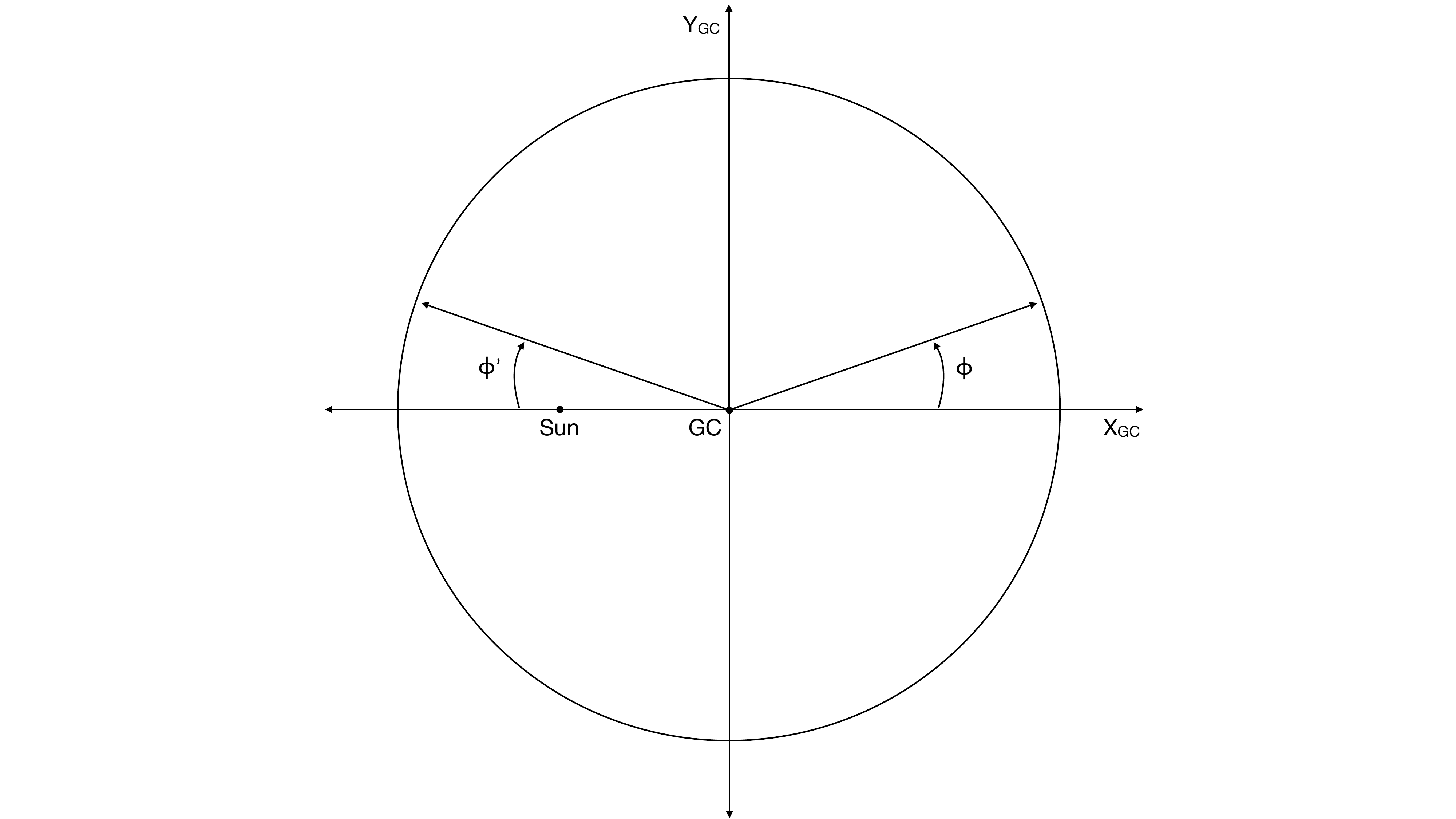} 
\caption{Coordinate system adopted in this paper.
} \label{fig:coordsys}
\end{figure}

Our adopted coordinate system is illustrated in Fig. \ref{fig:coordsys}. We first transformed our $(l,b)$ and distance, $d$, coordinates to heliocentric Cartesian coordinates, with the $X$-axis pointing towards the Galactic centre: 
\begin{equation}
 \begin{pmatrix}
X_{\rm hc} \\ 
Y_{\rm hc} \\ 
Z_{\rm hc}
\end{pmatrix}  = 
d \begin{pmatrix}
\cos b \cos l    \\ 
\cos b \sin l     \\ 
\sin b 
\end{pmatrix}.
\end{equation}
To transform these heliocentric coordinates to galactocentric Cartesian coordinates, we assumed $R_\odot = 8277 \pm 9$(stat) $\pm 30$(sys) pc from the ESO Gravity project's most recent measurement of the orbit of the star S2 around the Milky Way's supermassive black hole \citep{GravityCollaboration:2021}. The heliocentric Cartesian frame is then related to galactocentric one by a simple translation: $X_{\rm hc}= $\, \xgc{} $ +\, R_{\odot}$, $Y_{\rm hc}= $ \ygc{} and $Z_{\rm hc}=$ \zgc{} $-$ \Zsun{}. 
The galactocentric cylindrical radius $R$ is trivially found from $\sqrt{X_{GC}^2 + Y_{GC}^2}$, while the cylindrical coordinate angle $\phi={\rm tan}^{-1}(Y_{GC}/X_{GC})$ increases in the anti-clockwise direction, while the rotation of the Galaxy is clockwise, as seen from the north. The height of the Sun above the Galactic plane is adopted as \Zsun{}= 0 pc \citep[][hereafter Gaia23]{drimmel2023gaia}, though we found that our results remain unchanged even when we used the alternative 
\Zsun{}$=27$ pc \citep{Bennett:2019}.The uncertainties in our transformed coordinates are dominated by the uncertainties in distance estimates. We do not propagate these through the coordinate transformation, but instead generate multiple realisations of the data, each time sampling distances from the underlying distribution (see Sect. \ref{sec:RCdist}).

\subsection{3D Extinction maps}
\label{sec:dustmodel}
In order to compute the absolute magnitude, we need good estimates of the extinction A$_{\lambda}$ in passband ${\lambda}$. The widely used maps of reddening, $E(B-V)$, from \citet[][hereafter \schlegel{}]{Schlegel1998}, estimates the extinction at infinity given a star's 2D ($l,b$) sky coordinates. While these dust maps cover the entire sky, the 2D extinction values will overestimate the reddening for sources that are not outside the ISM layer. This can be improved upon by using 3D dust maps that require a prior for distances. All stars in our sample have a measured parallax, so in principle one could use this as a prior to estimate the distance, but as discussed earlier, instead we make use of the distances from \cbj{}, as our purpose here is only to use these initial distance estimates to arrive at a realistic first-estimate of the foreground extinction along a given line of sight. For the selected sample of \rc{} stars, we derived the distances independently, as described in Sect. \ref{sec:RCdist}.
 
At the time of running our algorithm, we had access to the following publicly available 3D dust maps: a) The \cite{Lallement:2019} map\footnote{\url{https://astro.acri-st.fr/gaia_dev/about}}, derived with \gaia{} and \twomass{} photometry and \gaia\ astrometry. This map provides the \gaia\ extinction parameter $A_{0}$, defined as the monochromatic extinction at 541.4 nm at a resolution of 5 parsecs within 3 kpc from the Sun and within 0.5 kpc perpendicular to the plane.
b) Further out, we make use of the \bayestar{} 3D extinction map by \cite{Green:2019}. This is based on a Bayesian scheme, combining \gdrtwo{} parallax with photometry from \twomass{} and \panstars{} surveys. The spatial coverage is limited to the sky north of declination ($dec$) of $-30^\circ$.  For each source, we multiply the \bayestar{} map output by a multiplicative factor to obtain the reddening, i.e. $E(B-V) = 0.884 \times E(Bayestar19)$, as recommended\footnote{\url{http://argonaut.skymaps.info/usage}}. c) For the remainder of the sky, we return to the \schlegel{} map, but in order to correct for the overestimation in the extinction, we estimate a dust fraction factor (following \cite{Koppelman:2021}) as

 \begin{equation}
     \frac{E_{B-V}(l,b,s)}{E_{B-V,\infty}(l,b)} =  \frac{\int_{0}^s \rho[x(s')]ds'}{\int_{0}^\infty \rho[x(s')]ds'} ,
 \end{equation} where $E_{B-V,\infty}(l,b)$ is the extinction estimate from \schlegel{}, $s$ is the heliocentric distance to a source, and $x$ is the position vector. We note that $\rho(R,z)$ is the dust density model, which we adopted in this case from \cite{Sharma:2011}:

\begin{equation}
\rho(R,z) \propto \exp\biggl(\frac{R_{\odot} - R}{h_{R}} - \frac{|z - z_{warp}|}{k_{\rm fl}h_{z}} \biggr), \\ 
\end{equation} with the disc warping and flaring (fl) modelled as
\begin{equation}
\begin{aligned}
k_{fl}(R) &= 1 +  \gamma_{fl}\text{Max}(0,R-R_{fl}) \\
z_{warp}(R,\phi) &= \gamma_{warp}\text{Max}(0,R -R_{warp})sin(\phi).
\end{aligned}
\end{equation} The parameters of the model are from \cite{Robin:2003} and listed in \autoref{tab:dustmodel}. In the \allwise{} bands, we adopted the following reddening coefficients: $A_{W1}/E_{(B-V)}=0.174$ and $A_{W2}/E_{(B-V)}=0.107$ \citep{Sharma:2011}. Using the \textit{dustapprox} code, we applied the mid-IR extinction curve from \citet{Gordon:2023} and adopted $R(V)=3.1$ to compute the ratios, $A_{W1}/A_{G}=0.07$, $A_{W1}/A_{Bp}=0.05$, and $A_{W1}/A_{Rp}=0.09$, which is appropriate for a typical red clump star ($4200< $\teff{}$<5400$ K, $1.8< $\logg{}$<3.2$). We find that these ratios are insensitive to \feh{}.

\begin{table}
\centering
\caption{Dust model parameters. \label{tab:dustmodel}}
\begin{tabular}{|l|l|l|}
\hline
Parameter  & Value & Unit  \\
\hline
\textit{$h_{R}$}  & 4.2 & kpc \\
\textit{$h_{z}$}  & 0.088 & kpc\\
\textit{$\gamma_{fl}$}  &  0.0054 & kpc$^{-1}$\\
\textit{$\gamma_{warp}$}  &  0.18 &kpc$^{-1}$\\
\textit{R$_{fl}$} &  1.12*\Rsun &kpc \\
\textit{R$_{warp}$} & 8.4 &kpc \\
\hline
\end{tabular}
\tablefoot{Parameters for the dust model adopted in this paper, with values taken from \cite{Robin:2003} }
\end{table}

\subsection{Red clump distance and uncertainty}
\label{sec:RCdist}

\begin{figure}
\includegraphics[width=1.\columnwidth]{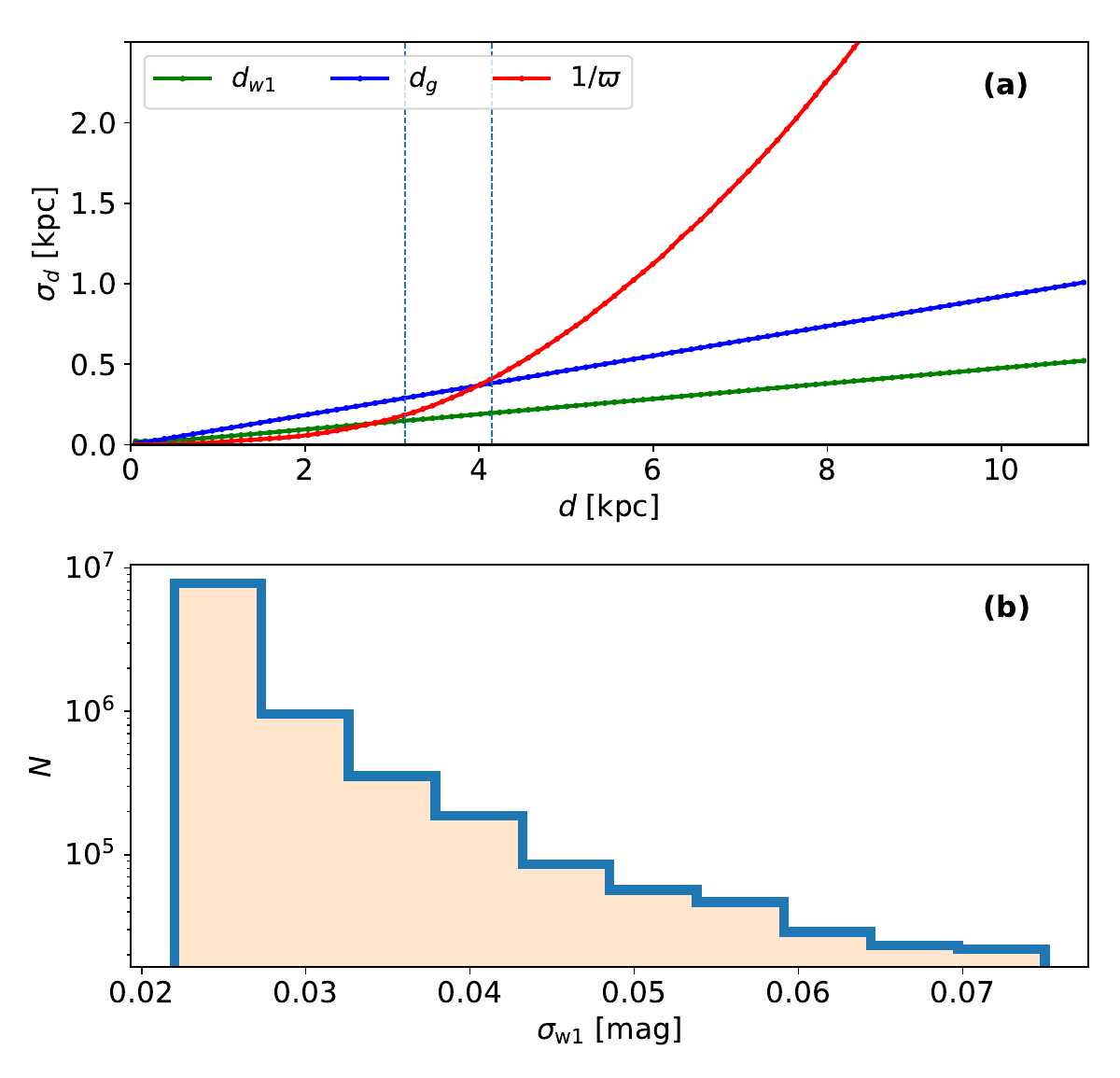} 
\caption{Panel (a): Typical distance uncertainty for the red clump as a function of distance. The red curve shows the expectation from a naive inverse parallax estimation, the blue curve shows the predicted uncertainties in the photometric distances in $G$ band and for $W1$ this is shown by the green curve. The two vertical dotted lines indicate roughly the distances beyond which the photometric distances become more informative than parallax inversion for the two bands. Panel (b): Distribution of photometric uncertainty in $W1$ for the \rc{} stars shows that $\sigma_{W1}<<0.1$, i.e. lower than the intrinsic dispersion in the absolute magnitude of \rc{} stars in $W1$.} \label{fig:distunc}
\end{figure}

For every star classified as \rc{} and assumed as having an absolute magnitude $M_{\lambda}$ as per table \ref{tab:tab_extinct}, we can invert Eq. \ref{eqn:absmag_cal} to calculate the distance modulus, but since the extinction \alambda{} itself depends on the distance, we follow an iterative procedure instead. First, we derived the maximum possible distance modulus for a star, assuming zero extinction along the line of sight,
\begin{equation}
\label{eqn:mumax_iterative}
    \mu\_max_{i} = m_{\lambda} - M_{\lambda}.
\end{equation} Using this equation, we could in turn derive the maximum distance ($d_{max}$), then compute the extinction, $A_{\lambda} (l,b,d_{max})_{i}$, and ultimately compute an updated distance modulus,

\begin{equation}
\label{eqn:mumax_dust}
    \mu_{i} = m_{\lambda} - M_{\lambda} - A_{\lambda} (l,b,d)_{i}.
\end{equation} 
Using $\mu_{i}$ we then recompute $A_{\lambda} (l,b,d)_{i}$, and repeat the procedure until the distance modulus is converged (typically within five iterations) to $\Bar{\mu_{\lambda}}$. For every star we sample the absolute magnitude from a \gaussian{} $\mathcal{N}$($\Bar{M_{\lambda}}$,$\sigma_{\Bar{M_{\lambda}}}$), using the values listed in \autoref{tab:tab_extinct}, to derive the PDF of the distance modulus.

From Eq. \ref{eqn:mumax_iterative}, we can consider two main sources of uncertainty in our distance estimate, a) $\sigma_{\Bar{M_{\lambda}}}$: The intrinsic dispersion in the absolute magnitude, and b) $\sigma_{m_{\lambda}}$: The photometric uncertainty. From \autoref{tab:tab_extinct} we see that $\sigma_{\Bar{M_{\lambda}}}$ is twice as large in the $G$ band (0.2mag) compared to that in $W1$ (0.1mag). For this reason, we decided to use only the distances based on $W1$ passband in order to minimise the distance uncertainty and the effects of extinction. As for photometric uncertainty, we find that about 99.5\% of our \rc{} sample has $\sigma_{W1}<0.1$. In fact Fig. \ref{fig:distunc}(b) shows that the overall $\sigma_{W1}<<0.1$, i.e. the photometric uncertainties are generally much smaller compared to the intrinsic dispersion in the absolute magnitude in $W1$. We exclude the remaining 0.5\% of the sources that have $\sigma_{W1}>0.1$, though given their insignificant number, their removal or not makes no noticeable difference in the results. For every star, we combined the two sources of uncertainty in quadrature to derive the uncertainty in the distance modulus as 
\begin{equation}
    \sigma_{\mu(W1)} = \sqrt{\sigma_{m_{W1}}^2 + \sigma_{\Bar{M_{W1}}}^2 }.
\end{equation}
Strictly, one should also consider the uncertainties in the extinction maps, but since in this contribution we use a combination of various maps, this is not a trivial exercise, and we ignore it for simplicity. However, since we use primarily the distances based on $W1$, we anticipate this to be a small effect. We assume the uncertainties in the distance modulus are \gaussian{}, i.e. $\mathcal{N}$($\Bar{\mu_{\lambda}},\sigma_{\mu(W1)})$. It can be shown then that the typical uncertainty in the photometric distance ($d$) is, $\sigma_{d_{RC, \lambda}} = 0.2 \ln(10) \sigma_{\mu_{\lambda}}  d $,
while a naive inversion of the parallax results in distance error given by $\sigma_{d_{\varpi}} = \sigma_{\varpi}d^{2}$, where $\sigma_{\varpi}$ is the parallax error.
Figure \ref{fig:distunc}(a) compares the profiles of the distance uncertainty with respect to distance from the two approaches, showing that beyond a heliocentric distance of about 3 kpc, the \rc{} distances are superior.

\subsection{RC catalogue and validation}
\label{sec:rc_catalogue}

\begin{figure}
\includegraphics[width=1.\columnwidth]{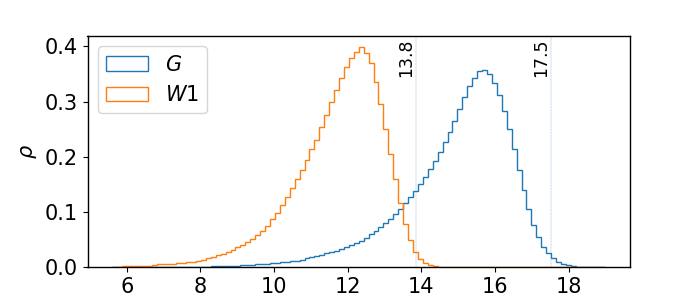} 
\caption{Apparent magnitude distribution of the \gaiawise{}[RC] sample in the $G$ and $W1$ bands. The vertical dashed lines at $G$=17.5 and $W1$=13.8 indicate the $99.5^{th}$ percentiles in the respective bands.} \label{fig:magdist}
\end{figure}

\begin{figure}
\includegraphics[width=1.\columnwidth]{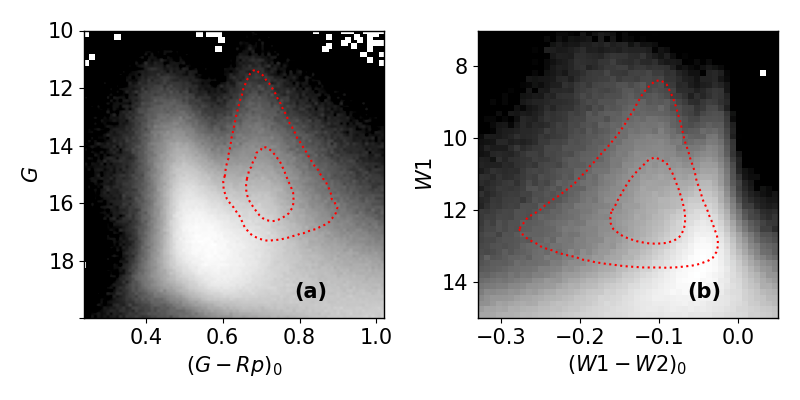} 
\caption{Colour magnitude diagram for the \gaiawise{} parent catalogue shown in gray for \gaia{} and \allwise{} colours. In each panel, we indicate the selected RC sample using the red contours ($1\sigma,2\sigma$).} \label{fig:cmd}
\end{figure}

\begin{figure*}
\includegraphics[width=2.\columnwidth]{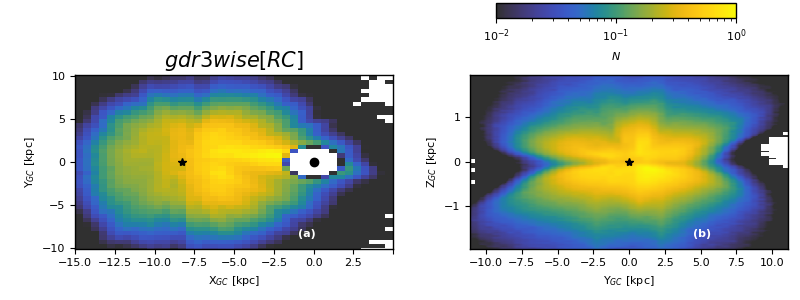} 
\caption{Spatial distribution of the \gaiawise{}[RC] sample for the region considered for model fitting. Panel (a) shows the galactocentric \xgc{}-\ygc{} projection, with the location of the Sun (black star) and the Galactic centre (black dot) also indicated. Stars within \rgal{} $<3kpc$ of the Galactic centre have been removed.  Panel (b) shows the corresponding galactocentric \ygc{}-\zgal{} projection. The number density is shown on a logarithmic scale.} \label{fig:data_sky_xy}
\end{figure*}

\begin{figure*}
\includegraphics[width=2.\columnwidth]{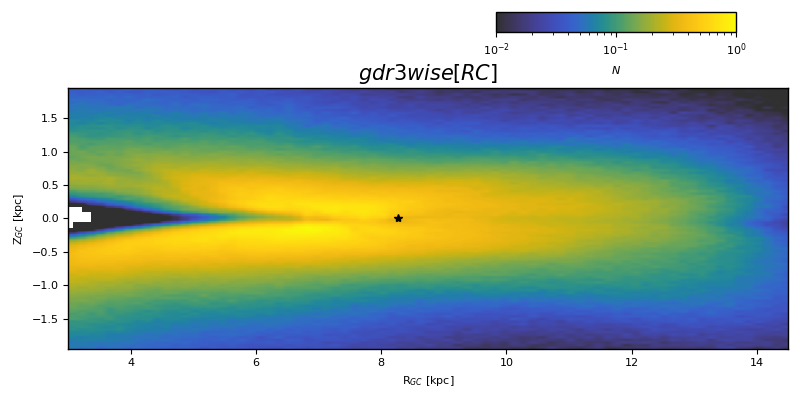} 
\caption{Galactocentric \rgal{}-\zgal{} projection for the \gaiawise{}[RC] sample over the region considered for our model fitting. The number density is shown on a logarithmic scale. The location of the Sun is indicated as a black star.} \label{fig:data_sky_RZ}
\end{figure*}

\begin{figure}
\includegraphics[width=1.\columnwidth]{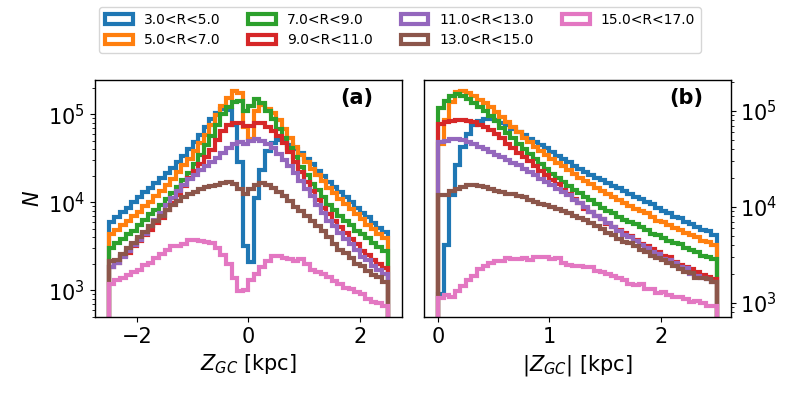} 
\caption{Vertical counts of \gaiawise{}[RC] for successive annuli in \rgal{} between 3 to 15 kpc. Panel (a) shows the counts as a function of \zgal{}, while panel(b) shows the same folded along the vertical axis, i.e. $|$\zgal{}$|$. In panel (b) a change of slope in the vertical counts is evident as we move from the inner to the outer disc region.} \label{fig:data_zprofiles_flare}
\end{figure}

In Fig. \ref{fig:allsky}, we show the full sky distribution ($l,b$) at healpix level 5, for the \gaiawise{} crossmatch (left panel), and for the selected \rc{} candidates (right panel). The apparent magnitude distribution of the selected \rc{} sample is shown in Fig. \ref{fig:magdist}, where we note the faint limits (99.5 percentile) at $m_{G,lim}$=17.5, and $m_{W1,lim}$=13.8. We also note that \allwise{} is essentially complete in this magnitude range \cite[see their Fig. 4]{Schlafly:2019}. We show in Fig. \ref{fig:cmd}, the colour-magnitude distribution (grey) for the \gaiawise{} parent sample in both \gaia{} and \allwise{} colours. In each panel, the bifurcation between Giants and dwarfs is apparent, and indicated in red contours is the region occupied by our selected \rc{} sample.

Most of the selected sources are confined within $|b| < 10^\circ$ of the plane, but our algorithm also picked out sources around the Large (LMC) and the Small (SMC) Magellanic clouds. The LMC ($RA,DEC = 81.28,-69.78$) is known be at a distance of about 50 kpc from the Sun \citep{Pietrzynski:2019}. Along this line of sight, we estimate extinction of $A_{W1}=0.15$, and $A_{G}=2.38$. From Eq. \ref{eqn:absmag_cal}, we would then expect the brightest \rc{} in the LMC to be visible at $G=21.3$ or $W1=16.9$, which lie outside the magnitude distribution of our \rc{} candidates. Instead, these sources enter the selection because the \cbj{} catalogue has no priors for LMC/SMC stars, and since the relative parallax uncertainty is high for these sources, they are assigned incorrect distances. Our goal is to primarily select \rc{} stars in the Milky Way disc, so for density modelling, we restrict our final sample to the region $|$\zgal{}$|<$2 kpc which also removes the contamination at $b > -30^\circ$ from the Magellanic clouds, as in Fig. \ref{fig:allsky}. Though we note that the results remain changed even without this cut, because of the very low relative contamination from these stars to the total number of about ten million.

In Fig. \ref{fig:data_sky_xy}(a) we show the density maps for this dataset in the galactocentric  \xgc{}-\ygc{} projection, showing clearly the region in the inner Galaxy (\rgal{}$<$3 kpc) that we mask. Our sample provides a good coverage between $3<$\rgal{}$<14$ kpc.
Figure \ref{fig:data_sky_xy}(b) shows the corresponding \ygc{}-\zgal{} projection to show the extent of the coverage vertically with regard to the midplane of the disc. This is also illustrated in Fig. \ref{fig:data_sky_RZ}, where it is clear that most of the sample lies within $|$\zgal{}$|<1.5$ kpc, and that towards the outer disc (\rgal{} $>10$ kpc) there is a hint of flaring in the vertical counts. To see this more clearly, in Fig. \ref{fig:data_zprofiles_flare}(a) we plot the number counts as a function of \zgal{} at successive \rgal{} annuli, spanning from 3 to 17 kpc. The slope of the profiles is a measure of the scale height of the disc at a particular annulus, while a change in the slope from the inner to the outer disc would indicate a change in the scale height, thus flaring if the scale height increases with \rgal{}.
Figure \ref{fig:data_zprofiles_flare}(b) shows the vertical number counts folded along the \zgal{} axis (i.e. $|$\zgal{}$|$), which makes it easier to spot the changing slope with \rgal{}, in particular beyond \rgal{}$=11$
kpc. We remind the reader, however, that Fig. \ref{fig:data_zprofiles_flare} shows the vertical counts in the data without accounting for the selection function, and therefore fitting a model (for example an exponential) directly to this figure is not appropriate. Here we are simply presenting the data for a visual inspection, and note the hints of a flared disc.

Any scheme for the selection of the \rc{} based purely on astrometry and photometry is bound to have contamination from neighbouring populations. This is mainly due to limitations such as the quality of extinction maps, and uncertainty in photometry and parallax. Also, in the case of the \rc{}, intrinsic colour and magnitude are not enough to separate such stars from red giant branch (RGB) stars occupying the same part of the HRD. In such cases, however, while not true \rc{} stars, their true photometric distances will not be too different than those of the true \rc{}. Since we are able to select a very large sample of \rc{} candidates, we carry out a comparison with a few spectroscopic (\glh{} and \apg{}), and asteroseismic catalogues for validation in Sect. \ref{sec:rc_validation}. For the stars that are found in common with these catalogues, we find that their distribution in the \kiel{} (Fig. \ref{fig:kiel_apogee_galah}), colour-magnitude (Fig. \ref{fig:cmd_ovplot}), and colour-absolute-magnitude diagrams (Fig. \ref{fig:camd}, is reasonable for our \rc{} sample.

\section{Method}
\label{sec:method}
\subsection{Modelling the Galactic disc}
\label{sec:model_descr}
The number density distribution, $N(R,z)$, in our model is described by two exponential discs, each of the form

\begin{equation}
    N(R,z) \propto exp\left(-\frac{R}{R_{d}} -\frac{R_{cut}}{R} -\frac{|z-z_{w}|}{h_{z}}\right), 
 \end{equation} with scale height (\hz{}), scale length (\rd{}), and an inner-cut radius (\rcut{}), which excludes the region not described by the disc, such as in the inner Galaxy \citep{Aumer:2017,agamaref18}. The disc is exponentially flared, i.e. \hz{} increases with \rgal{}, controlled by the flare scale length, \rflare{}, 
\begin{equation}
    h_{z} = h_{z,\odot}*exp\left(\frac{R - R_{\odot}}{R_{fl}} \right),
\end{equation} with scale height normalisation, $h_{z,\odot}$, at the solar radius \Rsun{}. The total density is thus given by
\begin{equation}
     N(R,z) \propto f_{d1} \times N(R,z)_{1}  + (1 - f_{d1}) \times N(R,z)_{2},
\end{equation} with \fdisc{} being the mass contribution of the first disc. The density function is normalised such that over the observed volume $\int_{V} \text{\fsel}\times N(R,z) dV= N_{observed}$, where the fraction of observable stars is in the range of \fsel{}$=[0,1]$ (see Sect. \ref{sec:sfdesc}). We also considered a warped disc in one of the components. This we parametrised as 
\begin{equation}
    z_{w} = h_{w0} \times(R - R_{w})^{a_{w}}  sin(\phi^{'} - \phi_{w}),  \\
\end{equation} where the onset of the warp is at \rwarp{} (i.e. \zwarp$=0$ inside this radius), and its amplitude is set by \hwarp{}. The warp is sinusoidal such that the line of nodes lies along \phiprime{}=\phiwarp{}, where, \phiprime{}= $tan ^{-1}  \left(\frac{Y_{GC}}{-X_{GC}}\right)$, being zero towards the anticenter and increasing in the direction of Galactic rotation. We distinguish this from the galactic azimuth, $\phi = tan ^{-1} (Y_{GC}/X_{GC})$, which instead increases anti-clockwise (see Fig. \ref{fig:coordsys}). We do so to both be consistent with standard coordinate frame conventions, as well as those preferred by the warp modelling papers.

\begin{figure}
\includegraphics[width=1.\columnwidth]{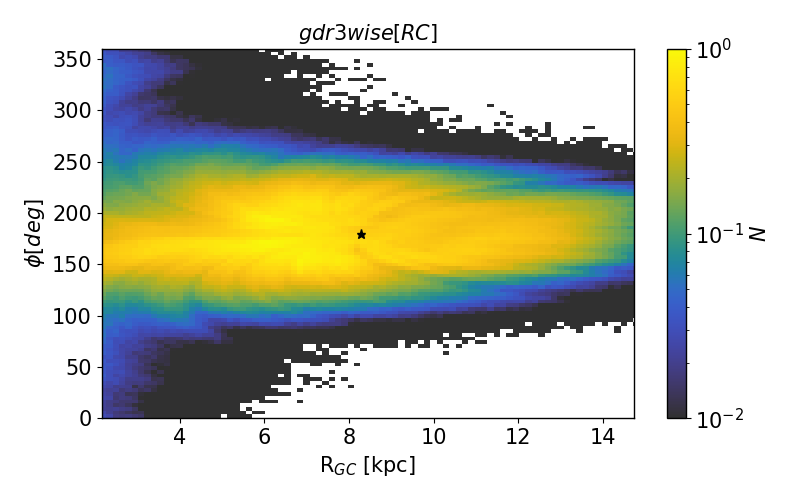} 
\caption{Spatial distribution in the $\phi-$\rgal{} projection of the observational red clump sample constructed from the \gaiawise{} parent dataset. The range shown here is the region over which we performed our model fitting. } \label{fig:data_sky_phiR}
\end{figure}

The Milky Way is known to host a bar, and over the past several years, the range of the bar length ($3 < R_{b} <5$ kpc) has been estimated using various tracers (Gaia23 and references within).  For our purpose we retained as much of the disc region as possible while removing the probable regions most affected due to the central bar. Recently, \cite{Vislosky:2024} estimated a value of $R_{b}=3.6$ kpc but suggest that it could be as low as 3 kpc. For this reason, we fixed the inner-cut radius, \rcut{}$=3$ kpc. Our tests showed that increasing \rcut{} to 3.5 kpc barely impacted the results, so we chose to stick to the lower bound of $R_{b}$. For our \rc{} sample, the $\phi-$\rgal{} projection is shown in Fig. \ref{fig:data_sky_phiR}. Our full model fits for a maximum of 10 parameters. These are the scale parameters of the first disc (\rd{}, \hzsun{}, \rflare{}), and its mass fraction (\fdisc{}), and four parameters describing the warp (\phiwarp{}, \rwarp{}, \awarp, \hwarp{}), and the  scale parameters of the second disc (\rdtwo, \hztwo). We only allow the first disc to be warped and flared. Our initial exploration of the observational data showed no major difference if both disc components were allowed to be both flared and warped, so in order to reduce the parameter space and degeneracy between parameters, we imposed this condition. All the parameters we fit for are listed in \autoref{tab:galmodel}, along with their bounds and units. 

\begin{table}
\begin{threeparttable}
\flushleft
\caption{Description of Galactic structure parameters. \label{tab:galmodel}}
\small
\begin{tabular}{|l |l |p{0.25\textwidth}|}
\hline
\textbf{Parameter}  & \textbf{Range}  & \textbf{Description}  \\
\hline
\rd{} & $[1,5]$ kpc & scale length of disc 1.\\
\hzsun{} &  $[0.1,1.5]$ kpc & scale height of disc 1 at \Rsun{}.\\
$log_{10}$ \rflare{}  & $[-3,3]$ & Logarithm of flare scale length (kpc) for disc 1. \\
\fdisc{} & $[0.05,1.]$ & Mass fraction of disc 1. \\
\rdtwo{} & $[1,5]$ kpc & scale length of disc 2. \\
\hztwo{} & $[0.1,1.5]$ kpc & scale height of disc 2 at \Rsun{}.\\
\hline
\multicolumn{3}{|l|}{Warp Parameters (disc 1 only):}  \\
\phiwarp{} & $[90^\circ, 270^\circ]$ &  \phiprime{} of the line of nodes. \\
\rwarp{} & $[1,15]$ kpc & radial onset of the warp. \\
\awarp & $[0,5]$ &  radial exponent of the warp.\\
\hwarp{} & $[0,5]$ kpc & amplitude of the warp. \\
\hline
\end{tabular}
\end{threeparttable}
\tablefoot{Parameters to be fitted and the range of values allowed. The first set of parameters describe the structure of the disc, such as the scale length, scale height and flare. The four parameters that describe the warp are listed separately at the bottom.}
\end{table}

\subsection{Selection function}
\label{sec:sfdesc}

\begin{figure}
\includegraphics[width=1.\columnwidth]{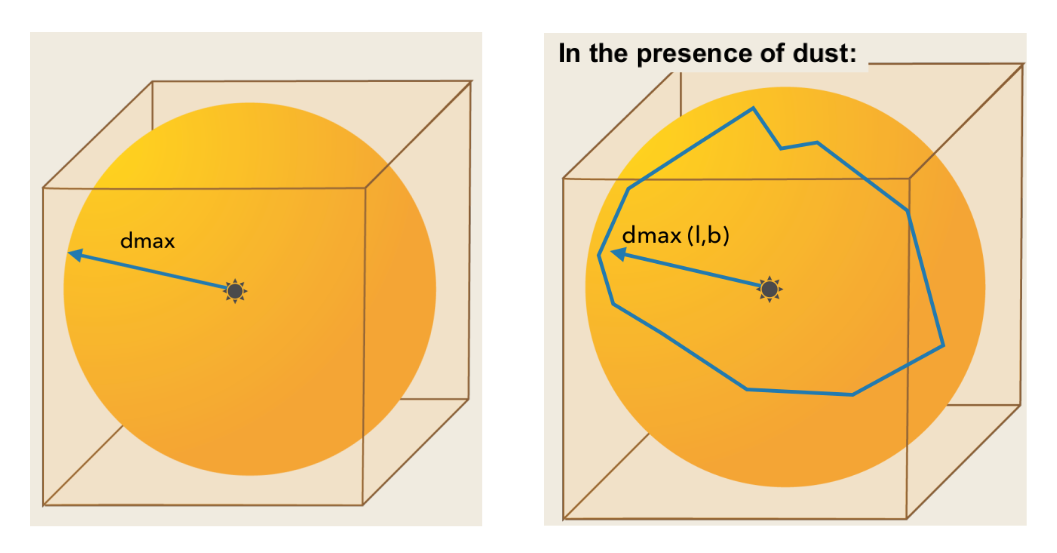} 
\caption{Illustration of the maximum observable distance ($d_{\rm max}$) for the \rc{} for a magnitude-limited survey. The left panel shows the ideal situation when there is no dust extinction, in which case the observable region traces out a sphere with radius $d_{\rm max}$. On the right is shown the case with extinction, which depends on the sky position ($l,b$) due to which $d_{\rm max}(l,b)$ varies across the sky and modifies the effective volume from that of a perfect sphere.} \label{fig:dmax_cartoon}
\end{figure}

Here, we discuss the various layers of the selection function that need to be accounted for in our modelling of the observational data. If we consider stars in a 3D grid in space, in this case in \xgc{} $\times$ \ygc{} $\times$ \zgc{}, our goal is to be able to predict the number counts in each voxel, $i$: 
\begin{equation}
    N_{i} = N_{i,true} \times S_i,
\end{equation} where $S_i[0,1]$ is the correction factor (selection function) to the counts predicted by the model, $N_{i,true}$. In the ideal case, $S_i$ would be equal to 1 if we are able to observe the entire ground truth, that is, all stars of our modelled population in a given voxel, $i$. In the worst case, it would be 0 if there was missing or poor quality data in some patches of the sky but also if the survey is not designed to observe certain fields. In our specific case, $S_i$ is the fraction of the \rc{} stars in voxel $i$ that can be seen  for our sample. However, we point out that in general, the selection function is dependent on observable quantities \citep{rixbovy13,frankel20}, and here we can only define an $S_i$ for a given voxel, $i$, and interpret it as the fraction of stars in that voxel because we are dealing with a standard candle, which allows us to translate the observed apparent magnitude into a distance once we assume an extinction map. We can set out two principle layers that determine $S_i$: 

\paragraph{a.) Effective volume:} For a magnitude limited survey, (\appmag{} $<$ \appmaglim{}), for a quasi-fixed absolute magnitude, such as is the case with the \rc{}, we can define a maximum distance modulus ($\mu$), 
\begin{equation}
    \mu\_max = m_{\lambda,lim} - M_{\lambda},
\end{equation} such that from our position at the Sun, the observable volume would be traced by a sphere with a radius ($d_{\rm max}$) corresponding to this maximum distance modulus. However, here we neglect the presence of interstellar dust, which is an unavoidable evil in the Galaxy. When accounting for it, the maximum distance modulus $\mu\_max$ is no longer a constant and instead varies for each voxel $i$:
\begin{equation}
\label{eqn:mumax_dust}
    \mu\_max_{i} = m_{\lambda,lim} - M_{\lambda} - A_{\lambda} (l,b,d)_{i},
\end{equation} 
Consequently, the observable or effective volume is crumpled and no longer a simple sphere, as illustrated in Fig. \ref{fig:dmax_cartoon}. Essentially, this is similar to integrating a density function over the whole sky, but with the limits on distance varying by line of sight. Hence, depending on the dust distribution, the limit of the volume sampled varies in different directions on the sky. Along some lines-of-sight one can only map to a few kiloparsecs, while along others it is possible to map out the edge of the Galactic disc. Thus we need to take into account this dust/\appmaglim{} induced distance limit.   

Since we are modelling the number counts on a fixed 3D grid, we must also take into account the finite size of our voxels. We can pre-compute the fractional effective volume, \fseli{}, for each volume element (ie. voxel), and then apply it to both simulated and observational data. For example, let us consider a cube shaped grid (\xgc{} $\times$ \ygc{} $\times$ \zgc{}), where every voxel has the same size, and is small enough such that the median heliocentric distance of all stars in that voxel can be approximated by the median heliocentric distance of that voxel. Then for every voxel ($l,b,d_{i}$) we can check if $d_{i}< d_{\rm max}(l,b)$ ( $d_{\rm max}$ condition), and retain only those that satisfy this condition. However, for nearby voxels the size of the voxel may be significant with respect to its heliocentric distance, or the extinction may vary significantly within a voxel. 
In this case the median quantities of the voxel ($l_{i},b_{i},d_{i}$) are not representative of the median quantities of the stars within it. This becomes more of an issue if the volume elements of the adopted grid do not have a fixed size. In this work, because the number of stars decrease exponentially with galactocentric radius, to increase the statistical significance of the counts per voxel in the outer disk, we adopt a cylindrical grid (\rgal{},$\phi$,\zgal{}), so that the volume of the voxels increase with \rgal{}, fixing the cylindrical grid to the binning ($\Delta$\rgal{},$\Delta\phi$,$\Delta$\zgal{})$= (0.25 \text{ kpc},10^\circ,0.25 \text{ kpc})$, from $R_{min}$=3 to 20 \kpc in radius, 0 to $360^\circ$ in $\phi$, and between $\pm2\kpc$ in \zgal{}.  

\begin{figure}
\includegraphics[width=1.\columnwidth]{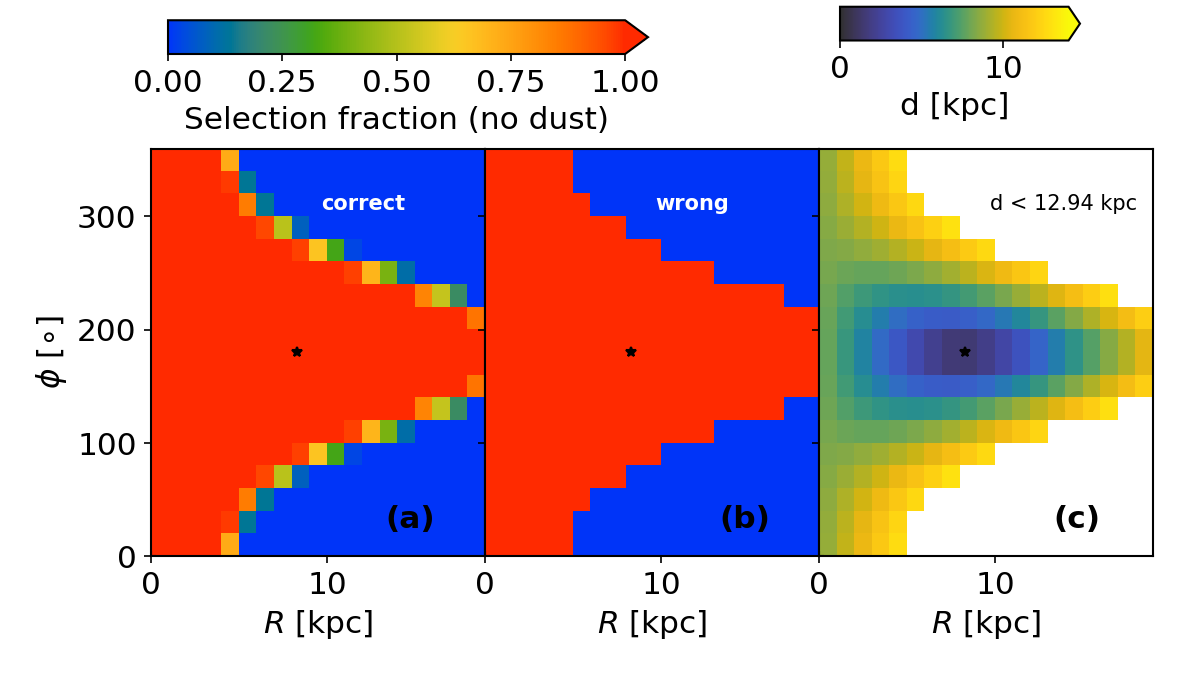} 
\caption{Illustration of the selection function projected in $\phi-$\rgal{} space for the case without dust at \zgal{}=0 kpc (midplane). Panel (a) shows the map of the selection fraction computed using the sub-binning method, panel (b) shows the same for the method without sub-binning, while panel (c) shows the distance to all voxels that are within a distance of $d_{max} < 12.94 \kpc$ of the Sun. Panel (a) also shows the additional voxels that would be missed by assuming the median values of observables ($l,b,G$). In all panels, the location of the Sun is indicated by the black star.} \label{fig:sf_illus_nodust}
\end{figure}

As an example let's first consider such a cylindrical grid where for every voxel we compute the heliocentric distance at its centre. 
Then we assume a single absolute magnitude \absg{}$=0.44$ for our RC stars, so that for a given magnitude limit (say $G=16$), we can estimate the $d_{\rm max}$. In Fig. \ref{fig:sf_illus_nodust} we consider the case without extinction, so $d_{\rm max}$ is a constant (12.94 kpc), and at \zgal=0 kpc, i.e. right in the midplane.
Figure \ref{fig:sf_illus_nodust}(a-c) show the $\phi$ versus \rgal{} projection of the voxels, mapped by different quantities.
Figure \ref{fig:sf_illus_nodust}(c) shows the projection mapped onto heliocentric distance, 
only for those voxels that satisfy $d_{i}< d_{\rm max}$ (using the centre of the voxel). So it follows that in Fig. \ref{fig:sf_illus_nodust}(b), all voxels outside of this condition have \fseli{}
set equal to 0; that is, no RC stars in this volume element will be in our magnitude limited sample. In contrast, those inside have \fseli{} set equal to 1; that is, the selection function resulting from the magnitude limit is binary, as expected.

\begin{figure}
\includegraphics[width=1.\columnwidth]{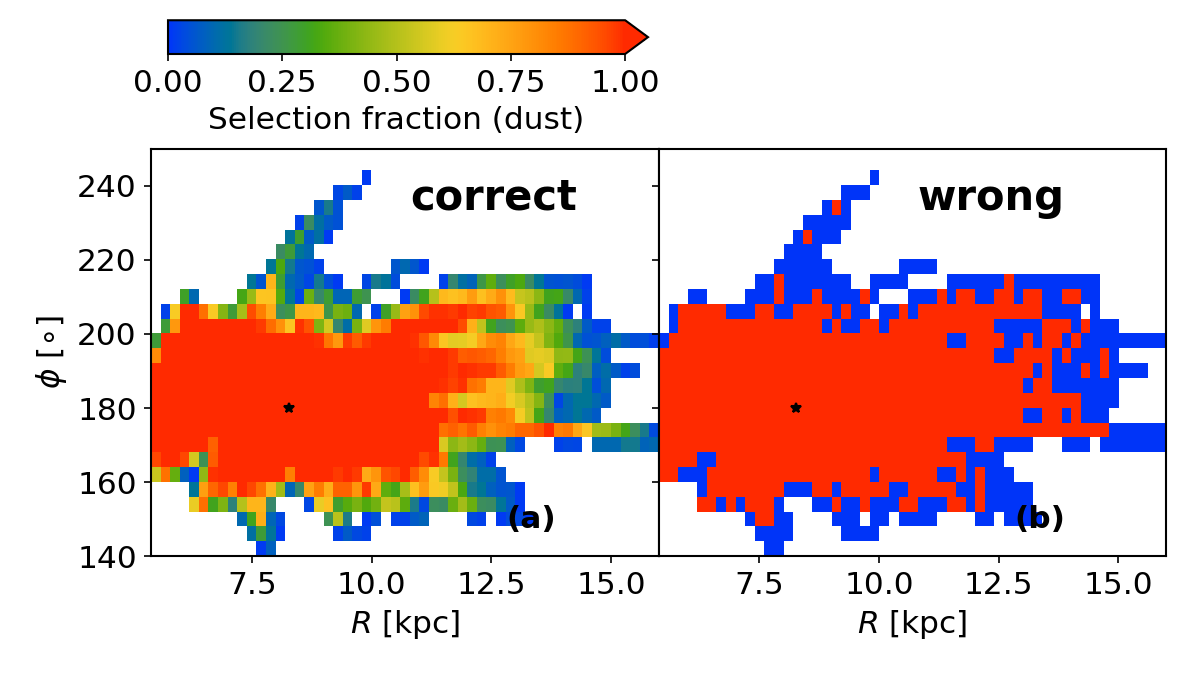} 
\caption{Illustration of the effective volume projected in $\phi-$\rgal{} space for the case with dust. Panel (a) shows the map of the selection fraction computed using the sub-binning method, and it shows a continuum of probability between 0 (unobservable) and 1 (fully observable). In panel (b), we show the same voxels as in panel (a), but for these the selection fraction is shown for the method without sub-binning. This shows that several voxels from panel (a) that have a probability of being observed would be considered unobservable or would be completely observed using the incorrect method of estimating the selection function. In both panels, the location of the Sun is indicated by the black asterisk (*). } \label{fig:sf_illus_dust}
\end{figure}

To check if this approximation is valid, we sub-bin our 3D cylindrical grid by a factor of 5 in each dimension (125 subvoxels) and then in each voxel we can count the number of sub-voxels that satisfy the $d_{\rm max}$ condition, to now obtain a fractional selection function ($N_{d < \rm dmax}/125$ instead of a binary 0 or 1 ).
Figure \ref{fig:sf_illus_nodust}(a) shows the $\phi$ versus \rgal{} mapped in the midplane by this corrected selection, which we can now compare it to the previous simple binary selection in Fig. \ref{fig:sf_illus_nodust}(b). We can see that while for most voxels the correct selection function is identical to that in Fig. \ref{fig:sf_illus_nodust}(b) (= 0 or 1), there is a set of voxels along the edge (where $d = d_{\rm max}$) where some fraction of the voxel is still observable. Thus the sub-binning approach allows us to account for estimating the effective volume of a voxel and correctly include this region in our model. 

This effect is drastic in the presence of dust, as is illustrated in Fig. \ref{fig:sf_illus_dust}, where again we show the projection of voxels in $\phi$ versus \rgal{}. In panel (b) we show all the voxels that should be observable but colour-coded by the incorrect binary selection function that does not take into account the finite size of the voxels. All the voxels coloured red have a selection probability of 1, while those in blue have a selection probability of 0 in the incorrect approach; thus only the red voxels would have been predicted to be observable. In panel (a), the same voxels are presented but instead colour-coded using the correct method, now also taking into account the extinction to each sub-voxel when checking if its distance is within $d_{\rm max}$. Now instead of having a binary 1 or 0, the voxels have a continuous selection fraction between 0 and 1. This approach would then correctly predict some of the blue voxels from panel (b) to be truly observable, while some of the red voxels are only partially observable.
Figure \ref{fig:sf_illus_dust} illustrates the non-intuitive pattern that the variation of Galactic dust imposes on the region observable in the Milky Way.

Henceforth for all our analysis, 
we use the sub-binning method described above to estimate the effective (fractional) volume of each voxel. For the observational data, we use the magnitude limit in the $G$ band 17.5 from Fig. \ref{fig:magdist}, and the absolute magnitude is sampled from a \gaussian{}, that is, the luminosity function of the RC is taken as $LF_{RC} = \mathcal{N}(\Bar{M_{G}},\sigma_{\Bar{M_{G}}})$, where $(\Bar{M_{G}},\sigma_{\Bar{M_{G}}})$ is taken from \autoref{tab:tab_extinct}. Thus compute the first layer of the selection function given as the fractional effective volume of each voxel as
\begin{equation}
\label{eqn:rc_sf1}
\begin{split}
    F_{i} = F_{i}(m_{\lambda}<m_{\lambda,lim}|LF_{RC}).
\end{split}
\end{equation}

\paragraph{b.) \gaia\ selection function:} Apart from the probability that an \rc{}-like star can be observed given the magnitude limit of our sample, we also need to consider the probability that a source can be observed by \gaia{} at all. We term this the top-level selection function (\fselitop{}). \cite{Cantat-Gaudin:2023} modelled the completeness of \gaia{} by comparing to the much deeper DECAPS  survey \citep{decaps2018}. They were able to devise a single parameter called $M_{10}$ that encapsulates the completeness based on only three observables, the sky position and magnitude ($l$,$b$,$G$). For our grid, we can compute the $M_{10}$ for every voxel using the aforementioned sub-binning approach and thus get the top-level selection function for every sub-voxel. We do this by using the \gunlim{} \textit{Python} package\footnote{\url{https://gaiaunlimited.readthedocs.io/en/latest/}}, and the class \textit{DR3SelectionFunctionTCG}, using the central values of ($l$,$b$,$G$) in each sub-voxel. In any case, given our $G$ magnitude limit of 17.5, \fselitop{} will be equal to 1, except in the most crowded fields.  

Our parent dataset \gaiawise{} is a cross-match between the \gaia{} and \allwise{} surveys. More specifically, every source in our parent catalogue is present in \gaia{}, and we are using the subset of stars that also have \allwise{} photometry. In addition, we require that our stars have a measured parallax, so in summary we require all stars to have a finite ($l,b,\varpi,G,W1$). Essentially, all these filters are introducing selection effects of their own. To account for this we can compare the number of stars in the catalogue before and after applying the selection criteria in bins of \gaia\ observables. This characterises the completeness as a function of quantities such as sky position and $G$ magnitude. This ratio-ing method allows us to build a sub-selection function (since we can safely assume every source we study here is observed by our principal survey \gaia{}). \cite{Castro-Ginard:2023} developed such a methodology to estimate the selection function for different subsamples of stars in the \gaia{} catalogue. This is done by comparing the number of stars in a given subsample to that in the overall \gaia{} catalogue, providing an estimate of the subsample membership probability (\fselisub) as a function of observables of choice. In the \gunlim{} code, the \textit{SubsampleSelectionFunction} class uses a \textit{beta-binomial} estimator to convert the number count ratios into a probability for a source to end up in the subsample. We choose to compute the ratios in bins of sky position and magnitude ($l$,$b$,$G$) only. In Appendix \ref{app:gaiawise_sfquery} we show the query that is run on the \gaia{} archive that allows us to compute this sub-selection function, and we show in Fig. \ref{fig:subsf_hpix}, the sky projection of the completeness for a bright (12$<G<$13) and a faint 16$<G<$17 magnitude bin at \hpix{} level 5.  The sub-selection function estimator relies on binning and computing number ratios, so naturally the more dimensions one would bin data in, the fewer stars there would be per bin, thus increasing the variance in the estimated selection function. We find that the selection function for the \rc{} hardly varies with colour ($G-Rp$), and so ignoring this dimension allows us to have a more robust number statistic on the sub-selection function. The overall selection function is then a product of the individual layers, 
\begin{equation}
\label{eqn:rc_sf3}
    S_{i} = F_{i}(m_{\lambda}<m_{\lambda,lim} | LF_{RC})  \times  S_{top}(l,b,G) \times S_{sub}(l,b,G) .
\end{equation}

\subsection{Generating mock data}
\label{sec:mock_descr}

For the purpose of validating our methodology, we generate mock datasets for which we know the true values. These are simplified models whose formulations are consistent with our model assumptions and thus ideal test cases for validating our procedure. 

\begin{figure}
\includegraphics[width=1.\columnwidth]{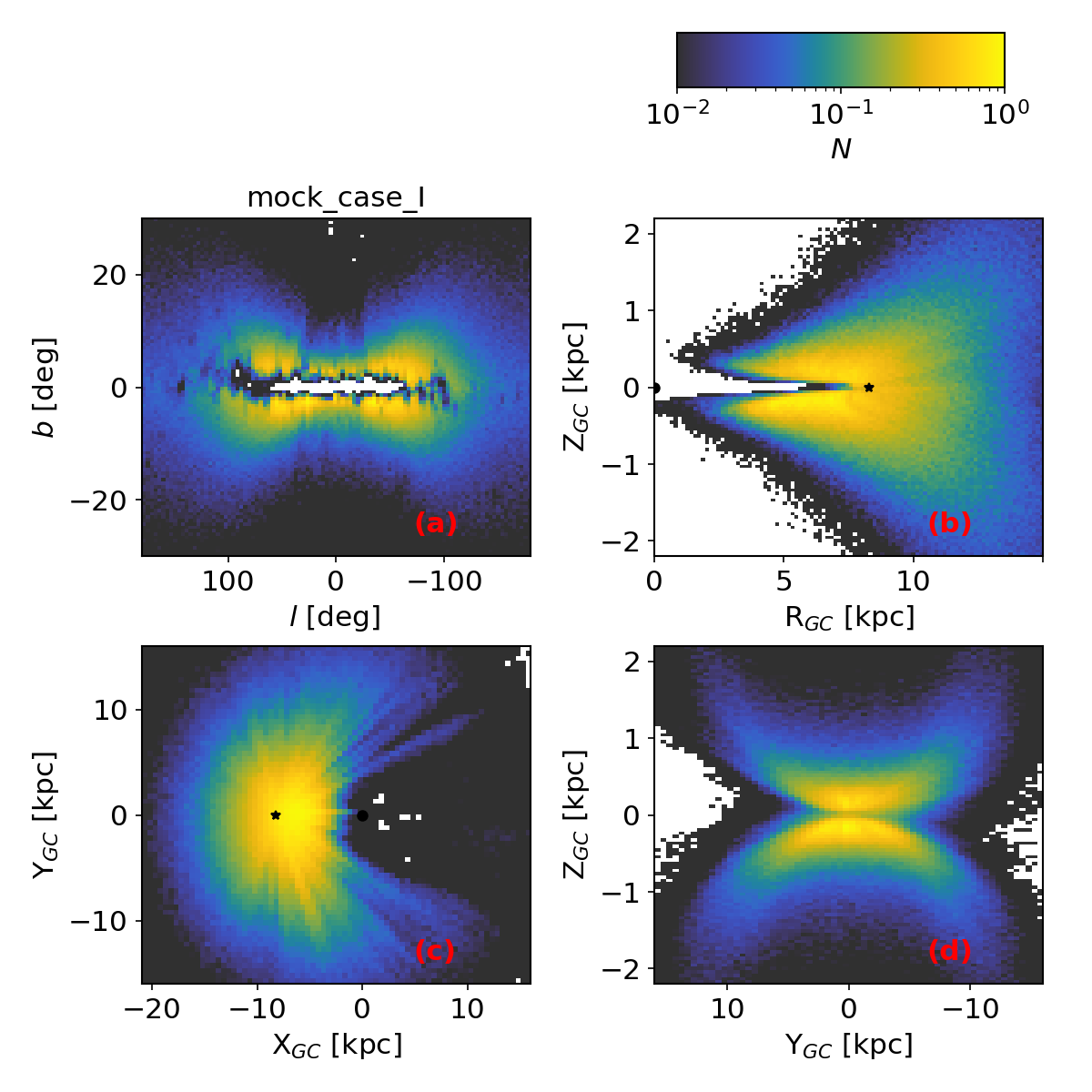} 
\caption{Mock case I: Density distribution for the example with a single exponential disc that is also flared for  magnitude limit $G=17.5$. A dust model was applied to the sample, as is evident in galactic coordinates ($l,b$) in panel (a) with the gap at very low latitudes. Panels (b), (c), and (d) show the corresponding spatial distribution in Cartesian galactocentric coordinates. The locations of the Sun (star) and Galactic centre (dot) are indicated by black points.} 
\label{fig:mock_case1_densityproj}
\end{figure}

\begin{figure}
\includegraphics[width=1.\columnwidth]{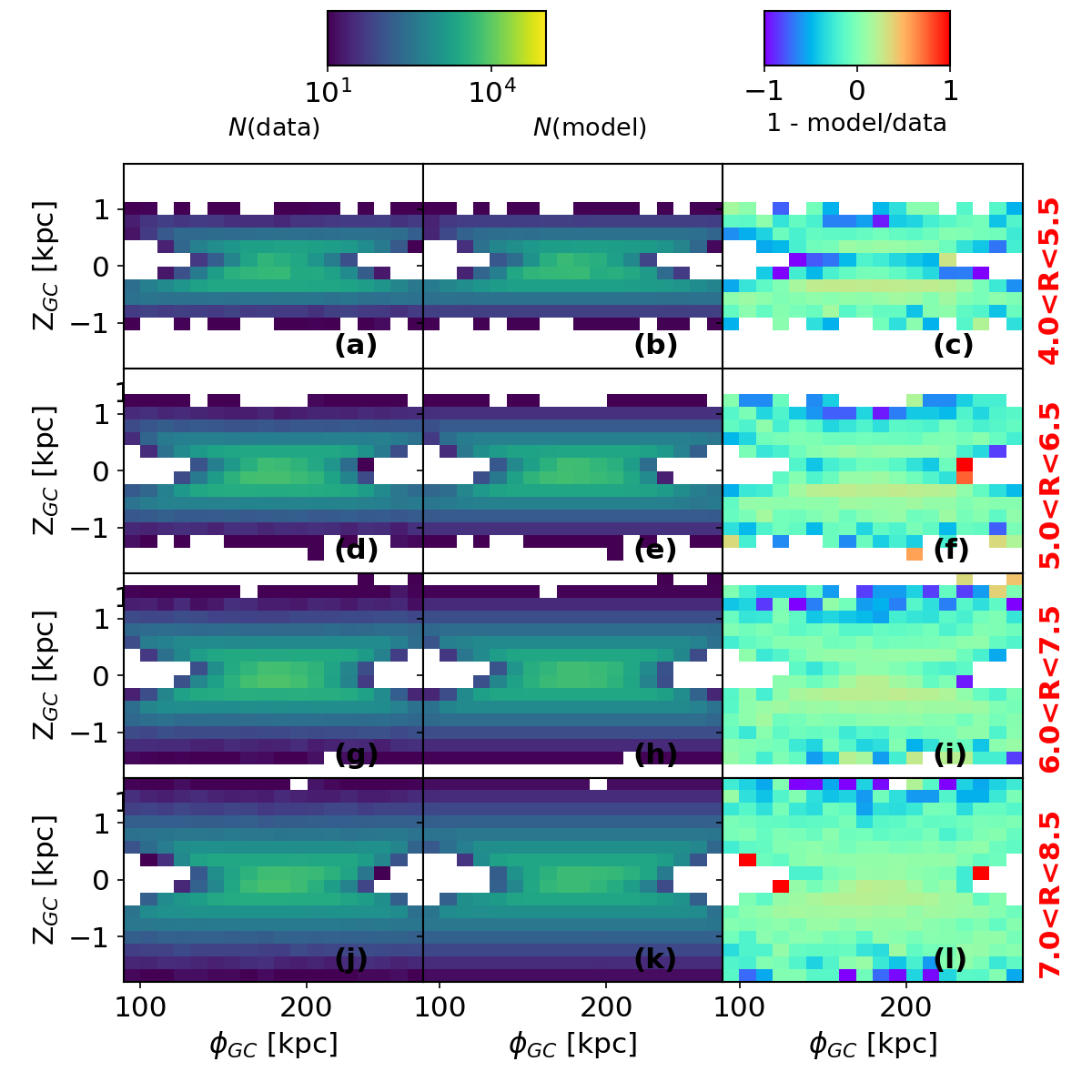} 
\caption{Mock case I: Fitting residuals shown in the $\phi$-\zgal{} projection. Panel (a) shows the number density (logarithmic scale) of the mock data, panel (b) shows the predicted number density of the best-fitted model, and panel (c) shows the residual (relative to mock data). Panels (a-c) are restricted to $4<$\rgal{}$<5.5$ kpc. The subsequent rows show the same information but for selected successive bins in \rgal{}.} 
\label{fig:mock_case1_innerdisc}
\end{figure}

For a given density model, the total number of stars can be computed by integrating over the volume of the survey, 
\begin{equation}
    N = \int N([R,z](l,b,\mu)|\Theta) \times \frac{d_{hc}^{3}log(10)}{5} \times S_{i}\quad d\mu dldb,
\end{equation} where $ N([R,z](l,b,\mu)|\Theta)$ is the number density as before but for a given line of sight ($l,b$) and at distance modulus $\mu$. To generate a mock catalogue, we sample ($l,sin(b),\mu$) from the density models and thus need the Jacobian $|j(XYZ:lb\mu)| =\frac{d_{hc}^{3}log(10)}{5}$.  The $S_{i}$ is the effective selection function of the mock catalogue. For example, if we generated a mock that is magnitude limited, has a sky dependent selection function, and is restricted only to the red clump, the net selection function would be
\begin{equation}
\label{eqn:rc_sf1}
    S_{i} = F_{i}(m_{\lambda}<m_{\lambda,lim} |LF_{RC}), 
\end{equation} where \lfrc{} is the luminosity function (absolute magnitude distribution) of the red clump. In Fig. \ref{fig:glx_rc_camd}, we showed the \camd{} distribution of the \rc{} from \galaxia{}. The marginalised distribution of absolute magnitude (right insets) shows that both \absg{} and \absw{} can be approximated as quasi-Gaussian. Therefore, for the rest of the paper, we adopt a \gaussian{} \lfrc{}, $\mathcal{N}(\Bar{M_{\lambda}}, \sigma_{\Bar{M_{\lambda}}})$, using values from \autoref{tab:tab_extinct}. One could also simplify the method by adopting a \textit{delta} function, $\delta$(\absmag{}-$\Bar{M_{\lambda}}$), in which case we would assume all \rc{} to have a singular absolute magnitude. 

We use the multidimensional sampler, \textit{sampleNdim}, from the \agama{} code \citep{Vasiliev:2019} to then generate our mock catalogue of one million sources. We convert from the heliocentric to the galactocentric frame, as described in Sect. \ref{sec:coord_trans}, but for the mock we here assume \Zsun{}$=0$ kpc. In Sect. \ref{sec:mock_results} we generated several versions of the mock catalogue, to demonstrate how our model fitting methodology performs for a range of Galactic parameters, magnitude limits as well as with and without having applied extinction.

\subsection{Fitting}
\label{sec:model_fitting}

We first grid all data in 3D space ($\Delta$\rgal{},$\Delta\phi$,$\Delta$\zgal{})$= (0.25 \text{kpc},10^\circ,0.25 \text{kpc})$ kpc. Then, using the model described in Sect. \ref{sec:model_descr}, we predict the counts in each bin $N_{i,raw}$. These "raw" counts predicted by the model need to be corrected, taking into account the selection function, i.e. $N_{i}$ =  $N_{i,raw} \times S_{i}$. Then we compared the predicted and the observed counts per bin by maximising a \textit{Poisson} log-likelihood \citep{Bennett:2019},
\begin{equation}
    ln p(N_{obs}|N) = \Sigma_{i} [-N_{i} + N_{obs,i}ln(N_{i})],
\end{equation} with the \textit{emcee} package \citep{Foreman-Mackey:2013}. We used max$(25,7\times N_{dim})$ walkers and up to 5000 iterations. We performed the model fitting only over those voxels with $N_{obs,i}>$ \nmin{}, i.e. with counts above a minimum threshold. For the mock data, we generated one million particles in all and used \nmin{}=10, while since the observational dataset is of the order of ten million stars, there we varied \nmin{}= (20,..,50) in order to estimate the uncertainties in our parameters (see Sect. \ref{sec:fitmodeldata}). In order to visualise our best-fit, we inspected two dimensional projections of the number counts. Specifically, we compared the surface density $\Sigma_{surf}$ versus \rgal{} for the data and the model in bins of $\Delta$\rgal{}$=0.25$ kpc in order to inspect the radial profile of number counts. Similarly, we compared \nzproj{} versus \zgal{} for the data and the model in bins of $\Delta$\zgal{}$=0.25$ kpc at different \rgal{} in order to inspect the vertical profile. Finally, we considered the projection in $\phi$-\zgal{} at various \rgal{}, moving from the inner to the outer disc. This was done to inspect if our model is able to fit azimuthally varying features in the data.

\section{Results: Density modelling}
\label{sec:results}
\subsection{Fitting mock data}
\label{sec:mock_results}

\begin{figure}
\includegraphics[width=1.\columnwidth]{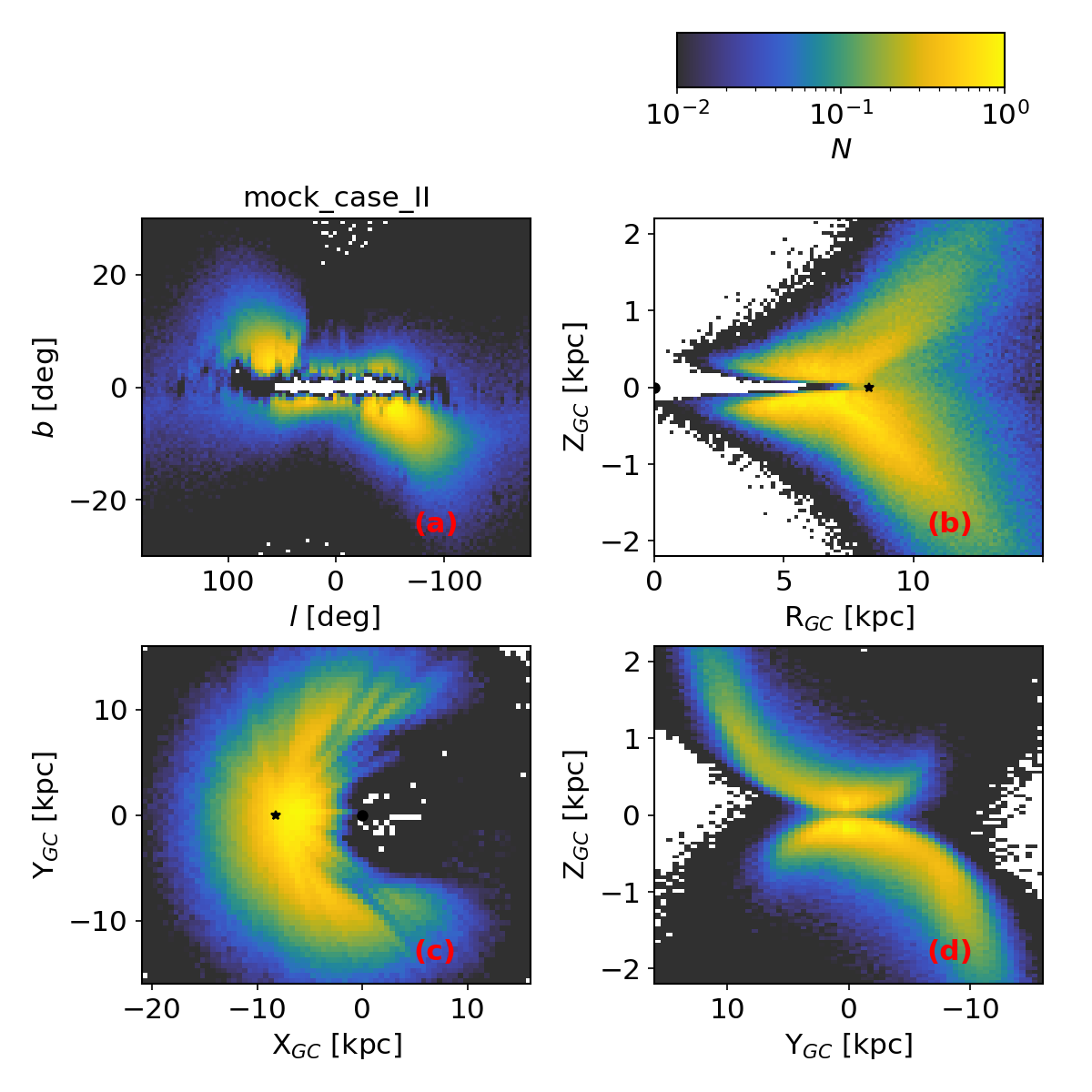}  
\caption{Mock case II: Density distribution for the example with a single exponential disc that is also both flared and warped for magnitude limit $G$=17.5. A dust model has been applied to the sample, as is evident in galactic coordinates ($l,b$) in panel (a) with the gap at very low latitudes. Panels (b), (c), and (d) show the corresponding spatial distribution in Cartesian galactocentric coordinates. The locations of the Sun (star) and Galactic center (dot) are indicated by black points. } 
\label{fig:mock_case2_densityproj}
\end{figure}

\begin{figure}
\includegraphics[width=1.\columnwidth]{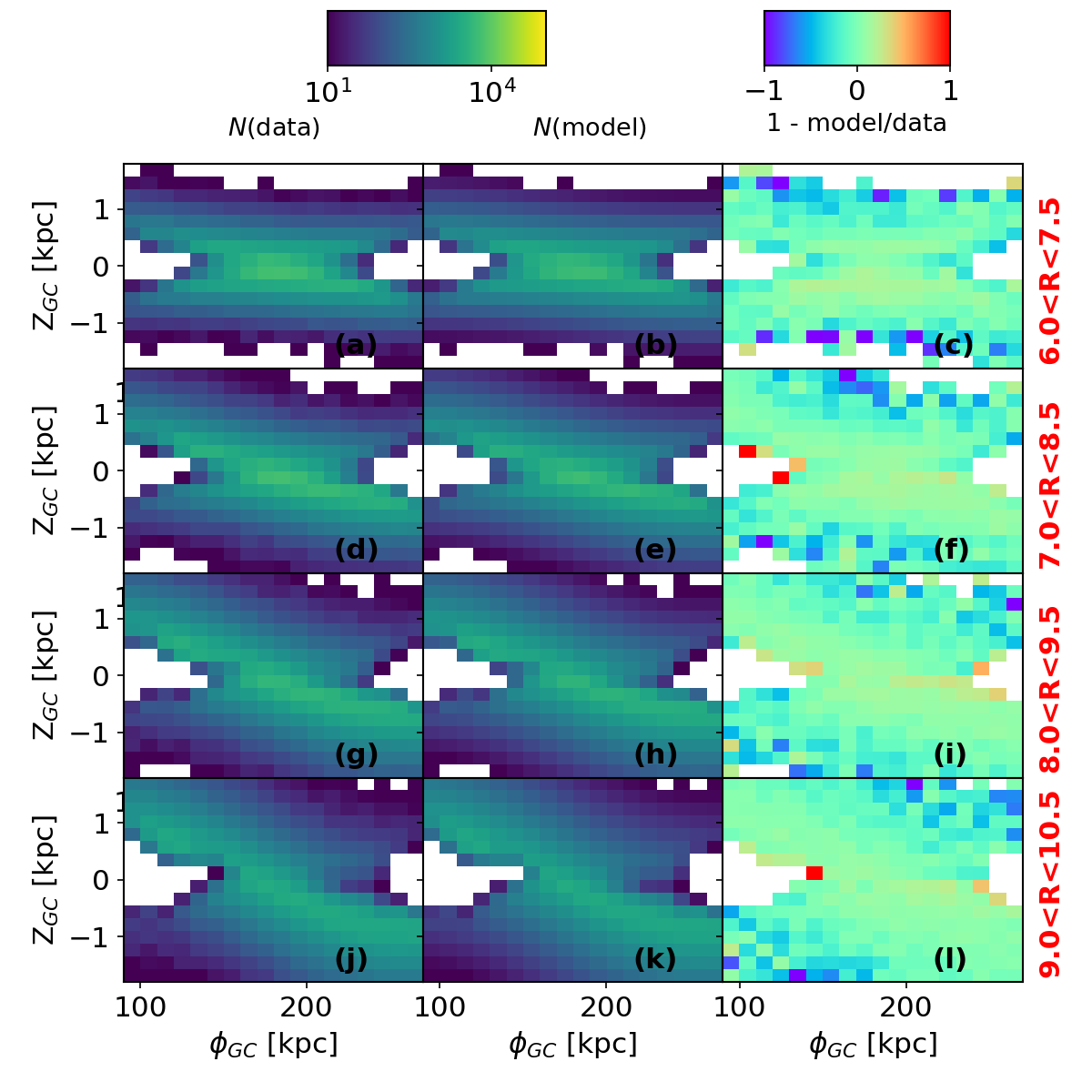} 
\caption{Mock case II: Fitting residuals shown in $\phi$-\zgal{} projection. Panel (a) shows the number density (logarithmic scale) of the mock data, panel (b) shows the predicted number density of the best-fitted model, and panel (c) shows the residual (relative to mock data). Panels (a-c) are restricted to 6$<$\rgal{}$<$7.5 kpc. The subsequent rows show the same information but for selected successive bins in \rgal{}.} 
\label{fig:mock_case2_fitresiduals}
\end{figure}

We begin this section by demonstrating our model fitting method on a mock catalogue of \rc{} stars that were generated as described in Sect. \ref{sec:mock_descr}. We tested our method on several different test cases, but for the demonstration we here restrict ourselves to the two cases discussed below. In both cases, we only consider the selection function layer due to extinction and the top level \gaia{} selection function since the sub-sample selection does not apply to our simple mock density model. The demonstration is also made available as a GitHub notebook.\footnote{\url{https://github.com/shouryakhanna/RedClumpSF_fit-demo}}

\paragraph{Case I:} We generated a mock galaxy of one million particles with a one-disc component and no warp. The density profile is described by an exponential in both \rgal{} and \zgal{}. Additionally, this galaxy has a flare, and we impose a magnitude limit of $G$=17.5. In order to exaggerate the effects of extinction we choose to use here the 2D dust model of \schlegel{} instead of the 3D model described in Sect. \ref{sec:dustmodel}. Finally, we convolve the simulated distance modulus ($\mu$), with typical \rc{} distance uncertainties (see Sect. \ref{sec:RCdist}). The true parameters for this disc are (\rd{}, \hzsun{}, $log_{10}$ \rflare{}[kpc])=(3.3 kpc, 0.3 kpc, 0.6). The density maps of the mock are shown in Fig. \ref{fig:mock_case1_densityproj}, with the projection in galactic coordinates ($l,b$) in panel (a) in \rgal{}-\zgal{} (to show the flare) in panel (b), and in Cartesian galactocentric coordinates (\xgc{},\ygc{},\zgc{}) in panels (c) and (d). In Fig. \ref{fig:mock_case1_fitmcmc} we show the corner plot from the MCMC fitting carried out for this example disc, where we had set \nmin{}=10. The best-fit values are indicated with vertical lines, and agree very well with the input values for the mock. To better judge the fit, we also show the projection in $\phi$-\zgal{} space in Fig. \ref{fig:mock_case1_innerdisc} of the mock data and the fitted model for galactocentric radial bins between $4<R<8.5$ kpc (columns 1 \& 2). Also shown in Fig. \ref{fig:mock_case1_innerdisc} 
are the residuals between the mock data and model (column 3). The colour map of the residuals shows general good agreement except at the edges of the density distribution where the number of stars drops rapidly. 

\paragraph{Case II:} The second example we include is that of a disc which is warped, described by (\phiwarp{}, \rwarp{}, \awarp{}, \hwarp{})=(170$^\circ$, 6.5 kpc, 1.0, 0.3 kpc). This is a rather extreme warp ideally oriented for the purpose of illustration and testing, and is not meant to represent the warp of the Milky Way. The rest of the parameters are identical to the disc in case I.
Figure \ref{fig:mock_case2_densityproj} shows the density maps for this mock data, where a tilt (top left to bottom right) is apparent in the ($l,b$) map in panel (a). This can be compared to Fig. \ref{fig:mock_case1_densityproj}(a) where the disc was not warped. In Cartesian coordinates, the sense of the warp here is such that the disc bends up (towards positive \zgal{}) at positive \ygc{}, and bends down at negative \ygc{}, as shown in  Fig. \ref{fig:mock_case2_densityproj}, 
as expected for the line-of-nodes being at $\phi=170^\circ$.
Figure \ref{fig:mock_case2_fitmcmc} shows the corner plot of the fit, again with the best-fit values indicated as blue lines. The MCMC procedure converges to recover the input values, although we did notice that the warp parameters (\rwarp{}, \hwarp{} and \awarp{}) show strong correlations between them. The $\phi$-\zgal{} projection of this dataset is also shown in Fig. \ref{fig:mock_case2_fitresiduals}, 
again for three successive bins in \rgal{}, though here we note that the density map shows a diagonal feature beyond \rgal{}$>7$ kpc that is the imprint of the warp. The residual maps  are again mostly close to zero across the pixels, except at the very edges (low number statistics), and thus show that the fitting works well in this case too, and we are able to fit both for the structural (scale length and height) and shape parameters (warp) of the disc, though these later show significant correlations.

\subsection{Observational data}
\label{sec:data_results}
\subsubsection{Selection function \& Completeness map}
\begin{figure}
\includegraphics[width=1.\columnwidth]{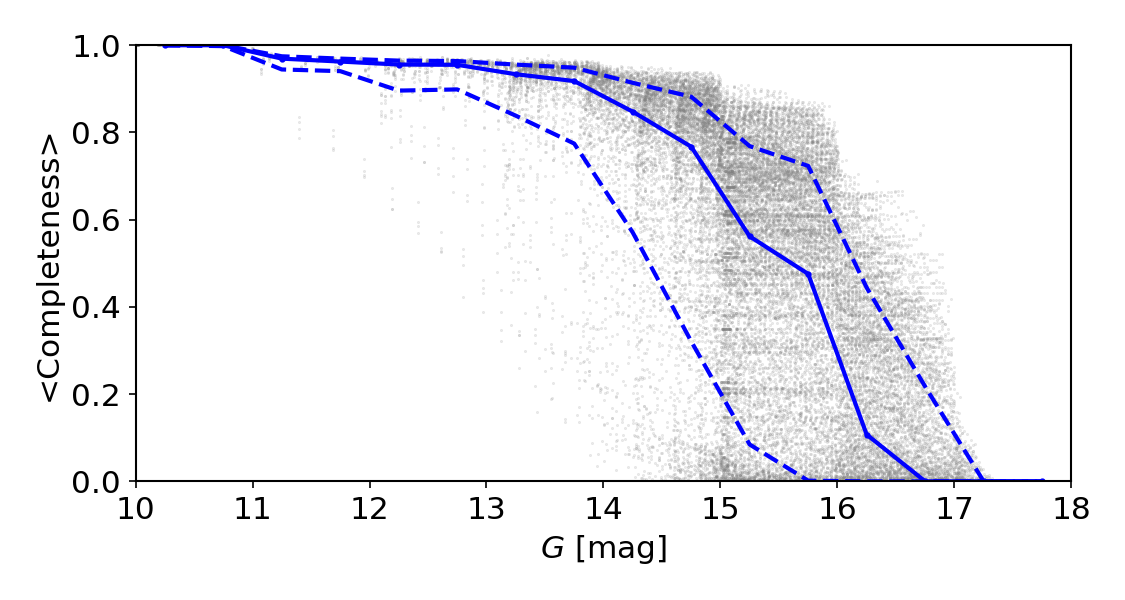} 
\caption{Completeness for the \rc{} sample for \rgal{}$>$3 kpc, shown in gray, as a function of $G$ magnitude (0.5 mag bins). The blue solid line shows the median profile, and the dotted lines show the 84$^{th}$ and 16$^{th}$ percentiles.} \label{fig:sf_gmag}
\end{figure}

\begin{figure}
\includegraphics[width=.95\columnwidth]{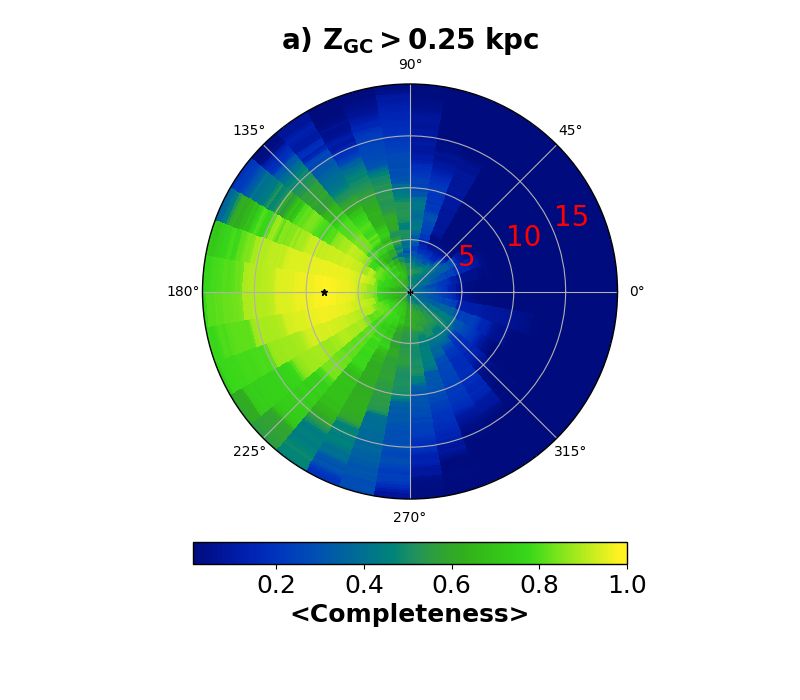}
\includegraphics[width=.95\columnwidth]{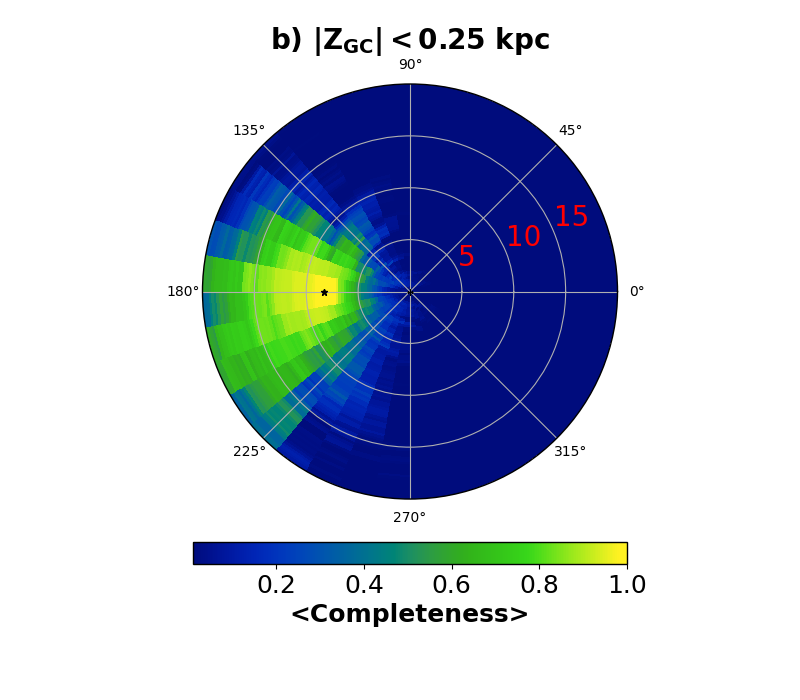}
\includegraphics[width=.95\columnwidth]{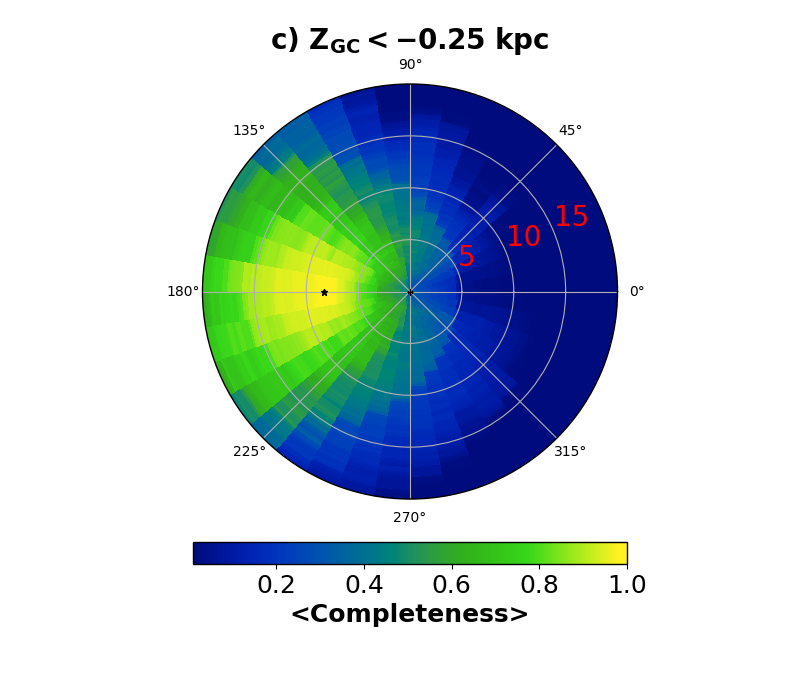}
\caption{Completeness for \rc{} stars over the entire grid shown in galactocentric polar coordinates $(\phi, R)$ for three slices in \zgal{} above the plane (panel a), in the midplane (panel b), and below the plane (panel c). 
The concentric circles indicate bins in \rgal{}, with values in \kpc denoted in red. The black star indicates the Sun's position.} \label{fig:sf_polar}
\end{figure}
We move on to fitting our model on the \gaiawise{}[RC] dataset. 
Following Sect. \ref{sec:sfdesc}, we apply all three layers of the selection function to the real data. The predicted completeness of the \rc{} sample as a function of $G$ magnitude is shown in Fig. \ref{fig:sf_gmag}, with the solid blue line indicating the median profile and dotted lines showing the variance in completeness. As this profile (blue line) is averaged over the entire volume (grid), we expect to see variance at any given radius. Overall, the \rc{} sample is largely complete (close to 1) down to about $G=14$, falls to about 0.5 completeness at $G=16$, and then drops sharply to 0 by $G=17$.

As mentioned in Sect. \ref{sec:sfdesc}, taking advantage that RC stars are standard candles, we can determine the fraction of RC stars in our sample (completeness) for any particular volume. 
In Fig. \ref{fig:sf_polar}, we show the completeness in polar coordinates (\rgal{},$\phi$), averaged over three different slices in \zgal{}. Towards the inner Galaxy, the completeness drops sharply due to extinction, while in the anticentre direction the completeness extends much further. 
While overall the two maps at \zgal{}$>$0.25\kpc and \zgal{}$<$ -0.25\kpc exhibit very similar completeness, there are minor differences between the two due to the asymmetrical dust distribution above and below the disc.
Figure \ref{fig:sf_polar}(b) shows the completeness for the midplane region, $|\text{\zgal}|<$0.25 kpc. Compared to the maps at higher \zgal{}, the region of high completeness covers a much smaller region. Indeed, in the plane we are most affected by dust, which obscures our view towards the inner Galaxy and along other lines of sight with high extinction. In addition, we are also affected by high source crowding towards the Galactic centre, which makes it difficult for \allwise{} to resolve individual sources and causes these fields to be less well sampled. Nevertheless, all three maps show that we are able to map out a large portion of the disc out to at least \rgal{}$=15$ kpc with fairly high completeness. An interactive three-dimensional visualisation of completeness for our \rc{} sample is provided on the \gunlim{} webpage\footnote{\href{https://home.strw.leidenuniv.nl/~brown/unlimited-demos/3dapp/}{Interactive 3D visualisation of the \rc{} completeness.}}.

\subsubsection{Fitting model to data}
\label{sec:fitmodeldata}
Following the method used for the mock catalogues, we fit for the parameters listed in \autoref{tab:galmodel}. In the first instance, we ignore the warp and check if we can fit the distribution with only a single flaring disc (i.e. we set \fdisc{}$=1$). An example corner plot for this fit is shown in  Fig. \ref{fig:datafitmcmc_single}, where we find, \rd{}$=3.13\pm0.20$ kpc, \hz{}$=0.46\pm0.02$ kpc, and $log_{10}$ \rflare{}$=2.14\pm0.20$. These are typical scale parameters for the old thin disc population Milky Way \citep[e.g..][]{jbhreview2016}, but in this case we obtain a slightly higher \hz{} which compensates for a very weak flare. While the fit does converge, a closer look at the residuals shows that a single disc is not enough to account for the stellar density both in the inner ($3<$\rgal{}$<10$ kpc), and the outer disc ($10<$\rgal{}$<15$ kpc) simultaneously (see Fig. \ref{fig:residuals_phiz_singdisc}). We therefore allowed for an additional disc component and fit for all the parameters (barring the warp again) listed in \autoref{tab:galmodel}. An example corner plot in Fig. \ref{fig:datafitmcmc}, shows that the fit converges. Essentially, we find that the \rc{} sample is best described by a two disc component model, with one component having \rd{}$=4.08\pm0.32$ kpc, \hz{}$=0.18\pm0.01$ kpc, and $log_{10}$ \rflare{}$=0.36\pm0.04$. This is essentially a disc with a long scale length exhibiting a strong flare, and constitutes about $34\%$ of the total mass. The second component is dominant ($\sim66\%$) has a scale length of \rd{}$_{2}=2.66\pm0.11\kpc$  and constant (non-flaring) scale height of \hz{}$_{2}=0.48\pm0.11\kpc$. When we included a flare parameter for this second disc, the fit was unable to constrain this, but this did not affect the values obtained for the remaining parameters, so we can assume that the thinner component is contributing alone to the overall flare in our dataset. 

\begin{figure*}
\includegraphics[width=2.\columnwidth]{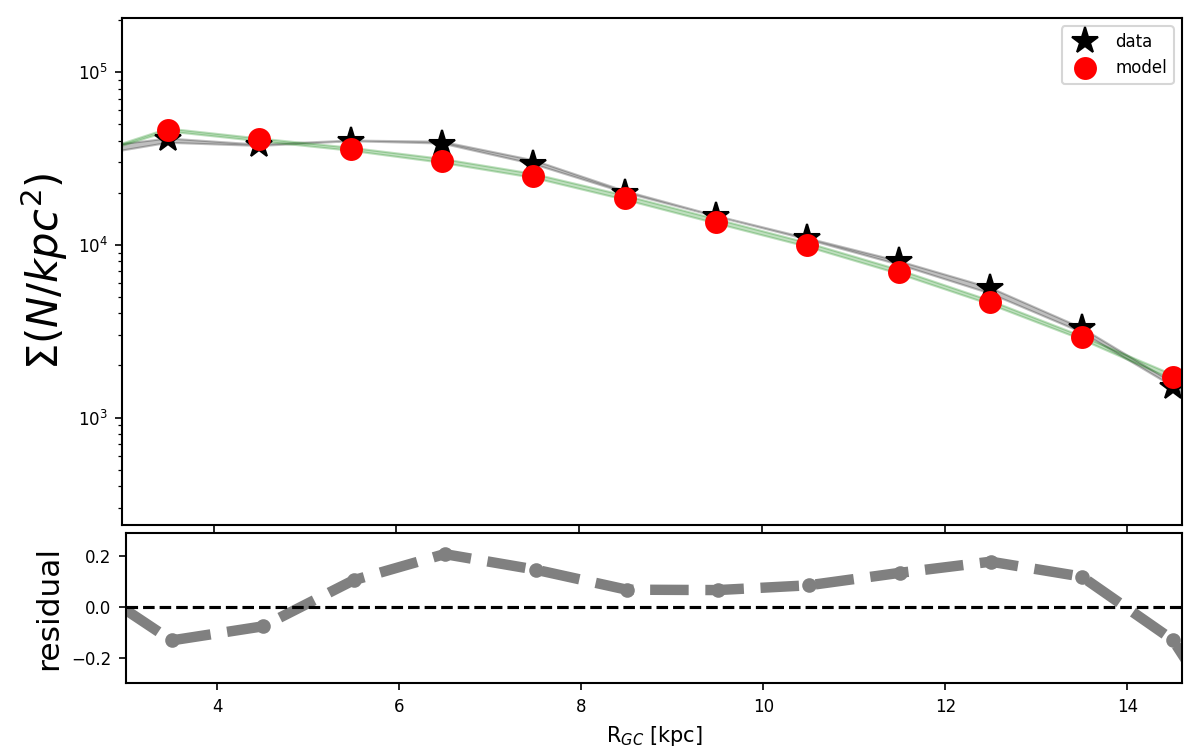}
\centering
\caption{Surface density of \gaiawise{}[RC] averaged over $|$\zgal{}$|<$1 kpc. Data are shown as black points, and the grey shaded area represents the two standard deviations over which each realisation of the data varies. The model predictions are shown in red and again the green shaded region represents the uncertainty in each bin. The residuals (1 -model/data) are shown in the lower panel.} \label{fig:residuals_surfdens}
\end{figure*}

\begin{table}
\caption{Best-fit parameters describing the Galactic disc.} \label{tab:bestfitdatamodel}
\centering
\resizebox{0.95\columnwidth}{!}{%
\begin{tabular}{c|c|c}
\hline
\hline
Parameter   & Model 1  & Model 2  \\
(unit) &   & \\
\hline
\rd{} (kpc) &  3.13$\pm$0.25 &  4.24$\pm$0.32 \\ [2pt]
\hzsun{} (kpc) &  0.46$\pm$0.03 &  0.18$\pm$0.01  \\ [2pt]
$\log_{10}$ \rflare{}[kpc]  & 2.14$\pm$0.20  & 0.36$\pm$0.04 \\ [2pt]
\rd{}$_{2}$ (kpc)& - &   2.66$\pm$0.11 \\ [2pt]
\hztwo{} (kpc) & -  &  0.48$\pm$0.11 \\ [2pt]
\fdisc{} & -  & 0.34$\pm$0.02 \\
\hline
\hline
\end{tabular}
}
\tablefoot{Best-fit values for the parameters fitted to observational data. Model 1 only fits for a single exponential disc that is also flared. Model 2 fits for two exponential disc components, with only one allowed to be flared. The units of the parameters are listed in parenthesis, except for $\log_{10}$ \rflare{}[kpc] which is dimensionless.}
\end{table}

In order to estimate the uncertainties of our estimated parameters, we adopted a Monte-Carlo approach and ran the fitting over 10 realisations of the data, each time sampling the distance modulus from the assumed uncertainty distribution, $\mathcal{N}$($\Bar{\mu_{\lambda}},\sigma_{\mu(W1)})$ as described in Sect. \ref{sec:RCdist}. Then we used the $16^{th}$ and $84^{th}$ percentiles in each parameter to estimate the spread in their median values. However, we found these uncertainties to be very small and therefore unrealistic. This is likely a consequence of using a very large sample of stars. Instead, just to estimate the uncertainty, we generated one realisation of the data and then varied the arbitrarily chosen parameter \nmin{} from 50 to 20; that is, we performed independent fits while varying the arbitrarily chosen minimum number of stars per voxel criterion. In general, the median values of the parameters remains very consistent for the different \nmin{} values, but we note that the scale length of disc 1 drops from \rd{}= 4.4 kpc (at \nmin{}=50) to \rd{}= 4.08 kpc (\nmin{}=20). We use the variation in the parameter values between these two fits as the uncertainty estimate. The best-fit values are then summarised for models 1 and 2 in \autoref{tab:bestfitdatamodel}.  As separate exercise, we also carried out the two-disc fit for the \gaiawise{}[RC] sample selected by setting \pearson=0, i.e. ignoring the correlations between $G$ and $W1$. Interestingly, nearly all the best-fit parameters were consistent with the values reported in \autoref{tab:bestfitdatamodel}, except for the scale length of disc 1 which we find to be about \rd{}=3.6$\pm0.32$ kpc. In both cases (\pearson=0 or \pearson=0.7); however, \rd{} is favoured to be much longer than \rdtwo{}.

\begin{figure*}
\centering
\includegraphics[width=2.\columnwidth]{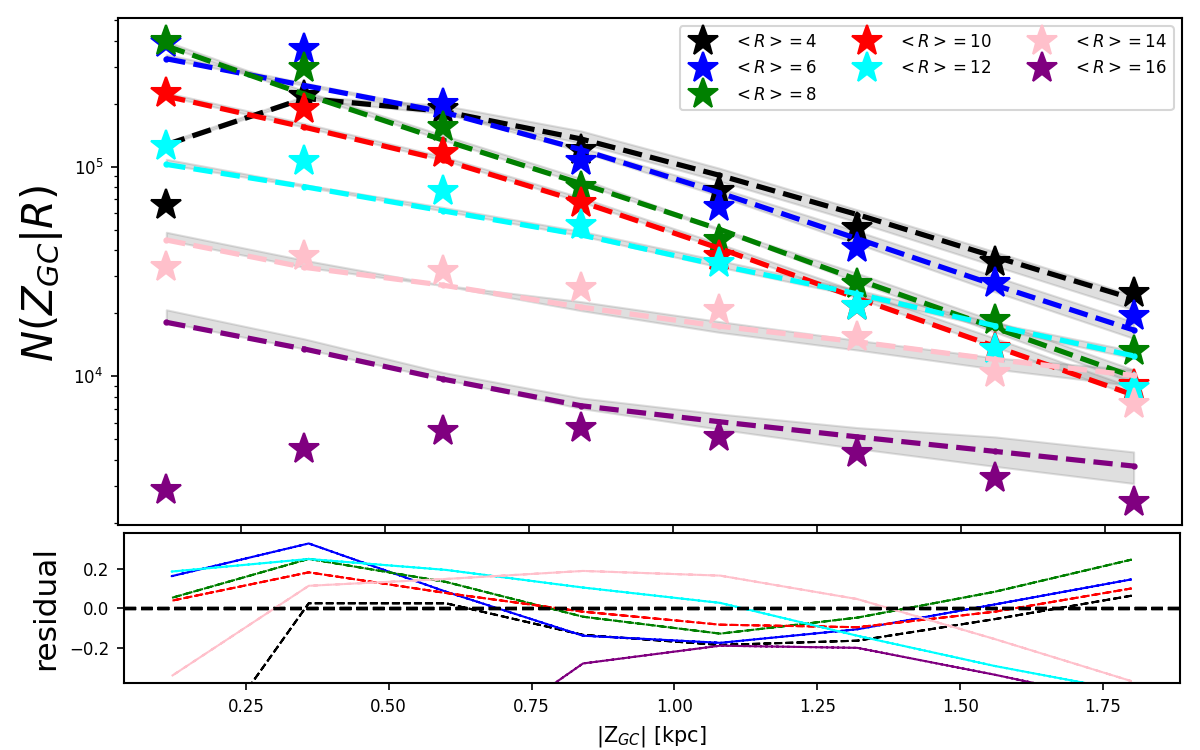}
\caption{Vertical counts as a function of $|$\zgal{}$|$ at progressive annuli in \rgal{} (2 kpc wide). In each case, the data are shown as stars, and the  predictions from Model 2 as dashed lines in the same colour. The uncertainties in the best-fit model are represented by the shaded grey area. The residuals between data and model are shown in the lower panel.} \label{fig:residuals_vert}
\end{figure*}

\begin{figure}
\includegraphics[width=1.\columnwidth]{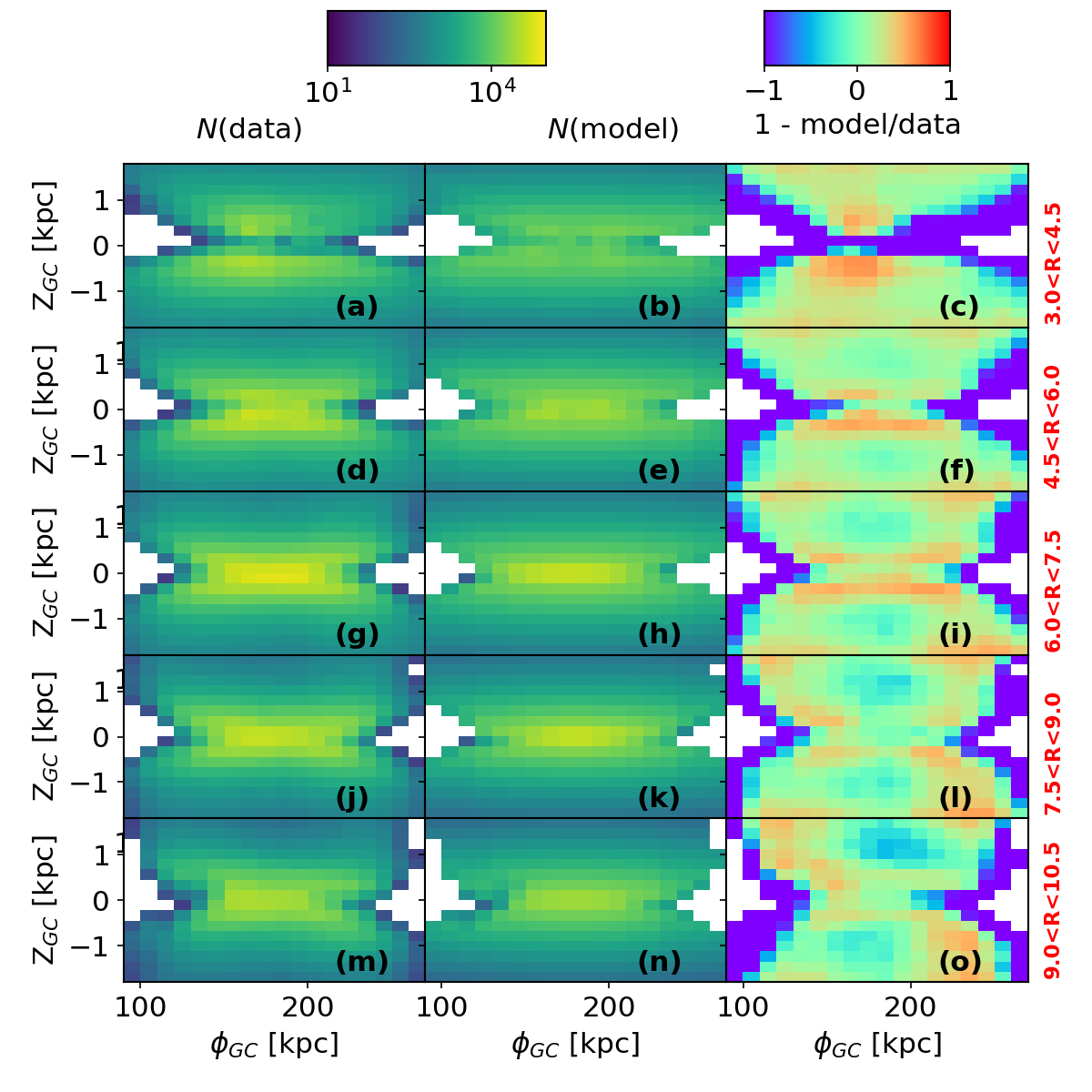}
\caption{Residuals between Model 2 from \autoref{tab:bestfitdatamodel} and data in the  $\phi$,\zgal{} projection, shown for the inner disc region ($3$ kpc$<$\rgal{}$<10.5$ kpc). In each row, the first column shows the number density of the data, the second row shows the number density predicted by the model, and the third row shows the residuals relative to the data. Each row represents a 1 kpc wide annulus in \rgal{}.} \label{fig:residuals_phiz_inner}
\end{figure}

\begin{figure}
\includegraphics[width=1.\columnwidth]{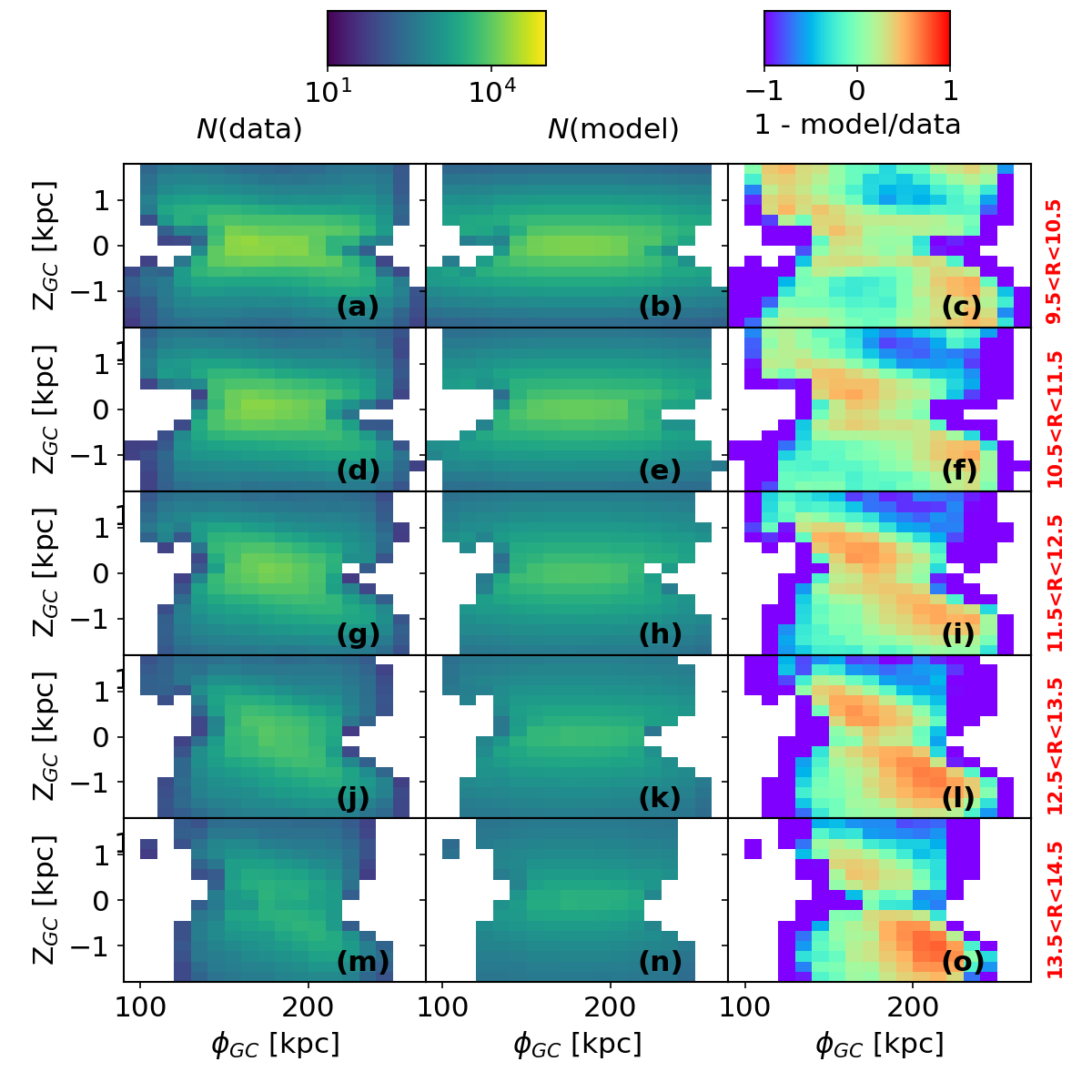}
\caption{Residuals between the model and data in the $\phi$,\zgal{} projection. The setup is the same as Fig. \ref{fig:residuals_phiz_inner} but for the outer disc ($9.5$ kpc$<$\rgal{}$<14.5$ kpc).} \label{fig:residuals_phiz_outer}
\end{figure}

Using these best-fitting parameters, we take a closer look at the residuals between data and model, by considering projections in 1d and 2d spaces. First, in Fig. \ref{fig:residuals_surfdens} (upper panel), we show the surface density $\Sigma(N/kpc^{2})$ as a function of \rgal{}, with the model in red, which overall, is able to capture the profile of the data (stars). The grey lines in the background indicate the variation in surface density (observed and model) for each bin for every realisation of the dataset. There is a slight bump in the data around \rgal{} $=6$ kpc, and another one around \rgal{} $=12.5$ kpc, which are clearly not fit by the model, seen more clearly in the residuals (lower panel). Next, we consider the vertical counts, $N(Z_{GC}|R)$ as a function of $|Z_{GC}|$ in Fig. \ref{fig:residuals_vert}, where we plot the number of stars in data (points) in 2 kpc wide annuli between $4<$ \rgal{} $<15$ kpc, and the model prediction as broken lines. Again, the model is largely able to capture the trends seen in the data profiles. We remind the reader that our fit was not performed in these projected spaces, so we do not expect one-to-one agreement for every bin, but judge the fit on its ability to capture the overall trends. As noted before, the slope of $N(z_{GC}|R)$ flattens for \rgal{} $\geq 12$ kpc due to the flare in the disc.

Finally, we inspected the residuals in the $\phi$-\zgal{} projection for bins in \rgal{}. We separately considered the inner disc ($3<$ \rgal{} $<10$ kpc), and the outer disc ($10<$ \rgal{} $<14.5$ kpc) in Figs.
\ref{fig:residuals_phiz_inner} and \ref{fig:residuals_phiz_outer} respectively.
Figure \ref{fig:residuals_phiz_inner} shows the $\phi$-\zgal{} projection for the inner disc, where each row corresponds to a radial bin, the first two columns show the number density of the data and model respectively, the third column shows the relative residual [1-(model/data)]. Considering the last column, we can see that overall our model is able to describe the data well, as most of the region in these panels is fairly uniform, with the exception of the edges of our grid due to low number counts and unaccounted for distance uncertainties, as mentioned before.  There are other regions of high residuals, however. In the inner-most annulus of Fig. \ref{fig:residuals_phiz_inner} we see the model is under-predicting the counts.  This is likely because we don't include a separate component for the Galactic bar/bulge, which may be contaminating this bin. 
While the remaining panels show smaller residuals, we note in the panels in the range corresponding to $6.0 < R < 7.5 \kpc$
the model under-predicts the counts close to the Galactic plane. This feature gets weaker beyond 7.5 kpc. This shows that the excess of stars at this radius, noted above in Fig. \ref{fig:residuals_surfdens}, is restricted close to the plane. 
When moving outwards in the 9$<$ \rgal{}$<$10.5 kpc annulus, there is a weak presence of residuals along the diagonal; that is, the data predicts larger counts with regard to the model at positive \zgal{} and $\phi<180^\circ$, then at negative \zgal{} and $\phi>180^\circ$. This becomes more pronounced as we move to larger radii, as seen in Fig. \ref{fig:residuals_phiz_outer}, where we see the model is severely under-predicting the counts along this diagonal with increasing \rgal{}, particularly beyond \rgal{}$>$10.5 kpc. These residuals are clear evidence that the distribution of \rc{} stars is warped in the outer disc. 

\begin{figure}
\includegraphics[width=.9\columnwidth]{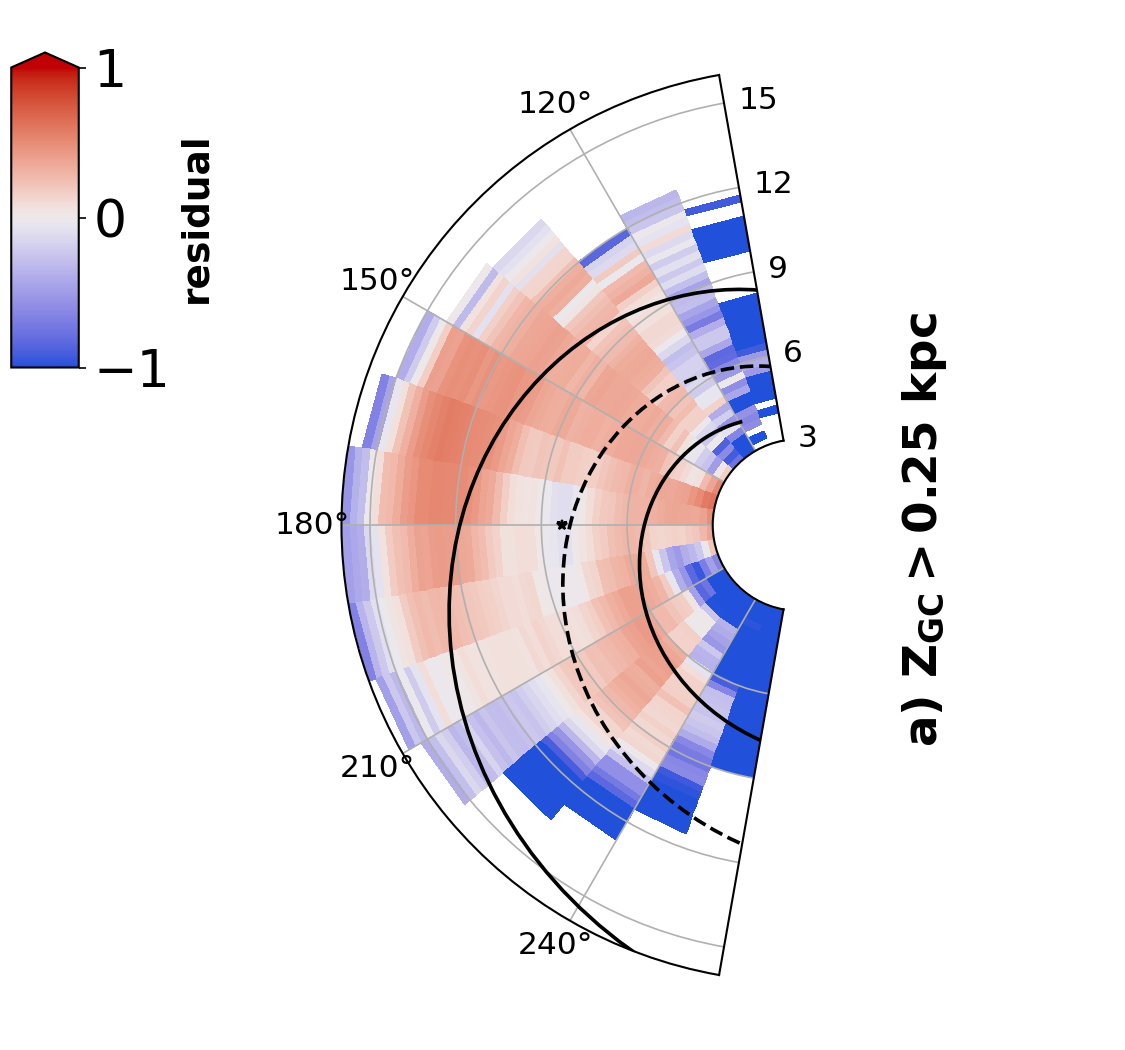}
\includegraphics[width=.9\columnwidth]{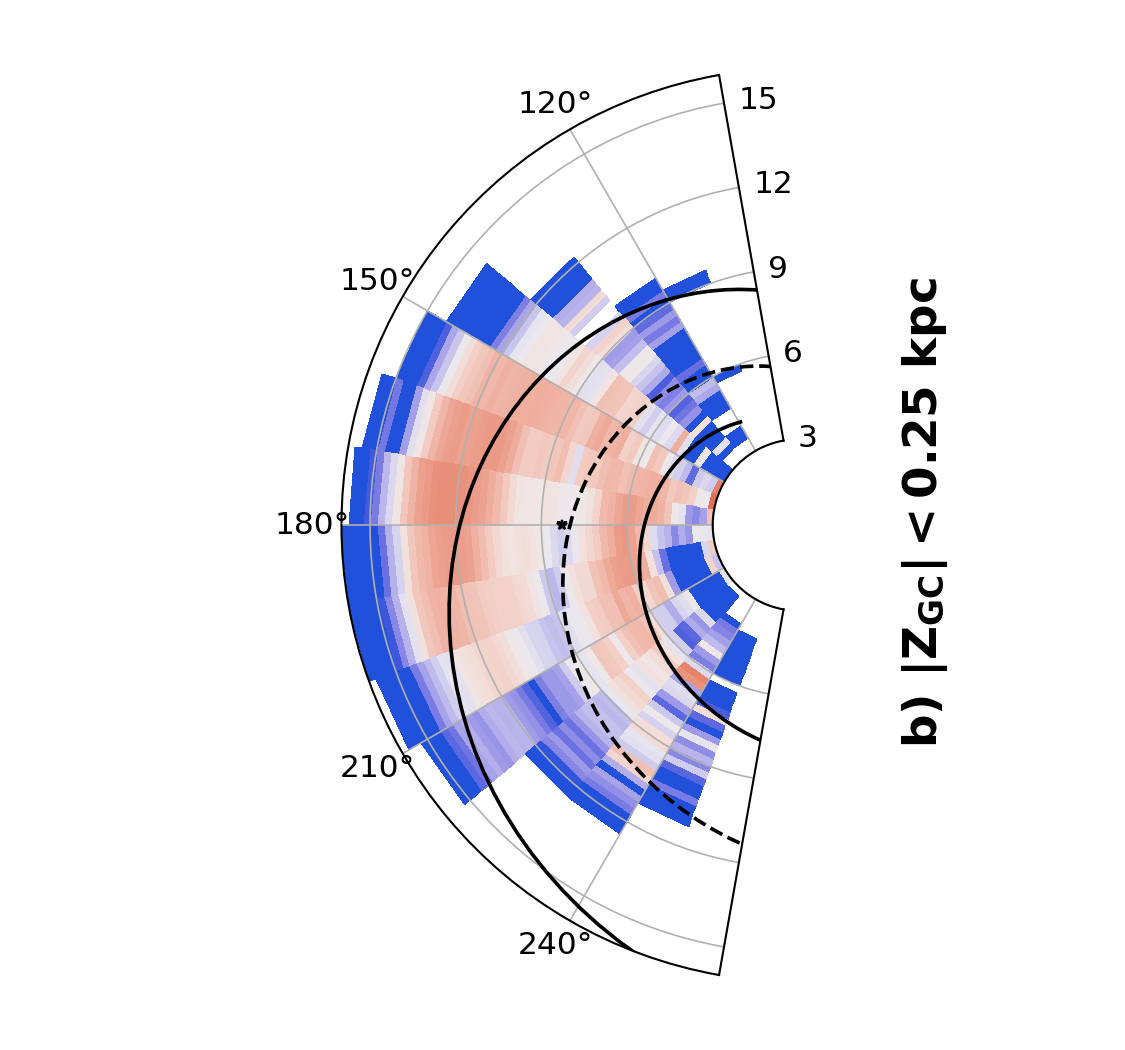}
\includegraphics[width=.9\columnwidth]{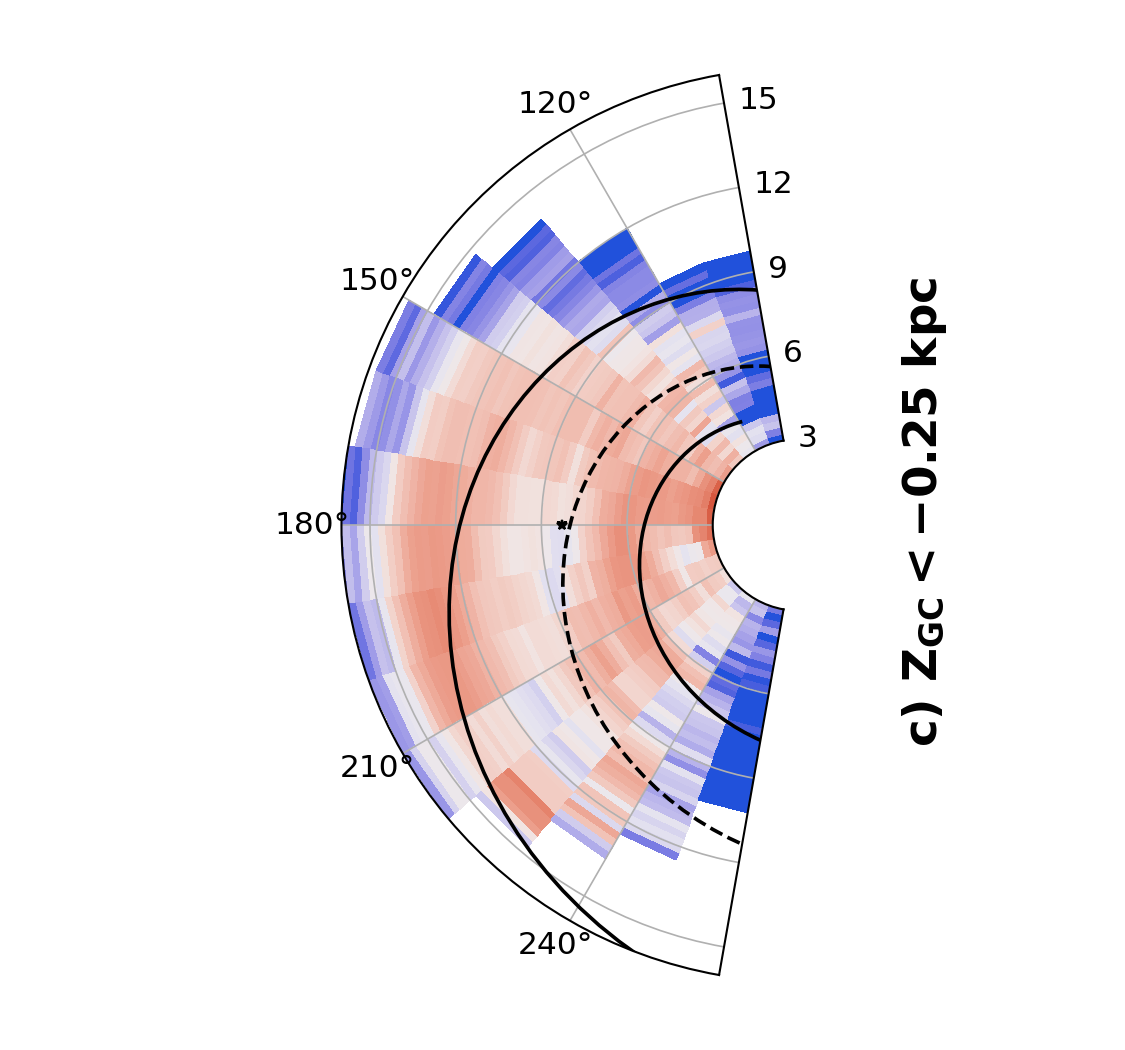}
\caption{Relative residual (1-model/data) for Model 2 from \autoref{tab:bestfitdatamodel}, applied to \gaiawise{[RC]}, and shown in polar coordinates. The residuals are shown for three slices in \zgal{} above the plane (panel a), in the midplane (panel b), and below the plane (panel c). The two-arm NIR spiral model from \cite{Drimmel:2000} is overplotted as black curves.} \label{fig:residuals_polar}
\end{figure}

In Fig. \ref{fig:residuals_polar}, we show the relative residuals in polar coordinates (\rgal{},$\phi$), averaged over three different slices in \zgal{}. On these maps, white regions indicate where the data and the model are in close agreement, while red (model underpredicts) and blue (model overpredicts) highlight the  discrepancy. We can identify three regions with discrepancy from the model prediction where the model is underpredicting counts. First, there is a small patch around \rgal{}$<$3 kpc along $\phi=180^\circ$ seen in the slices above and below the Galactic mid-plane. This is likely to be due to the contribution from the bulge that we do not model for. The next feature is in the outer disc beyond \rgal{}$>$9 kpc. We note that this outer feature appears to shift to higher azimuth when comparing the residuals in the upper slice (\zgal{}$>$0.25 kpc) with respect to the lower slice (\zgal{}$<$0.25 kpc), shifting from 160$^\circ<\phi<$180$^\circ$ to 
170$^\circ<\phi<$210$^\circ$. This is consistent with the residuals in the outer disc shown in Fig. \ref{fig:residuals_phiz_outer}, where the residuals are seen to lie along a diagonal, that is, at positive \zgal{} the residuals are higher than the corresponding negative \zgal{} slice below the plane for $\phi<$180$^\circ$. And again the residuals are higher below the plane for $\phi>$180$^\circ$, compared to the corresponding \zgal{} slice above, which we interpret to be the signature of the Galactic warp. 
However, we also see these residuals also shift towards higher radii with increasing azimuth, consistent with an overdensity with a spiral geometry similar to the outer arm of the two-arm spiral model (solid black curves) of \citet[][, hereafter D00]{Drimmel:2000} based on near-infrared (NIR) data.

For the inner disc(\rgal{}$<$9 kpc) the residuals show a clear and distinctive pattern only near the Galactic midplane ($|\,$\zgal{}$\, |<0.25$ kpc, Fig. \ref{fig:residuals_polar}(b)). Here we see that the data has an under-density (shown in blue) with respect to the axisymmetric model, seen approximately at the Sun's position and following the dotted curve that indicates the inter-arm position where the density is expected to be a minimum between the two arms. Inside this curve we again see positive residuals, though it is less clear whether these residuals follow a spiral geometry that is expected from the NIR-based model. Indeed, the inner edge of the positive residuals, approximately delimited by the inner spiral arm, is probably largely determined by extinction severely limiting how far we can reliably map the RC sample.  In any case, any hint of spiral geometry in the inner-disc residuals is not clearly evident above (\zgal{}$>$0.25 kpc) and below (\zgal{}$<$-0.25 kpc) the Galactic plane (upper and lower panels of Fig. \ref{fig:residuals_polar}).

\subsubsection{Testing warp parameters with real data} \label{sec:warped}
As discussed in the previous paragraphs, the best-fit of our assumed axisymmetric model exhibits diagonal residuals in the $\phi$-\zgal{} projection. It is natural to wonder then whether the warp of the Milky Way can account for this outer feature. 

To check this we tried to simultaneously fit for a model which included all the parameters from Model 2 (\autoref{tab:bestfitdatamodel}) as well as the additional four warp parameters (\phiwarp{},\awarp{},\rwarp{},\hwarp{}). However, the MCMC routine did not converge for these warp parameters, but interestingly, the remaining best-fit parameters did not significantly differ from Model 2. We then locked the best-fit parameters from Model 2, and only ran a fit for the four warp parameters, but again the MCMC had trouble converging.
Our tests showed that the fit was in particular unable to constrain two parameters, \awarp{}, and \phiwarp{} which indicates the line-of-nodes (LON) for the warp. However, recent works have shown that the Milky Way's warp does not have a constant LON, and that it curves towards lower $\phi$ beyond about \rgal{}=12.5 kpc \citep{Dehnen:2023,CabreraGadea:2024,Jonsson:2024,Poggio:2024}.
This could possibly explain in part why we are unable to constrain the \phiwarp{} as our warp parametrisation assumes a single LON for simplicity.

\begin{figure}
\includegraphics[width=.49\columnwidth]{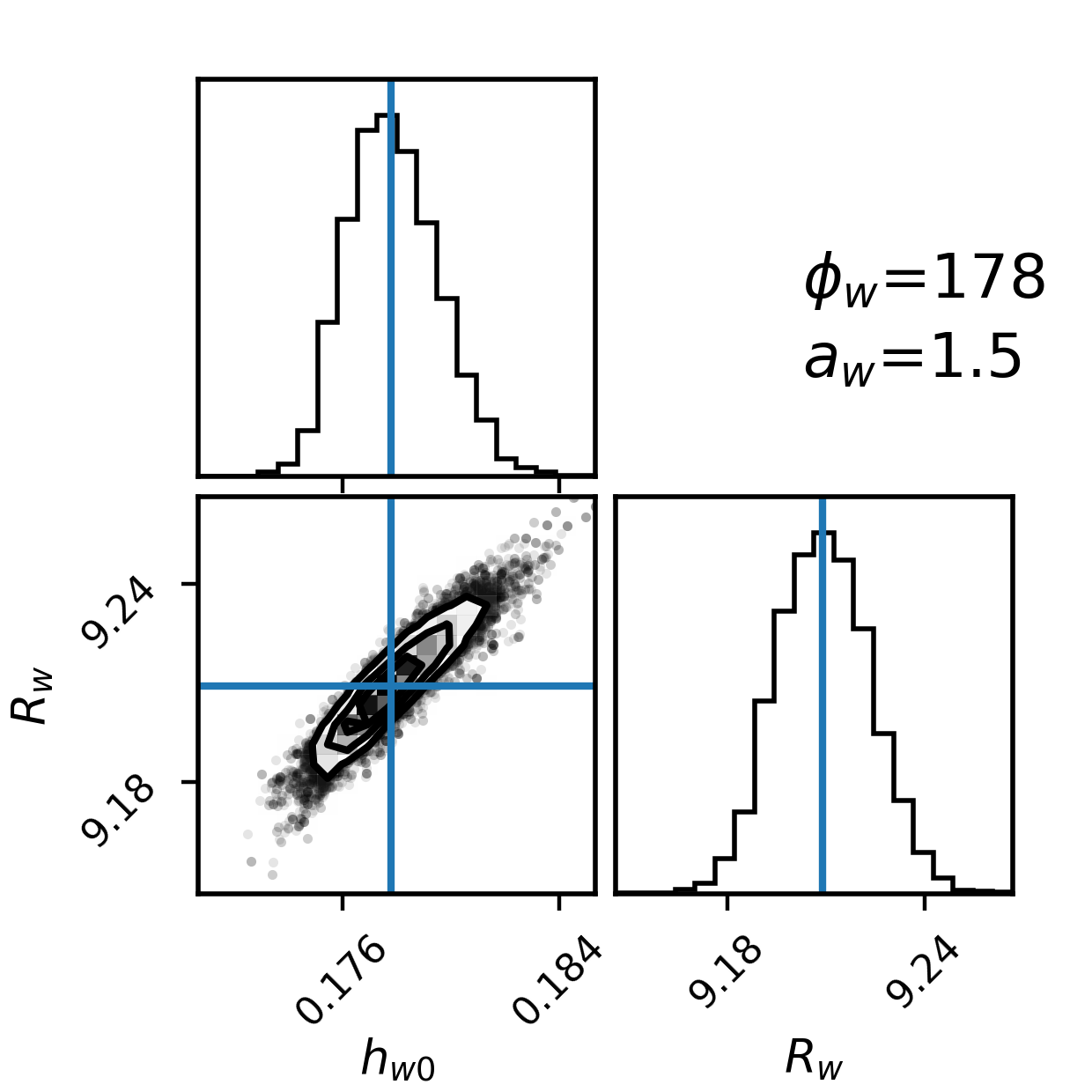}
\includegraphics[width=.49\columnwidth]{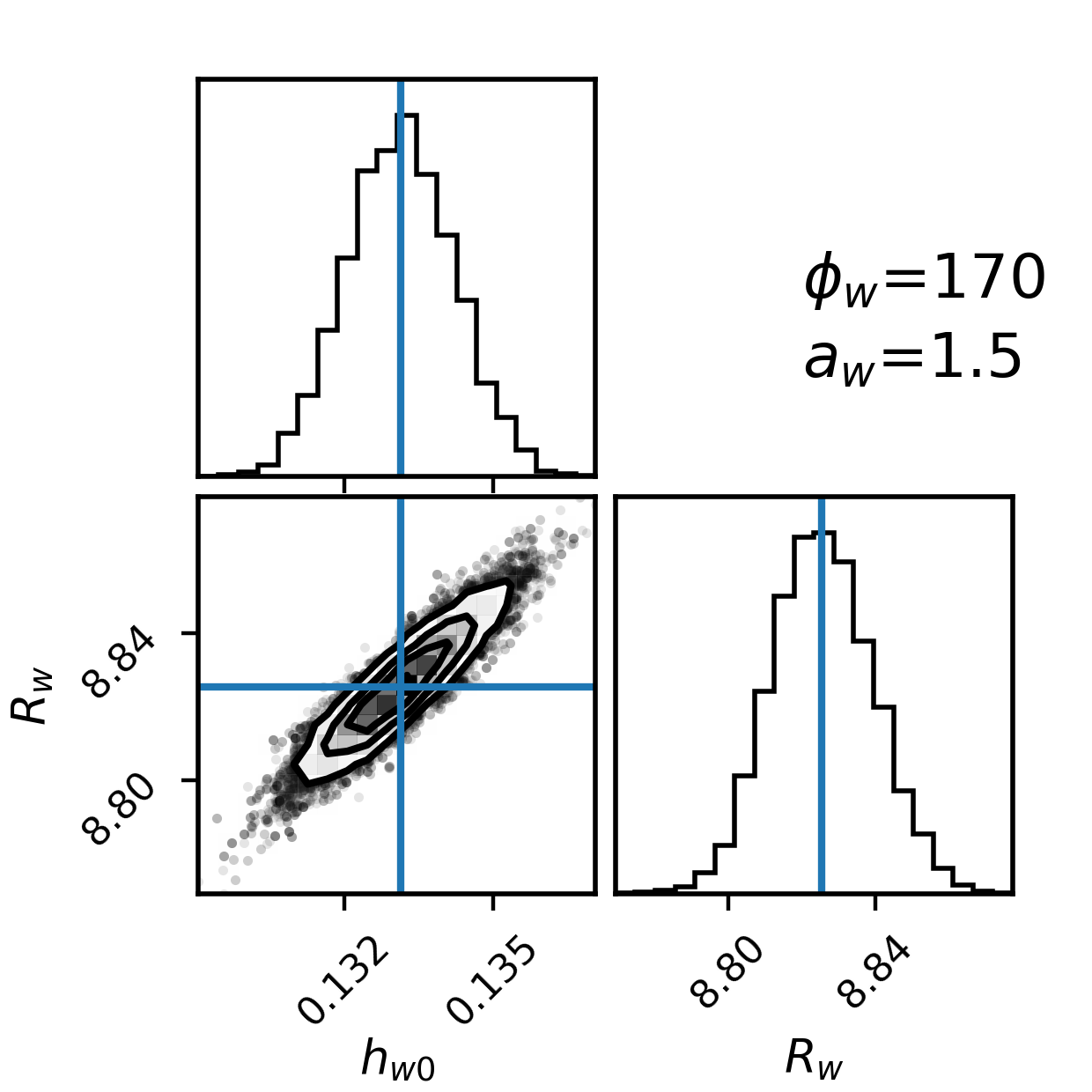}
\includegraphics[width=.49\columnwidth]{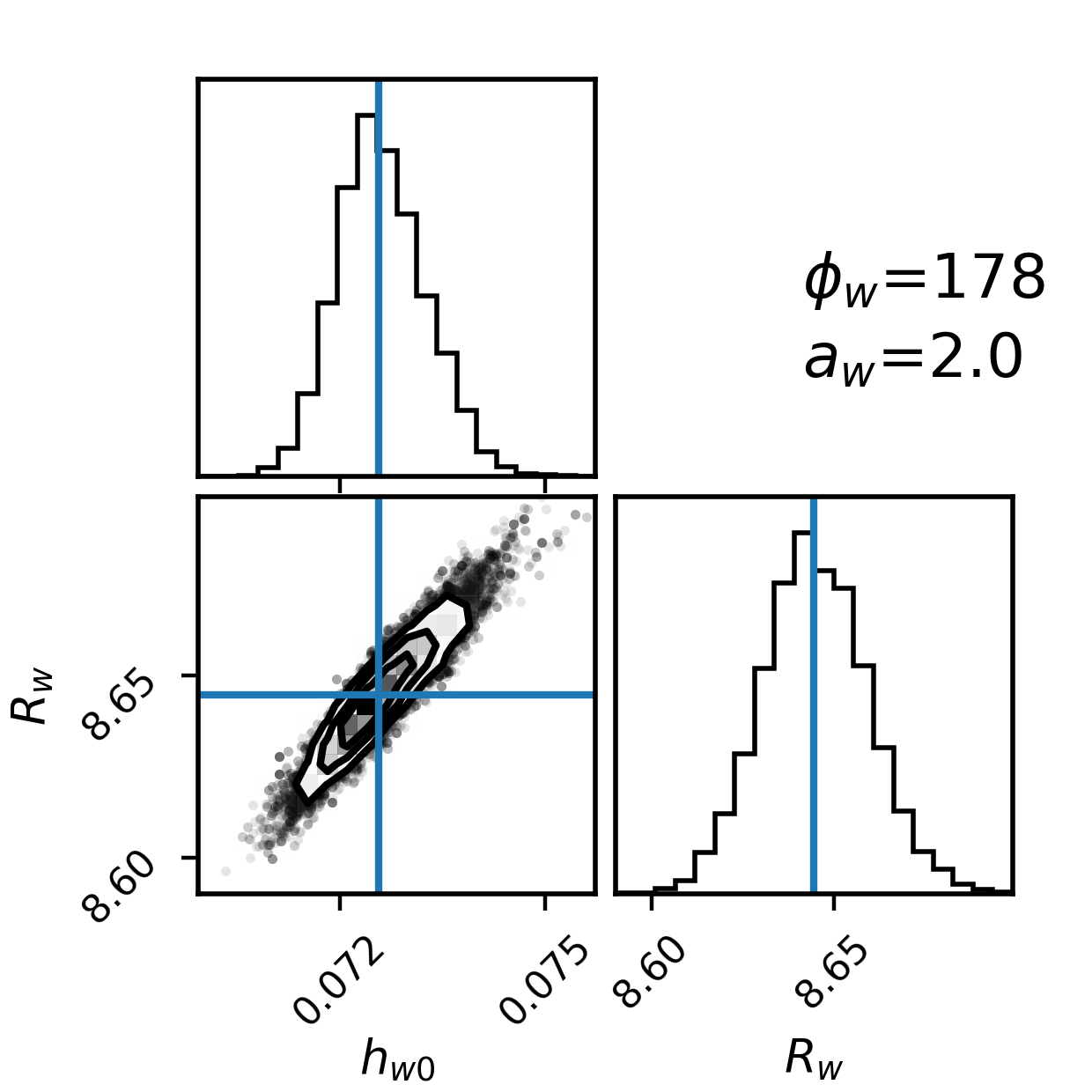}
\includegraphics[width=.49\columnwidth]{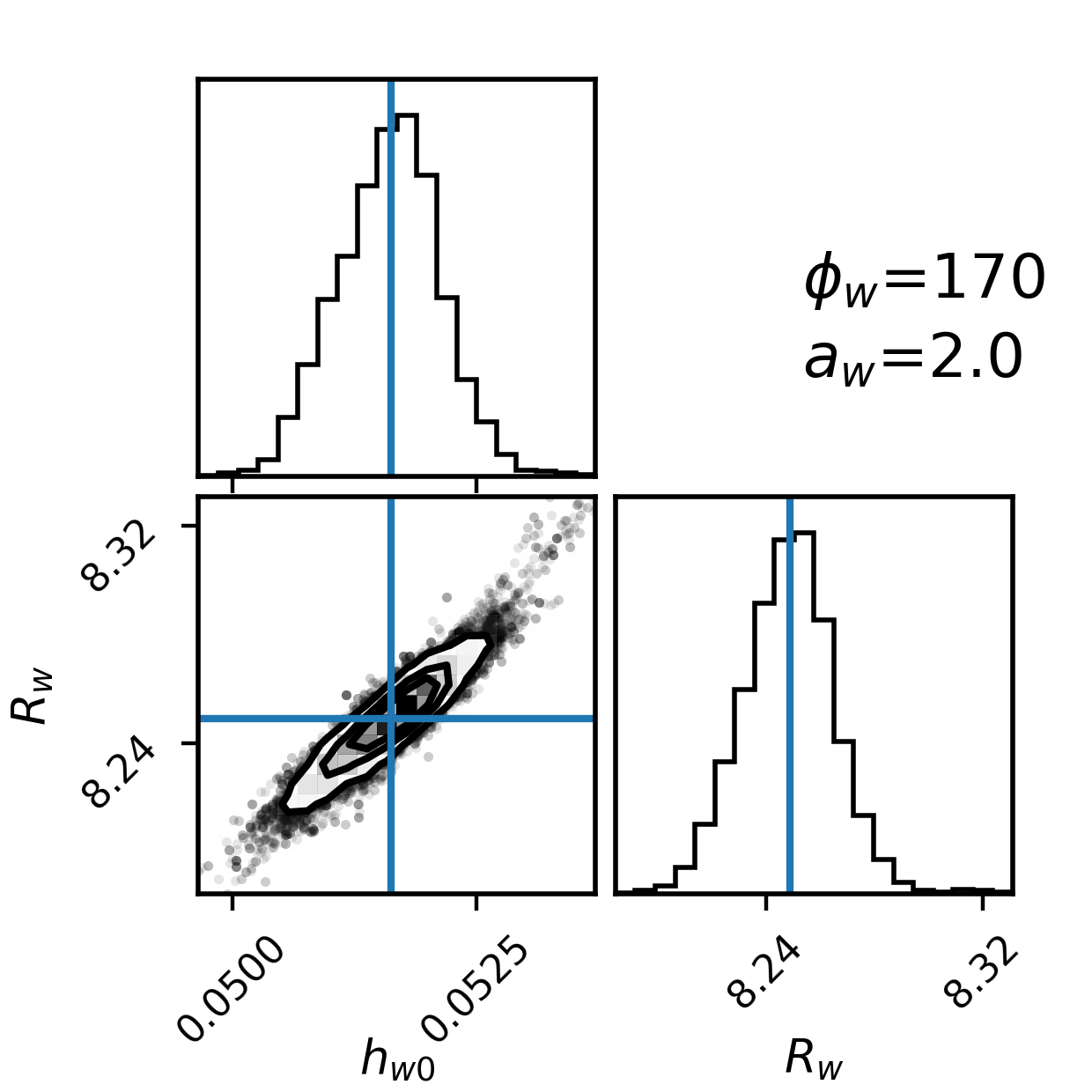}
\caption{Posterior probability distributions for two warp parameters (\rwarp{},\hwarp{}) assuming Model 2 (\autoref{tab:bestfitdatamodel}) applied to \gaiawise{[RC]}. We show four cases where the other two warp parameters (\phiwarp{},\awarp{}) have been fixed. The residuals for these are shown in Fig. \ref{fig:warp_fit_res_test}.} \label{fig:warp_fit_test}
\end{figure}

Figure \ref{fig:residuals_phiz_outer} shows that the diagonal feature already starts appearing around \rgal{}=9.5 kpc, i.e. well inside the regime where a constant LON is reasonable to assume. So, instead of adjusting all warp parameters simultaneously, we fit for only two (\rwarp{}, \hwarp{}), while keeping the other two fixed within the range of values found in the literature, at \phiwarp{}$=(178^\circ$,$170^\circ)$, and \awarp{}$=(1.0,2.0)$.
Figure \ref{fig:warp_fit_test} shows the corner plots for four of these configurations, where we find \rwarp=8.25 kpc for (\phiwarp, \awarp)$=(170^\circ,2.0)$, and \rwarp$=9.21$ kpc for (\phiwarp, \awarp)$=(178^\circ,1.5)$. Essentially, assuming that the LON of the warp varies between 2 to 10 degrees from the Galactic anti-centre allows us to place constraints on the onset of the warp at between $8.25<$\rwarp$<9.21$ kpc. In Fig. \ref{fig:warp_fit_res_test} we show the relative residuals in $\phi-$\zgal{} projection in the outer disc for these four configurations. Each column corresponds to one configuration, covering the $9.5<$\rgal{}$<14.5$ kpc part of the disc.
Figure \ref{fig:warp_fit_res_test} shows that, after having allowed for a warp, the residuals along the diagonal are much weaker. We summarise our best-fit warp parameters from this exercise in \autoref{tab:bestfitwarpmodel}, where the uncertainties in \hwarp{} \& \rwarp{} are estimated by varying the \nmin{} between 50 and 20 as before for the Model 2.

\begin{figure}
\includegraphics[width=.95\columnwidth]{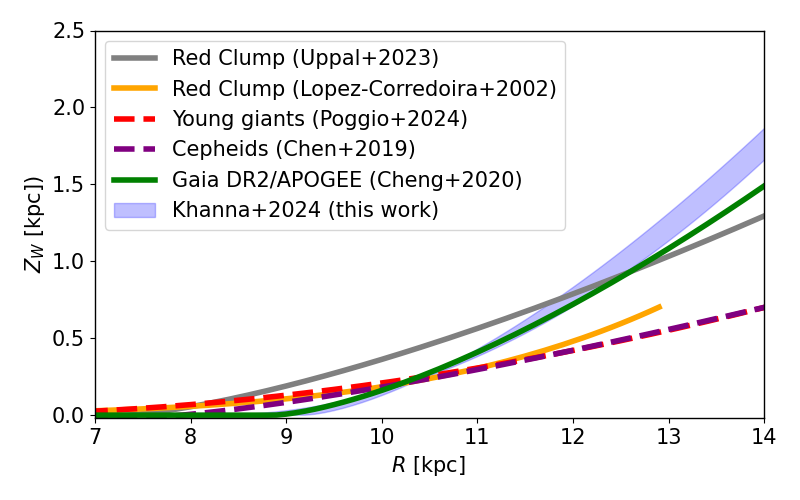}
\caption{Comparison between the warp amplitude obtained in this work and other parametrisations available in literature based on different stellar tracers. The blue shaded area shows the warp amplitude obtained in this work in the region where our dataset and our adopted approach allow us to reasonably constrain the warp shape. At larger galactocentric radii ($R \gtrsim 12$ kpc, grey shaded area), it is not possible for our approach to give a reliable estimate of the warp amplitude based on our dataset due the lack of convergence of the fitting routine (see discussion in the text).} \label{fig:warp_amplitude}
\end{figure}

Figure \ref{fig:warp_amplitude} shows different estimates of the warp amplitude available in the literature, together with the amplitude obtained with our sample, shown as a cyan shaded area for different assumed values of \phiwarp{} (between $170^\circ$ and $178^\circ$) and \awarp{} (between 1 and 2). As we can see, for $R\gtrsim$ 12 kpc the difference between the obtained warp amplitudes for different assumed \awarp{} and \phiwarp{} becomes prominent; however, the variation is still well below than 0.5 kpc at $R\sim$ 14 kpc.

For R$\lesssim$ 12 kpc, we note that our obtained warp amplitude is in broad agreement with the parametrisations by \cite{Uppal:2024warp}, based on \rc{} stars, and in reasonable agreement with \cite{Cheng:2020}, based on general stellar populations in \textit{Gaia} DR2/APOGEE, but larger  than the estimate from \cite{Lopez-Corredoira:2002}, again based on a selection of red clump stars.
This latter estimate is in better agreement with the amplitude seen in the young giants \citep{Poggio:2024} and Cepheids \citep{Chen:2019} \citep[See also other warp parametrisations based on Cepheids, e.g. ][whose estimated warp amplitudes are consistent with the Cepheid's warp amplitude shown in Fig. \ref{fig:warp_amplitude}.]{Skowron:2019,Dehnen:2023,CabreraGadea:2024}

\begin{table}
\caption{Best-fit parameters describing the warp. } \label{tab:bestfitwarpmodel}
\centering
\resizebox{0.95\columnwidth}{!}{%
\begin{tabular}{c|cccc}
\hline
\hline
Parameter   &  case I & case II  & case III & case IV   \\
(unit) &   & \\
\hline
\phiwarp{} (deg) &  178 &  170 &  178 &  170 \\ [2pt]
\awarp{} &  1.5 &  1.5  &  2.0 &   2.0 \\ [2pt]
\hwarp{} (kpc$^{1-a_{w}}$) & 0.194$\pm$0.017  & 0.136$\pm$0.006 &  0.083$\pm$0.011 &  0.052$\pm$0.002  \\ [2pt]
\rwarp{} (kpc) & 9.30$\pm$0.10 &  8.88$\pm$0.04 &  8.79$\pm$0.15 &  8.28$\pm$0.03  \\ [2pt]
\hline
\hline
\end{tabular}
}
\tablefoot{Best-fit parameters for the warp using the disc parameters from Model 2 from \autoref{tab:bestfitdatamodel} for four different cases where we keep \phiwarp{} and \awarp{} fixed.}
\end{table}

\section{Discussion}
\label{sec:discussion}

Our best-fit model suggests that the \rc{} stars are found in a two component disc (\autoref{tab:bestfitdatamodel}, Model 2), with a long (\rd{}$\sim4.24$ kpc) and flaring disc (disc1) that is also potentially warped. This disc describes the stellar distribution from the Solar circle and beyond, while a second thicker and shorter (\rd{}$\sim2.66$ kpc) component that constitutes about 66\% to the disc mass, describes well the distribution from $3<R<8$ kpc.  A natural question to ask then, is how our model compares with those of past works.

\paragraph{Scale length:} There have been several studies mapping the scale parameters of the disc, and using a wide variety of tracers. Using flux maps in the near-infrared from the COBE satellite, \cite{Drimmel:2001} found a scale length of about 2.31 kpc, which is close to what we find for the dominant component, possibly tracing the old thick disc. 

Using high fidelity spectroscopic parameters from \apg{}, \cite{Lian:2024} studied the surface density profiles of mono-abundance populations in the disc in bins of ages. They find that in the inner disc $3.5<\ $\rgal{}$\ <7.5$ kpc, the surface density profile is very flat, in agreement with high redshift Milky-Way type galaxies, while the region outside is best described by a single exponential with \rd{}$\ =2.6$ kpc. For our sample, the measured surface density profile shown in Fig. \ref{fig:residuals_surfdens} also shows a flatter profile inside of \rgal{}$<7$ kpc; however, this is in the observed data-space and so does not necessarily correspond to a flatter profile in the underlying model once selection effects are taken into account. 

On the other hand, \citet[][BV24]{bv24}, recently combined \gdrthree{} with chemistry from \apg{}-DR17, and distances from the  \textit{StarHorse} catalogue  \citep{Queiroz:2023}, to produce a chemodynamical model of the Galaxy using action-based distributions \textit{f(\textbf{J}}). In their analysis they treat the Galaxy as a composite of several disc components, and in their Table 3 list the  best-fit parameters for components of the Galactic disc, split by age. They provide $J_\phi/v_{circ}(R_0)$ = \rd{}, where $J_\phi$ is the angular-momentum of stars. They find that the old disc has a scale length of about \rd{}=2.12 kpc, while the younger discs are at about \rd{}=4 kpc. Essentially, scale length was found to decrease with age. Without an age dissection, they find \rd{}=3.6 kpc, but this also includes the much older bulge component. 

Our parameters are broadly in agreement with their analysis in that our model favours two components with long and short scale lengths respectively, and indeed our \rc{} sample is expected to be populated by a mixture of ‘young' and ‘old' stars. In fact, the age distribution of the \rc{} is not clearly determined. Using the APO-K2 asteroseismic catalogue, \cite{Warfield:2024} recently derived the age of RGB stars, and showed that the \rc{} in the traditional thin disc ($\alpha$-poor) has a broad age distribution with a peak around 3-5 Gyr, while the age distribution of the \rc{} in the traditional thick ($\alpha$-rich) disc seems to be much flatter. This blurs the distinction between the old and young disc because, as they point out, the Galaxy seems to have many young  $\alpha$-rich \rc{} stars that may have gained 
$\alpha$-rich material through mass accretion in close binaries \citep{Grisoni:2024,Yu:2024}.

\paragraph{Vertical counts (scale height):} In their best-fitting model, BV24 also provide the scale heights of the numerous disc components in terms of their vertical action as \hzsun{}= 0.012 $\times J_{z,0}$ \citep{bv23}. They find that their old disc has a scale height at the Sun of \hzsun{}$\sim0.28$ kpc, while the high-$\alpha$ component has \hzsun{}$=0.78$ kpc. Interestingly, however, their young and the middle disc components have very short scale heights of \hzsun{}$\sim$40 pc, suggesting that the youngest populations in the Milky Way are confined to a very thin sheet-like distribution. Therefore, though our `thin'-flaring disc may not be as young as the young counterpart of BV24, that it has a short scale height at the Sun is perhaps not too surprising. Meanwhile the thicker disc component in our \rc{} sample has \hzsun{}$=0.48$ kpc, and this could be due to some overlap with both the old and high-$\alpha$ component of BV24. Indeed, using extremely precise stellar parameters from \gdrthree{} (\textit{GSP-Spec}), \cite{Recio-Blanco:2024} showed the presence of a bimodality in the \rc{} on the \kiel{} diagram, that is suggested to map to the typical bimodality seen in \alfe{}-\feh{} separating the traditional thin and thick disc. However, due to the flare, our `thin' disc with a longer scale length has a larger scale height than the second `thick' disc which has a larger scale height at the Sun.  Indeed, the traditional distinction of the two disks as `thin' and `thick' is misleading. (For further discussion see \cite{2019MNRAS.486.1167B}).

\begin{figure}
\includegraphics[width=1\columnwidth]{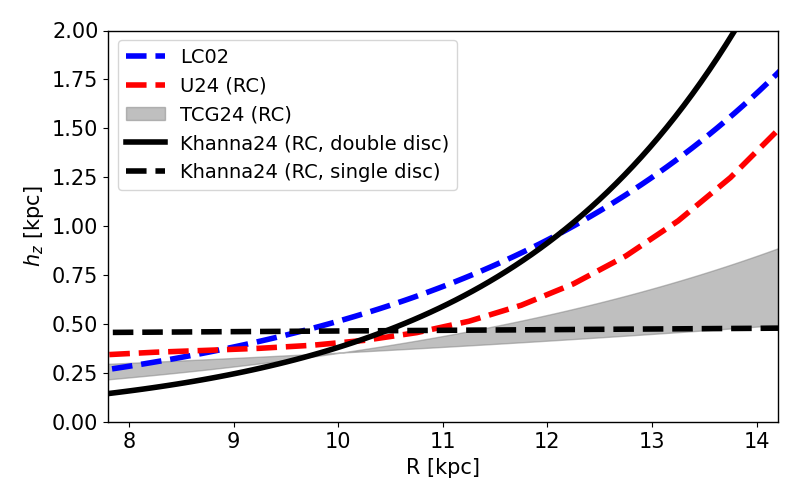}
\caption{Profile of scale height as a function of \rgal{} predicted from our best-fit model with two discs (Model 2) for \gaiawise{}[RC], shown as a black solid curve (disc1), and for the single disc model (Model 1), shown as a dashed black curve. The red and blue dashed curves and grey shaded area show the profiles from other studies of \rc{} stars in the literature. (See text for further details.)} \label{fig:flare_comp}
\end{figure}

The vertical counts in the \rc{} disc are generally well traced by the parameters of our Model 2 with two discs, as shown in Fig. \ref{fig:residuals_vert}. In the observed counts (starred points) the flare is seen as a change in slope of \nzproj{} as a function of $|$\zgal{}$|$. Specifically, we find $\log_{10}$ \rflare{}[kpc]$\sim0.36$ for the flared disc component.
Figure \ref{fig:flare_comp} shows the predicted profile of the scale height as a function of \rgal{}, shown as black curves for both the single disc (Model 1, dashed curve) and disc1 of Model 2 (solid curve). Also plotted are a few profiles of the scale height from the literature:  \citet[][LC02 hereafter]{Lopez-Corredoira:2002} studied the disc in the NIR using data from \twomass{} where the sample is dominated by \rc{}-like stars. They found the normalisation \hzsun{}$=0.31$ kpc, and a flare parameter of $\log_{10}$ \rflare{}$=0.53$ (blue curve). More recently, combining both \twomass{} and \gaia{},  \citet[][U24 hereafter]{Uppal:2024warp} also constructed an all-sky \rc{} sample and found \hzsun{}$=0.35$ kpc, but a  weaker flare parameter of $\log_{10}$ \rflare{}$=0.85$ (red curve). \citet[][TCG24 hereafter]{Cantat-Gaudin:2024rc} fit for density profiles of a combined \gaia{}-\apg{} sample of low $[\alpha/Fe]$ \rc{} stars for which high quality spectroscopic parameters are available. 
This allowed them to study the profiles by population, specifically by metallicity. Crucially, their model fitting takes into account the complex selection function of the combined sample. In the range $-0.3<$\feh{}$<0.4$, where we expect most of our \rc{} stars to be populated, they find  $0.24<$\hzsun{}$<0.31$ kpc with the flare ranging between $0.64 < \log_{10}$ \rflare{}$ <1.09$. In Fig. \ref{fig:flare_comp} we include the results from TCG24 as a shaded grey region, showing the span of the flare parameter between -0.3$<$\feh{}$<$0.3. Except for U24, all studies use an exponential parameterisation for the flare similar to us. However, all the aforementioned studies only consider a single disc component in their fitting procedure. 

Our single disc model (Model 1) in \autoref{tab:bestfitdatamodel} for just a single disc component predicts \hzsun{}$=0.46$ kpc and $\log_{10}$ \rflare{}$=2.14$, and we have over plotted this in Fig. \ref{fig:flare_comp} as a dashed black curve. It is interesting to note that between 8$<$\rgal{}$<$11 kpc, the profile is very similar to that found by U24, and also quite similar to the single exponential discs of TCG24, but offset in the scale height. However, as noted earlier, this single disc model is unable to account for the stellar counts in the outer disc. 

Our value of $\log_{10}$ \rflare{}$=0.36$ for the flare is smaller than the aforementioned studies, which would indicate a stronger flare than previously found, though looking at Fig. \ref{fig:flare_comp}, the profile is not too dissimilar to that found by LC22, except we have a smaller normalisation of \hzsun{}$=0.18$ kpc. However, we note that our model up to this point does not account for the warp, which the model might be compensating for with a stronger flare.

\paragraph{Warp:} As discussed in Sec. \ref{sec:warped}, we see a clear signature of the warp in the outer disc residuals (Fig. \ref{fig:residuals_phiz_outer}), which we attempted to add to our model.  
While the resulting warp amplitude  suggests that it may differ between young and old populations (Fig. \ref{fig:warp_amplitude}), we should bear in mind that, in this study at least, we are able to explore the warp only over a limited portion of the Galactic disc, while the warp is a large-scale feature, and we therefore should avoid over-interpreting the warp amplitude obtained here. 
In any case, Fig. \ref{fig:warp_amplitude} shows that our knowledge of the warp amplitude is quite uncertain, especially for the old populations.
Moreover, we note that while the residuals in the outer disc (shown in Fig. \ref{fig:warp_fit_res_test}) are lower than those in the non-warped case, they do not disappear completely. This may be due in part to assuming an over-simplified warp geometry with a straight line of nodes. whereas the young populations clearly show the line-of-nodes remains straight only for $R \simeq 12\kpc$ \citep{Dehnen:2023,Poggio:2024}. Nevertheless, our tests represent a first step in the right direction, confirming the warp signature in our \rc{} sample, but there is still need for further exploration in future work.

\begin{figure}
\includegraphics[width=1.\columnwidth]{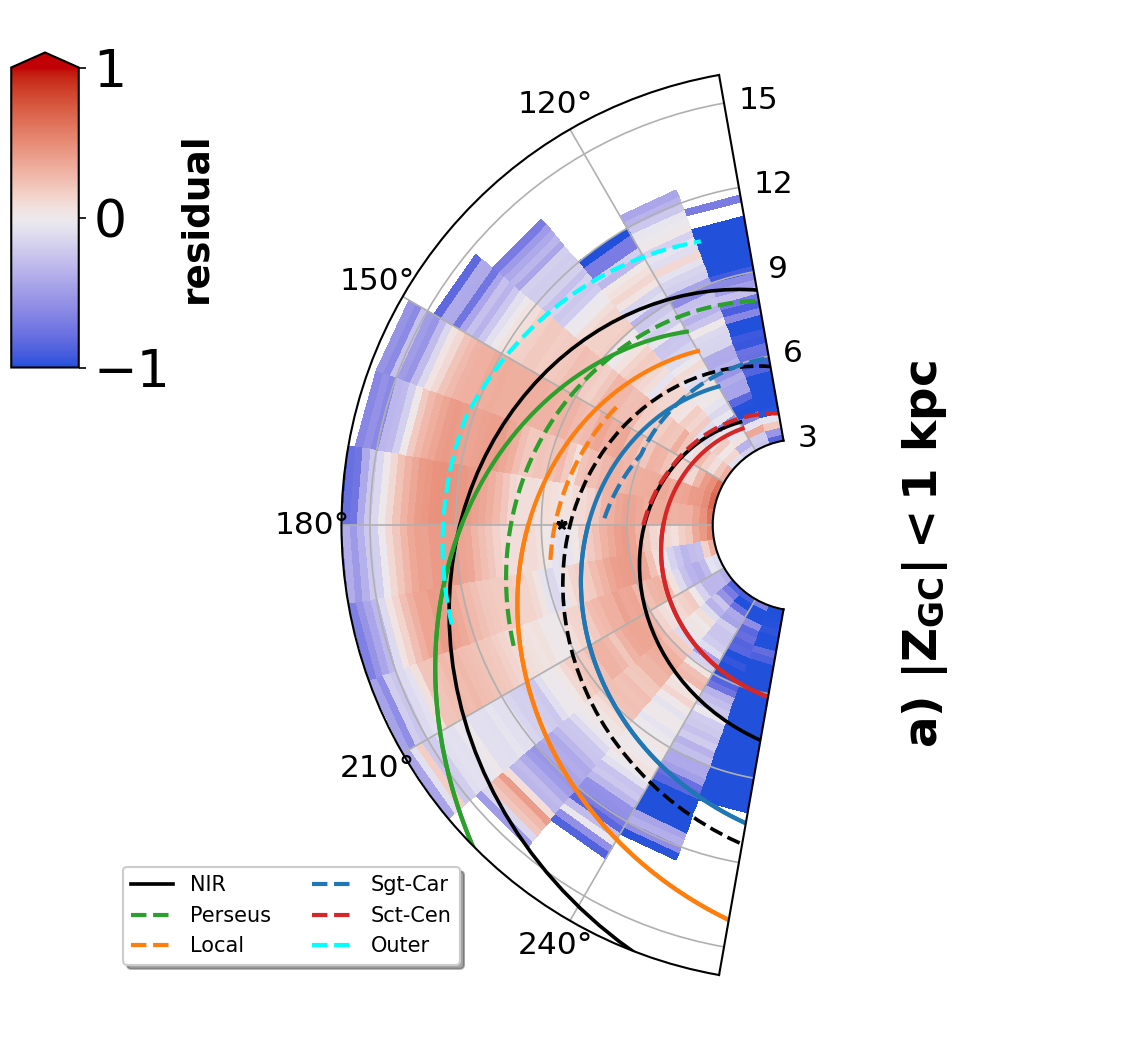}
\caption{Same as Fig. \ref{fig:residuals_polar} but for the $|$\zgal{}$|<$1 kpc slice. Various spiral arm models are overplotted: Coloured dashed lines show model based on masers from  \cite{Reid:2019}, coloured solid lines show model based on Cepheids \citep{skowron:2024,Drimmel:2024}, and in black is the two-arm NIR model from D00.} \label{fig:residuals_polar_1kpc}
\end{figure}

\paragraph{Two-armed Spiral:} In the previous section we also discussed the residuals between the two-disc model and the data projected on the Galactic plane (Fig. \ref{fig:residuals_polar}), noting that they seem to show evidence of a two-arm spiral pattern with geometry similar to that inferred from NIR data (D00). In Fig. \ref{fig:residuals_polar_1kpc} we show the residuals for $|Z_{\tt GC}|<1\kpc$ also with the geometry of spiral arm models based on masers \citep[][hereafter R19]{Reid:2019} and the Cepheids \citep{Drimmel:2024}. We find that the location of the outer residuals also roughly coincides with the location of the Outer Arm proposed by R19. 
The Perseus arm as mapped by the Cepheids and the HI \citep{Levine2006} also cross this region, though with an apparently higher pitch angle. 
Recently \citet{Uppal:2023spiral} also claim to have detected the Outer Arm in a sample of RC stars selected using \gaia\ and 2MASS photometry using the overdensity mapping technique of \citet{Poggio:2021}.  However, we note that the arm detected here in the RC residuals is much broader than the one they report. This is likely due to their choice of bandwidths for calculating the overdensity, restricting their sensitivity to features with scale lengths between 0.3 and 2\kpc. In addition, the overdensity mapping approach may be highlighting a small fraction of young stars transiting in their selected RC region of the colour-magnitude diagram as they evolve from the main sequence to the giant branch. Finally, though we note generally higher residuals in the first quadrant between $6<R<8\kpc$ from $130^\circ<\phi<160^\circ$,
we do not detect a feature in the residuals that might correspond to the Local arm, especially near the disc midplane ($|Z_{\tt GC}|<0.25\kpc$, see middle panel of Fig. \ref{fig:residuals_polar}), where the Local arm instead falls in an under-dense portion of the disc that lies along the expected inter-arm region of the NIR arms. This is in contrast with \citet{Lin2022}, who also use the overdensity mapping approach for a selection of RC stars, and may suffer the same contamination issues mentioned above.  

In summary, if our residuals are mapping a spiral perturbation in the old stellar population, the geometry and profile of this perturbation is more consistent with a broad two-armed spiral than the multiple arms traced by 
young populations and star formation products \citep[R19;][; Gaia23]{Zari:2021,Poggio:2021}. 
However, we should bear in mind that the residuals shown in Fig. \ref{fig:residuals_polar} might also be influenced by additional effects. For instance, any deviation of our best-fit parametrisation (especially in the case of two or more disc-like components) with respect to the unknown true underlying density distribution might result in radial over/under-densities, which might interfere with interpreting any possible spiral arm signature. In addition, our residuals are with respect to our model without a warp, which may introduce additional residuals in the outer disc.

Nevertheless, that the geometry and width of the spirals as traced by the RC stars might not be the same as that seen in the young populations should not be a surprise. Presumably the \rc{} population is tracing the old, kinematically relaxed stellar population that dominates the underlying mass distribution of the stellar disc, while young stars are a product of the response of the gas to a disc with multiple pattern speeds and, possibly, external perturbations \citep{Purcell:2011, Pettitt2016}. If the residuals from our axisymmetric disc model are tracing a spiral perturbation in the old stellar disc of the Milky Way, their magnitude suggests that this perturbation is at the level of 10 to 20\%.  These residuals in the outer disc extend to a galactocentric radius of at least 14-15 kpc, well beyond the outer Lindblad radius of the bar which is at a radius of 10-12 \kpc based on recent estimates of the bar's pattern speed \citep[][Gaia23]{Binney2020,Chiba2021,Dillamore2024,Lucchini2024}, so that any spiral perturbation responsible for these residuals must have a lower pattern speed than the bar, as a spiral arm perturbation is expected to be located between its own inner and outer Lindblad resonances. 

\section{Summary}
\label{summary}

In this paper, we have constructed an all-sky catalogue of \rc{} candidates for which we could derive reliable distances using \mir{} photometry from \allwise{}. This sample of nearly ten million stars allowed us to map the stellar number density for galactocentric radii between $3< $ \rgal{} [kpc] $<14$ and at a distance from the Galactic mid-plane of $|$\zgal{}$|<2$ kpc. Taking advantage of the near Gaussian luminosity function of the \rc{} and 3D extinction maps, we built a selection function tailored to our sample, i.e. the probability of observing an RC star at a given location on the sky at a given magnitude. The selection function then allowed us to build a three-dimensional completeness map of our sample and to fit a range of models to the observed stellar density. It is only by taking into account the selection function that we can use the RC sample to trace a large fraction of the Galactic disc. Indeed, limiting ourselves to a sample with 90\% completeness would limit our volume to within only 2 to 4 \kpc of the Sun (see Fig. \ref{fig:sf_polar}). 

Our best model consists of two discs, where each is exponential both radially and vertically. Specifically, the \rc{} population of the Galaxy seems to be well described by a long and flared disc component that makes up about 36\% of the stellar mass. It has a scale length of \rd{}$=4.24$ kpc and scale height at the Sun of \hzsun{}$=0.18$ kpc, flaring to 2 kpc at $R=14\kpc$ 
The remaining 66\% of the mass is in a disc with a scale length of \rd{}$=2.66$ kpc and a scale height of \hztwo{}$=0.48$ kpc without a flare. So, while our model is purely geometrical, it is interesting that our two-component disc model recovers two disc components similar to those seen when mapping the \alfe\ distribution as a function of $R$ and $z$, namely, a disc with a short scale length and thicker scale height dominating the inner Galaxy and a second flaring disc with a longer scale length dominating the outer Galactic disc \citep{Hayden:2015}. This suggests that while we have not made a chemical separation of our RC sample, we apparently recover the same two-disc morphology as evidenced in the \alfe\ distribution. 

By subtracting the best-fit model from the data, we found residuals in both the inner and the outer disc. These residuals show a clear warp signature beyond \rgal{} $>10$ kpc, and in the Galactic plane, they show a signature that seems to coincide with the expected location of the spiral arms inferred from the NIR \citep[D00; ][]{Drimmel:2001, Benjamin2005, Hou2015}, corroborating the two-armed geometry implied by the observed spiral arm tangents at these wavelengths and implying a perturbation in the mass surface density of 10 to 20\%. Confirming these results will require better modelling of the outer disc to fully take into account the warp and/or a deeper catalogue towards the inner Galaxy, where extinction still limits the volume we are able to sample. 

We have modelled the density of the \rc{} stars with a purely geometrical model, which requires sufficient knowledge of the extinction in three dimensions as well as the combined selection function of the photometric surveys used.  Because of these stringent conditions for mapping the three-dimensional density in the Galactic plane, earlier attempts to verify the existence of a spiral perturbation in the mass density of the stellar disc have relied on the measured velocities of tracers \citep{Grosbol2018,Eilers2020}. Recently, \citet{Palicio:2023} detected spiral-like features in the radial action, $J_R$, of disc stars. However, since \gdrthree{} it has become clear that the bar of the Milky Way dominates the velocity field of the disc well beyond co-rotation (Gaia23), thus complicating any possible interpretation of the disc's velocity field.  Nevertheless, we intend to study the velocities of the RC sample in the future. Indeed, including this additional information may allow us to better constrain the warp in this old stellar population. 

\section*{Data availability}
The \rc{} catalogue corresponding to Fig. \ref{fig:data_sky_RZ}, is available at the CDS via 
\url{https://cdsarc.cds.unistra.fr/viz-bin/cat/J/A+A/701/A270}.

\begin{acknowledgements}
We thank the anonymous referee for their constructive comments.
SK \& RD acknowledge support from the European Union's Horizon 2020 research and innovation program under the GaiaUnlimited project (grant agreement No 101004110). SK acknowledges use of the INAF PLEIADI@IRA computing resources, and would like to thank Hai-Feng Wang \& Paul McMillan for useful suggestions. RD \& EP are supported in part by the Italian Space Agency (ASI) through contract 2018-24-HH.0 and its addendum 2018-24-HH.1-2022 to the National Institute for Astrophysics (INAF). We also thank Kevin Jardine for visualisation of the selection function.\\

This work presents results from the European Space Agency (ESA) space mission Gaia. Gaia data are being processed by the Gaia Data Processing and Analysis Consortium (DPAC). Funding for the DPAC is provided by national institutions, in particular the institutions participating in the Gaia MultiLateral Agreement (MLA). The Gaia mission website is https://www.cosmos.esa.int/gaia. The Gaia archive website is https://archives.esac.esa.int/gaia. \\

This work made use of the \url{https://gaia-kepler.fun} crossmatch database created by Megan Bedell.
This work has used the following additional software products:
\href{http://www.starlink.ac.uk/topcat/}{TOPCAT}, and \href{http://www.starlink.ac.uk/stilts}{STILTS} \citep{Taylor:2005};
Matplotlib \citep{Hunter:2007};
IPython \citep{PER-GRA:2007};  
Pandas \citep{reback2020pandas}; 
Astropy, a community-developed core Python package for Astronomy \citep{AstropyCollaboration:2018}; NumPy \citep{harris2020array}; Vaex \citep{vaex2018A&A...618A..13B}; and \href{https://spiralmap.readthedocs.io/en/latest/?badge=latest}{\texttt{SpiralMap}} \citep{spiralmap}. 
\end{acknowledgements}

\bibliographystyle{aa} 
\bibliography{mybib}

\begin{appendix}
\label{sec:app}

\section{Intrinsic correlation plot}
\label{app:intrinsic_corr}

\begin{figure}
    \includegraphics[width=1.\columnwidth]{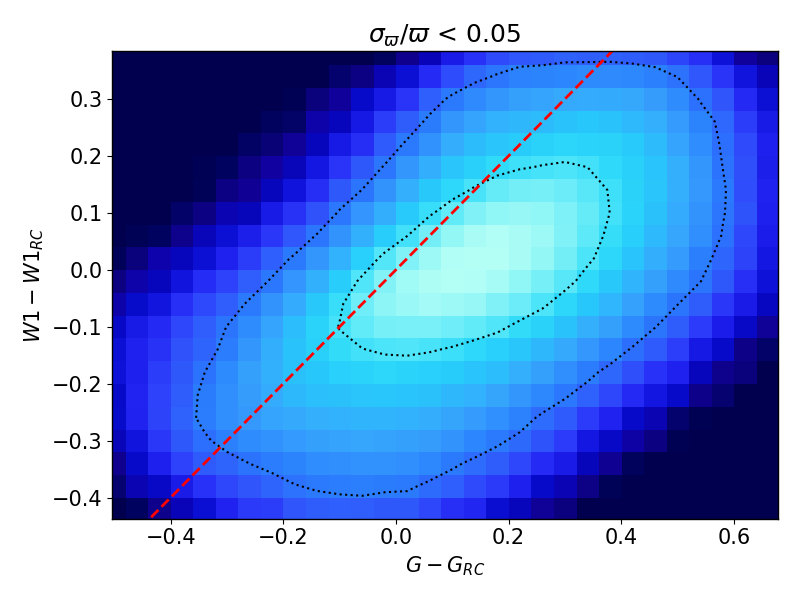}
     \caption{Intrinsic correlation between the $G$ and $W1$ bands for \rc{} stars with \errorovparallax{}$<0.05$. The (1,2)$\sigma$ contours of this distribution are marked in black, the red dotted line is the 1:1 relation, while the \textit{Pearson} correlation coefficient for this set is 0.43.}
      \label{fig:intrinsic_corr}
\end{figure}

For the sample of stars from \gaiawise{} where \errorovparallax{}$<0.05$, we can infer their absolute magnitudes from their inverse parallax distances ($d_{\varpi^\prime} $). Retaining those with inferred absolute magnitudes within $1\sigma$ values from literature, we can assume these are consistent with being on the red clump phase. To these stars then, we assign an absolute magnitude from the Gaussian $\mathcal{N}(\Bar{M_{\lambda}}, \sigma_{\Bar{M_{\lambda}}})$, and using $d_{\varpi^\prime} $, can predict their apparent magnitudes in both passbands ($G_{RC},W1_{RC}$).
Figure \ref{fig:intrinsic_corr} shows the distribution of difference between the true and predicted apparent magnitudes, for which we find the \textit{Pearson} correlation coefficient $\sim0.42$.

\section{Distance posterior PDF from \cbj{}}
\label{app:bailerjones}
For clarity, we reproduce here the relations to obtain the posterior PDF from \cbj{}. The geometric distance prior is written in the form of a generalised gamma distribution, 
\begin{equation}
    P(r|hpix) = \frac{1}{\Gamma(\frac{\beta + 1}{\alpha})} \frac{\alpha}{L^{\beta+1}}r^{\beta}e^{-(r/L)\alpha},
\end{equation} valid for distance, $r\geq0$. Given a HEALpixel (level 5), $hpix$, one can obtain the parameters $\alpha,\beta$, and $L$ from the \href{https://www2.mpia-hd.mpg.de/homes/calj/gedr3_distances.html}{auxiliary file} provided by \cbj{}. Then, using the likelihood function, 
\begin{equation}
    P(\varpi^\prime|r,\sigma_{\varpi}) = \mathcal{N}(\varpi^\prime - \frac{1}{r},\sigma_{\varpi}), 
\end{equation} where $\varpi^\prime = \varpi + 0.017$, we can write the posterior PDF as a product of the likelihood and the prior as
\begin{equation}
    P_{g}(r|\varpi^\prime,\sigma_{\varpi},hpix) = P(\varpi^\prime|r,\sigma_{\varpi}) P(r|hpix)
\end{equation}

\section{\rc{} validation}
\label{sec:rc_validation}
\subsection{Selecting red clump from stellar parameters}
\label{sec:rcsel_spectro}
\begin{table}[h!]
\small
\centering
\caption{Best-fit coefficients for Eq. \ref{direct_calib_eqn}.\label{tab:calib_table}}
\begin{tabular}{l|l|l|l|l|l|l}
\hline
\hline
Population & $a_{0}$ & $a_{1}$ & $a_{2}$ & $a_{3}$ & $a_{4}$ & $a_{5}$ \\
\hline
Giants & -0.957 & 0.000 & -0.006 & -0.020 & 1.489 & 0.002 \\
Red clump & -0.800 & 0.046 & 0.008 & -0.060 & 1.199 & 0.132 \\
\hline
\end{tabular}
\tablefoot{Used to derive \jkzero{} for Giants and the red clump in K19. The fitting was carried out over the temperature range 4200$<$\teff$<$8000.}
\end{table}

In this paper, our \gaiawise{}[RC] sample is selected by combining \gaia{} astrometry \& photometry with \allwise{} photometry. Ultimately, however, a purer sample of \rc{} stars can be selected using spectroscopic stellar parameters (\logg{}, \feh{}, \teff{}). 
In K19, we had developed such a scheme to select high fidelity \rc{} stars, and we provide a short summary of this below:

\begin{eqnarray}
1.8 \leq \logg \leq 0.0018\, \mathrm{dex}\, \mathrm{K}^{-1}\,\,\Big(\teff-\teff^{\mathrm{ref}}(\feh)\Big)+2.5\ , \label{logg_cut}\\
Z  >  1.21 [(J-K)_{0} -0.05]^{9} + 0.0011 \label{Z_Cut1},\\
Z  <  {\rm Min}\left(2.58 [(J-K)_{0} -0.40]^{3} + 0.0034, 0.06\right) \label{Z_Cut2},\\
0.5  < (J-K)_{0} <0.8 \label{clr_cut},
\end{eqnarray}

where

\begin{equation}
\teff^{\mathrm{ref}}(\feh) = -382.5\,\mathrm{K} \, \mathrm{dex}^{-1}\, \feh+4607\,\mathrm{K}\
\end{equation} and $Z = Z_{\odot}10^{\feh}$, with $Z_{\odot} = 0.019$. Here, \clr{} is the intrinsic colour in \twomass{} passbands for RC stars. One can derive the intrinsic colour solely from stellar parameters as

\begin{equation}
\label{direct_calib_eqn}
(J-K)_{0} = a_{0} + a_{1}X + a_{2}X^{2} +a_{3}XY + a_{4}Y + a_{5}Y^{2} ,
\end{equation}

where $X =$[Fe/H] and $Y=5040\, \mathrm{K}/T_{\rm eff}$, and the coefficients as listed in \autoref{tab:calib_table}. 

\begin{figure}
\centering
\includegraphics[width=1\columnwidth]{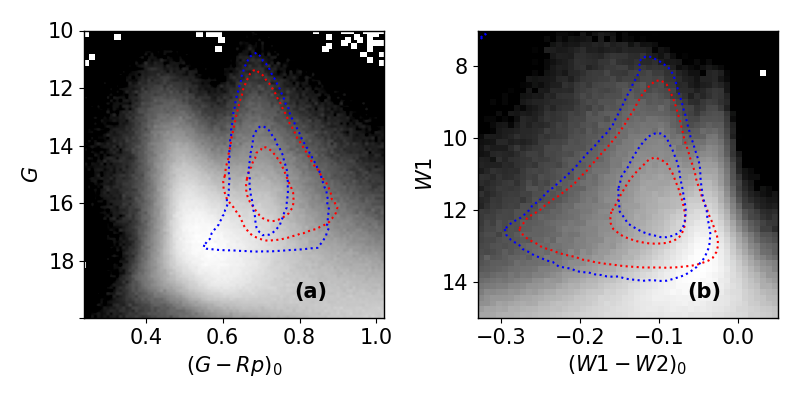} 
\caption{Colour magnitude diagram for the \gaiawise{} parent catalogue shown for \gaia{} and \allwise{} colours. In each panel, we indicate the selected RC sample using the red contours ($1\sigma,2\sigma$). Also, shown in blue contours are the spectroscopically selected RC from \andrae.} \label{fig:cmd_ovplot}
\end{figure}

\begin{figure}
\includegraphics[width=1.\columnwidth]{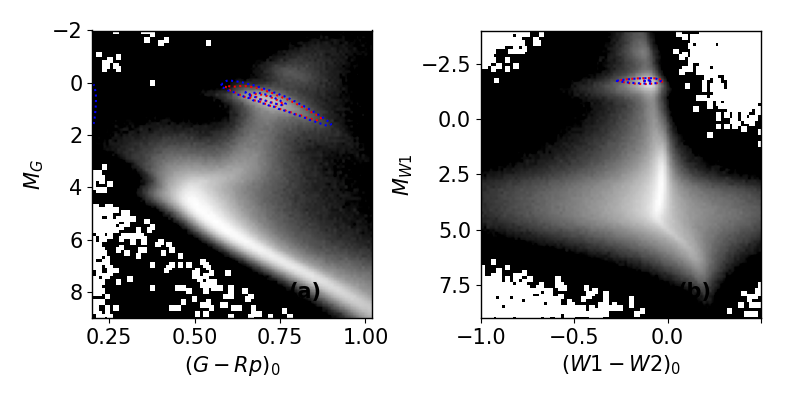}
\caption{Colour-absolute magnitude diagram for the \gaiawise{} dataset, with \absg{} vs (\grp{})$_{0}$ shown in panel (a), and \absw{} vs $(W1-W2)_{0}$ shown in panel (b). The red contours in each panel represent the selected \gaiawise{}[RC] stars. Also, shown in blue contours are the spectroscopically selected RC from \andrae. } \label{fig:camd}
\end{figure}

\paragraph{\andrae:} In this paper, we only use the relations above for validation. In particular, we select \rc{} candidates from the recently published stellar parameter catalogue of \cite{Andrae2023SP}(\andrae) who derived these for \gaia{} stars with XP spectra and \allwise{} photometry (about 125 million in all). We are able to select 6,840,662 \rc{}-like stars from their catalogue, and find that 63\% of these are present in \gaiawise{}[RC].
Figure \ref{fig:cmd_ovplot} shows the CMD in both \gaia{} and \allwise{} colours for the \gaiawise{} parent sample in grey, with the \gaiawise{}[RC] shown as red contours and the sample selected from \andrae{} in blue contours. In Fig. \ref{fig:camd} we use the distance priors from \cbj{} to show the \camd{} for these in both \gaia{} and \allwise{} colours, with the two \rc{} samples once again indicated by red and blue contours. Our photo-astrometrically selected \rc{} and that spectroscopically selected from \andrae{} trace very similar distributions in these figures, though our selection is a bit more conservative, covering a smaller area in $(\log g, \teff)$ space.

\begin{figure}
\includegraphics[width=1.\columnwidth]{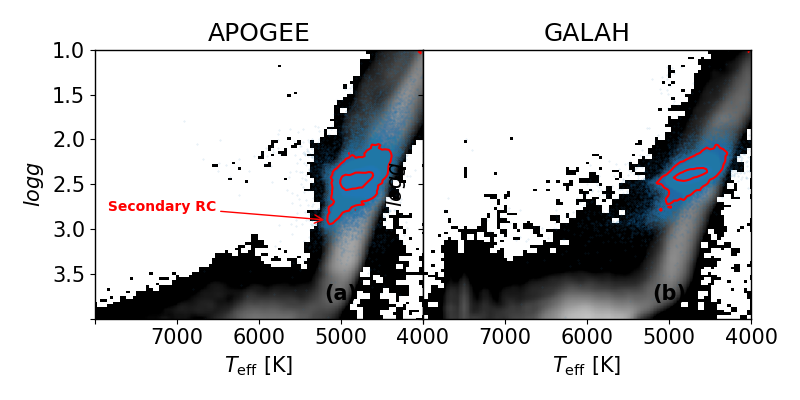}
\caption{External validation of \gaiawise{}[RC] against the spectroscopic catalogues, \apg{} (panel a) and \glh{} (panel b). For each survey we show the overall \kiel{} diagram, with the common sources in our catalogue overplotted (blue points). Also, shown is the (1,2)$\sigma$ density contours for the common sources. Note: the presence of an overdensity around (\teff{},\logg{})$=(5100$ $K ,2.9)$, which is likely the secondary red clump population.  \label{fig:kiel_apogee_galah}}
\end{figure}

\paragraph{\apg{} and \glh{}:} Next we consider the two main contemporary high resolution spectroscopic surveys \apg{} \citep[][DR17]{apogeedr17} and \glh{} \citep[][DR3]{galahdr3}. For each, we use their internal crossmatches to \gedrthree{} source\_id. We find about 150,000 of our \rc{} sources in common with the two surveys combined. The \kiel{} diagram for \apg{} \& \glh{} surveys is shown in Fig. \ref{fig:kiel_apogee_galah}, with our \rc{} selection overlaid as contours at (1,2)$\sigma$, which lie within 1.8$<\log g<$3.0. Interestingly, in both comparisons, we also note the presence of an overdensity around (\teff{}, \logg{})$=(5100$ $K ,2.9)$. This would be the secondary red clump population, which are more massive and younger counterparts to the typical \rc{}, burning helium in their cores after leaving the RGB without having experienced a helium flash \citep{Girardi:2016}.

\subsection{Selecting \rc{} in Galaxia and metallicity CDF}
\label{app:rc_feh_cdf}
\begin{figure}
\includegraphics[width=1.\columnwidth]{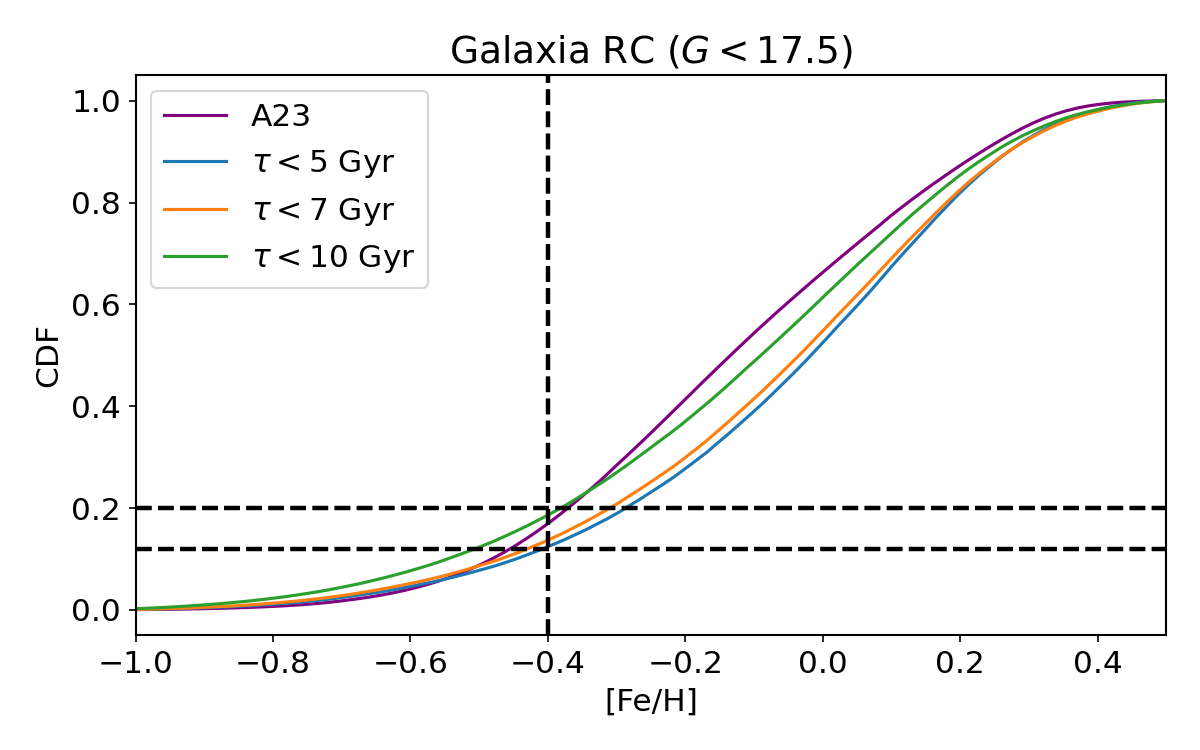}
\caption{Cumulative distribution function (CDF) of the metallicity of \rc{} population selected from \galaxia{} for a magnitude limited survey at $G<17.5$. Three CDF curves are shown, each for a cut in maximum stellar age at $\tau=$(5,7,10) Gyr. The horizontal black lines indicate CDF=(0.12,0.2), while the vertical black line indicates \feh=$-0.4$. Also shown is the CDF for the spectroscopically selected \rc{} from \andrae{}. } \label{fig:rc_feh_cdf}
\end{figure}

In order to look at properties of \rc{} stars, we generated an all-sky mock catalogue with magnitude limit of $G<$17.5, using the \galaxia{} population synthesis code \citep{Sharma:2011}, which implements the \textit{Besancon} Galaxy model \citep{Robin:2003}. The \textit{warpflare} option was set to 1.  
\galaxia{} allows for sampling of stars in a continuous fashion across the sky, and also provides photometry, in most of the commonly used bands.
From the mock, we selected \rc{} stars using the scheme laid out in Sect. \ref{sec:rcsel_spectro}. Additionally, we select only those stars that have an initial stellar mass greater than their RGB tipping mass, and that are present in the Galactic disc. The \camd{} in both \gaia{} and \allwise{} colours is shown in Fig. \ref{fig:glx_rc_camd}, with the \rc{} population indicated as black contours. For this sample, the Cumulative Distribution Function (CDF) of their metallicity in Fig. \ref{fig:rc_feh_cdf} for three cuts in stellar age at $\tau=$(5,7,10) Gyr is shown in Fig. \ref{fig:rc_feh_cdf}, where we see that in general the fraction of stars with \feh{}$<$-0.4 ranges around 15-20\% for the typical ages of \rc{} stars ($\tau<$5 Gyr). For completion, Fig. \ref{fig:rc_feh_cdf} also shows the metallicity CDF of \rc{} that we select from \andrae{} (purple curve), and this is consistent with that seen in \galaxia{}.

\subsection{Comparison with asteroseismology}
In general the selection scheme laid out in Sect. \ref{sec:rcsel_spectro} should be able to select high fidelity \rc{} stars. However, given the close overlap with the RGB and secondary \rc{} populations in both photometric and spectroscopic spaces, it is easy to also select contaminants. The gold-standard for distinguishing between these phases of stars is asteroseismology - the study of stellar oscillations. This technique is sensitive to the internal structure of stars and thus allows one to select pure \rc{} stars. In red giant stars, these oscillations are excited by near-surface convection, and are broadly characterised as two dominant modes:  a) acoustic ($p-$mode): the propagation zone of such modes is the outer envelope of the star, with pressure acting as the restoring force. Observationally, these modes are approximately equally spaced in oscillation frequency; b) gravity ($g-$mode): such modes propagate in the interior of the core, and buoyancy acts as the restoring force here. Pure $g-$modes are approximately equally spaced in their period. In evolved stars, coupling between the $p$ and $g$ modes results in the so called mixed modes. In their seminal work, \cite{Bedding:2011} showed that such mixed modes in red giant stars can help distinguish between the \rgbp{} and the \rc{} phases. 

This relies on their different distributions in the $\Delta \nu$-$\Delta P$ plane. Here, $\Delta \nu$ is the frequency separation between adjacent p−modes with the same angular degree  but at different radial order, while $\Delta P$ is  the period spacing between adjacent mixed modes. Fundamentally, this is due to the differences in the structure of the cores of the two types. While, \rgbp{} have a radiative core, \rc{} stars have a fully convective core. It turns out that mixed modes cannot stably propagate though convective cores, but can do so in the radiative regions above the core in \rc{} stars. Since, the radiative region in \rc{} stars is smaller compared to \rgbp{} stars, this has the consequence of larger \periodspacing{}, thus making it a key discriminant between the two stellar phases \citep{Hekker:2017}.
We now compare our sample to three independent publicly available catalogues that use asteroseismology to classify stellar phases. For each of the following, we provide the precision, recall and accuracy, as quality markers, defined as Precision = (True positive/Results), Recall = (True positive/ True positive + False Negative), and Accuracy = (True positive + True Negative) / (Total - Unclassified).

\citet[][\jie{} {\rm hereafter}]{Yu:2018} carried out an extensive asteroseismic analysis on a sample of 16,094 red giants with light curves from the \kepler{} Space Telescope \citep{Borucki:2010,Koch:2010}. In addition to global seismic parameters, their catalogue also provides a classification for the stellar phase, given as [0:`unclassified' (\textbf{706}), 1: \rgbp{} (\textbf{7685}), 2: \rc{} (\textbf{7703})], with the number of each quoted in bold. We make use of the $1$ arcsec spatial crossmatch performed for \kepler{}\footnote{\url{https://gaia-kepler.fun/}}  stars with \gdrthree{} in order to grab the \gaia{} source identifier (ID) for the \jie{} sample. We find 8327 stars in common between our catalogue and \jie{}, out of which $78\%$ were classified as phase 2 by the latter. Taking into account the `impurity' (RGB) in our sample, with respect to \jie{}, our selection achieves an accuracy of about $80\%$.

In the literature, one finds a diverse range of techniques employed in the classification of the stellar phase of oscillating giants, and the results are also varied. \citet[][\elsworth{} {\rm hereafter}]{Elsworth:2019} 
compiled a `consensus' evolutionary state for 6000 red giant stars, from four different methods of analysing the frequency-power spectrum. This sample has been observed by \kepler{}, and also by the \apg{} and/or \sdss{} surveys. In their catalogue, they assign the following primary classifications, 'RGB' (red giant branch), 'RC' (red clump), '2CL' (secondary red clump), 'AGB' (asymptotic giant branch), and, 'U' (unclassified). Additionally, some stars are assigned a combination of these classifications, i.e. '2CL/U', 'AGB/U', 'RC/2CL', 'RC/U', 'RGB/AGB', 'RGB/U'. We obtain the \gedrthree{} source IDs for \elsworth{}, following the exact same crossmatching procedure as we did for \jie{}. Next, we select three populations, namely, a) \textbf{\rc{}}: where 'consensus'$==$'RC'$|$'RC/2CL'$|$'RC/U'$|$'2CL'$|$'2CL/U', b) \textbf{\rgbp{}}: where 'consensus'$==$'AGB/U'$|$'RGB'$|$'RGB/AGB'$|$'RGB/U'$|$'U', and, c) \textbf{Unclassified}: where consensus$==$'U'. We find 3047 stars in common with \elsworth{}, out of which, 72\% are classified as \rc{} by the latter. Taking into account the impurity (RGB), we still achieve an accuracy of about 83\%, though we of course note the lower completeness compared to \jie{}. 

\begin{figure}
\includegraphics[width=1.\columnwidth]{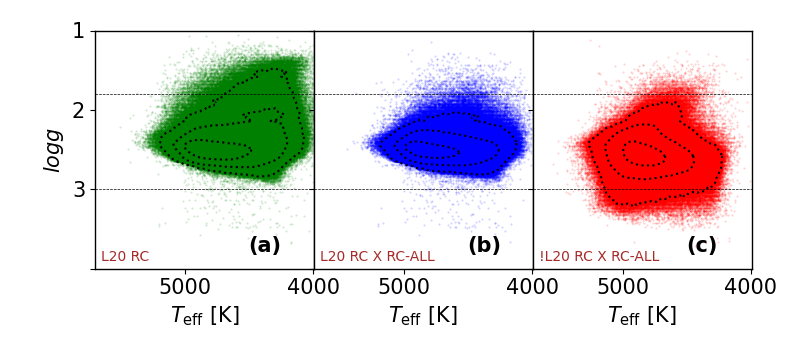} 
\caption{External validation of our catalogue (RC-ALL here) by crossmatching against the catalogue of \lucey{}, shown on the \kiel{} diagram using parameters from the latter. We also overplot density contours at (1,2,3)$\sigma$, shown in black. The typical crude boundary for \rc{} stars, 1.8$<logg <$3.0 is indicated by black dotted lines. Panel (a) shows that the distribution of \lucey{} falls mostly (1,2 $\sigma$) inside the crude boundaries, and is also broad in \teff{}. Panel (b) shows the crossmatch between \lucey{} and our sample. Nearly all of these overlapping sources lie between the crude boundary, and also display a tighter distribution in \logg{}. Panel (c), finally shows the stars that are present in \lucey{}, and are not classified as \rc{}, but are present in our sample. Again, the stellar parameters from \lucey{} for those that lie within the (1,2)$\sigma$ contour are consistent with expected boundaries, although there are possible contaminants around the region \logg{}$>$3 and \teff{}$>$5000 K.  \label{fig:xmatch_lucey}}
\end{figure}

\citet[][\lucey{} hereafter]{Lucey2020} put out a photometric catalogue of 2.6
million \rc{} candidates. Their method involves predicting asteroseismic parameters (\periodspacing{}, \deltanu{}) and stellar parameters ($logg$, \teff{} ) from spectral energy distributions (SED), using a neural network trained on \rc{} stars from \cite{Ting:2018}. They combined photometry from \panstars{}, \allwise{}, \twomass{}, and \gdrtwo{}. In their catalogue, they provide two sets of \rc{} samples: a) `Tier 2': 2,210,769 candidates with a contamination rate of $\approx 33\%$, and b) `Tier 1': A superior set of 405,231, with a lower contamination rate of $\approx 20\%$. For our comparison we restrict to their stricter `Tier 1' sample. As before, we make use of the \textit{gaiaedr3.dr2\_neighbourhood} table to recover the corresponding \gedrthree{} source IDS for all stars in the \lucey{} catalogue. In Fig. \ref{fig:xmatch_lucey}, we illustrate the comparison between \gaiawise{}[RC] and \lucey{} on the \kiel{} diagram, using stellar parameters from the latter. In Fig. \ref{fig:xmatch_lucey}(a) we show all the red clump candidates (Tier 1 and Tier 2), from \lucey{}, and the 1,2, and 3 $\sigma$ density contours are shown in black. Also, marked are typical crude boundaries, 1.8$<\log g<$3.0, used to select \rc{}, shown as a guide to the reader. We noticed the presence of a skew in the density distribution at lower \logg{} values beyond the lower end of this crude boundary. Next, in Fig. \ref{fig:xmatch_lucey}(b), we show the stars in common between \lucey{} and \gaiawise{}[RC], where we note that the common selection lies well within the crude boundary and in a tight locus in \logg{}. It is quite clear that our selection is a bit more conservative, and at the cost of accuracy, we miss out on possible contaminants in the \lucey{} catalogue. Finally, in Fig. \ref{fig:xmatch_lucey}(c), we plot those stars which are classified as 'nan' in \lucey{}, but are present in our selection scheme. It is interesting to note that the stellar parameters for these from \lucey{} are consistent with these being part of the red clump, although there are possible contaminants present, especially in the region with \logg{}$>$3 and \teff{}$>$5000 K. Since our selection method relies on both updated astrometry and photometry from \gedrthree{}, as well as \allwise{}, we are fairly confident that a large fraction of these are consistent with being members of the \rc{}.

\section{Sub-selection function queries: \textit{Gaia} X AllWISE}
\label{app:gaiawise_sfquery}

\begin{figure}
\includegraphics[width=1.\columnwidth]{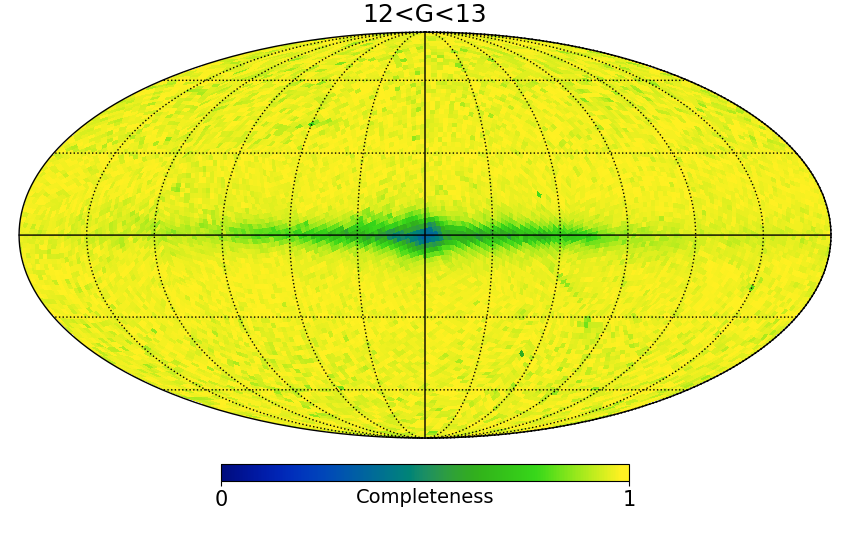}
\includegraphics[width=1\columnwidth]{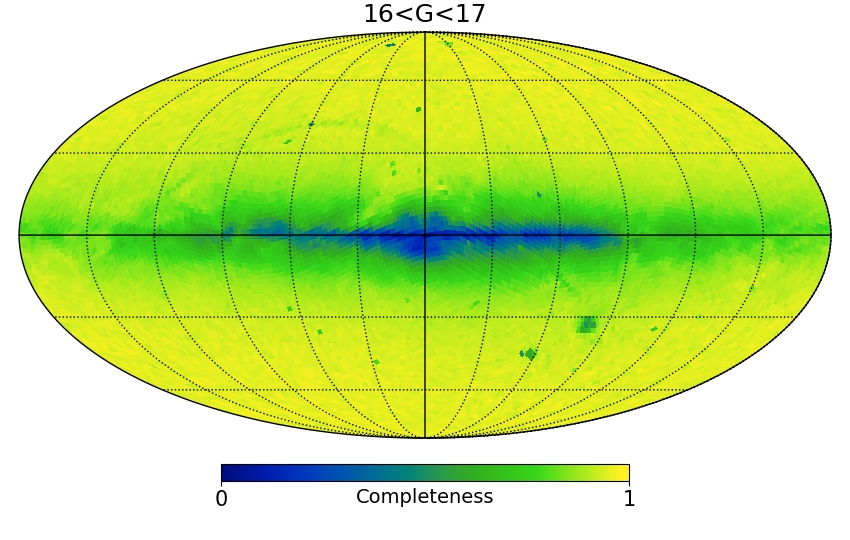}
\caption{Completeness map for the sub-selection layer for \gaiawise{} dataset at \hpix{}  level 5, shown for a bright (top) and a faint magnitude bin (bottom).} \label{fig:subsf_hpix}
\end{figure}
To build the sub-selection function layer of the selection function, we compute the ratio of sources that end up in the \gaiawise{} crossmatch versus all sources in \gaia{}. We compute this ratio as a function of sky position (at \hpix{} level 5), and $G$ (1 magnitude wide bins). To this end, we ran the following query on the \gaia{} archive: 

\begin{lstlisting}
SELECT healpix_, phot_g_mean_mag_, COUNT(*) AS n, SUM(selection) AS k
FROM (SELECT to_integer(GAIA_HEALPIX_INDEX(5,source_id)) AS healpix_, to_integer(floor((phot_g_mean_mag - 3)/1.)) AS phot_g_mean_mag_,
to_integer(IF_THEN_ELSE(wisenb.source_id >0,1.0,0.0) ) AS selection
FROM gaiadr3.gaia_source  as gdr3                                                       left outer join 
gaiadr3.allwise_best_neighbour as wisenb USING (source_id)              
WHERE phot_g_mean_mag > 3 AND phot_g_mean_mag < 20 AND g_rp > -1 AND g_rp < 3.0 and parallax is not null) AS subquery
GROUP BY healpix_, phot_g_mean_mag_         
\end{lstlisting}

In Fig. \ref{fig:subsf_hpix} we show the on sky projection of this sub-selection function for two magnitude bins, one at the bright end (12$<G<$13) where most of the sky has a high degree of completeness (though with noise due to small number statistics), and another at the faint end (16$<G<$17) where it is much less so. The reduced completeness near the Galactic plane and towards the inner Galaxy, as well as towards the Large and Small Magellanic Clouds, is mainly due to crowding and the low resolution of \allwise{}. 

\section{Corner plots for mock cases}
\label{app:corner_plots}

\begin{figure}
\includegraphics[width=1.\columnwidth]{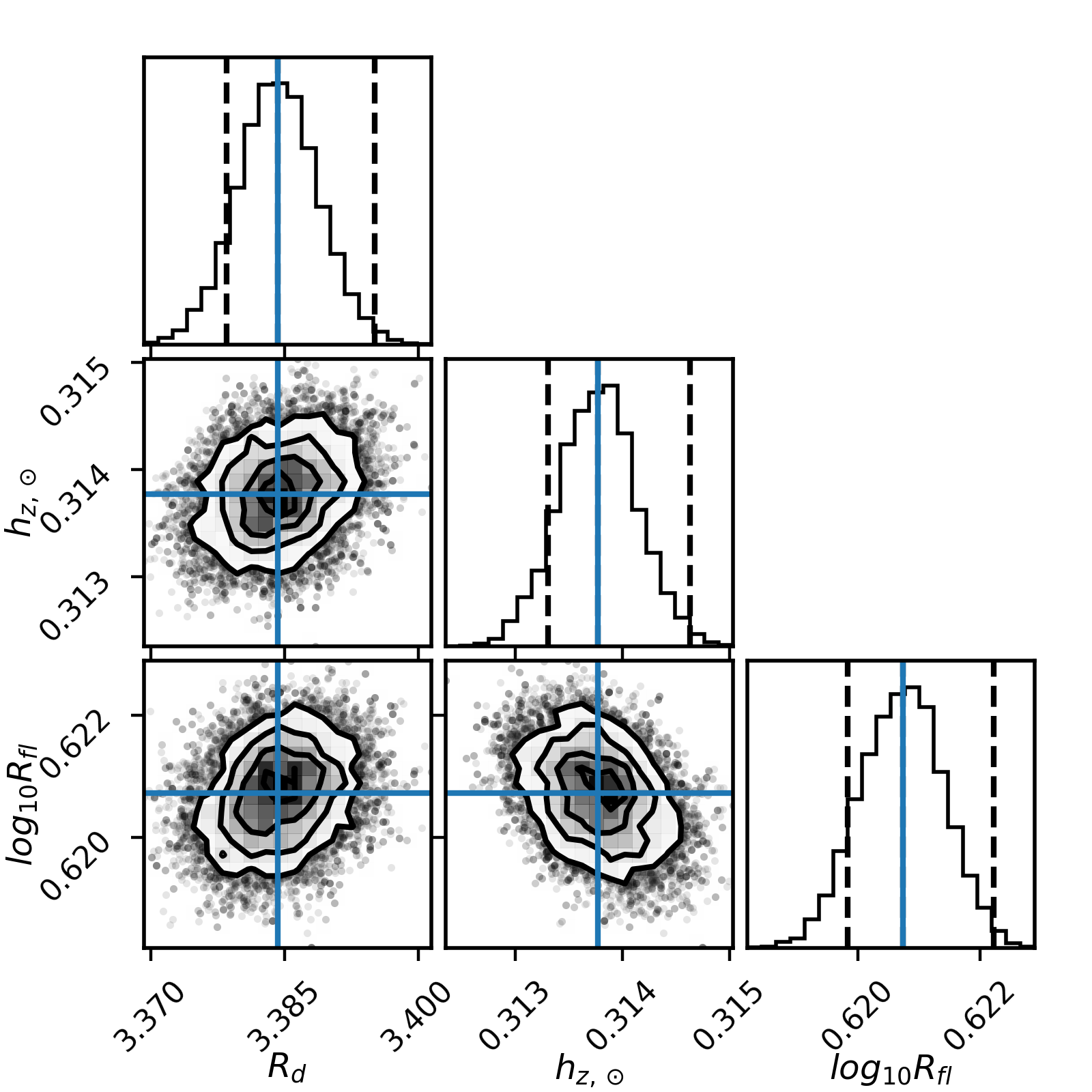}
\caption{Mock case I: Corner plots from the MCMC fitting for the three parameters describing mock case I, with the best-fitting values indicated as vertical lines. The true values for this example are: \rd{}$=3.3$ kpc, \hzsun{}$=0.3$ kpc, $\log_{10} \rflare{}=0.6$}. 
\label{fig:mock_case1_fitmcmc}
\end{figure}

\begin{figure*}
\includegraphics[width=2.\columnwidth]{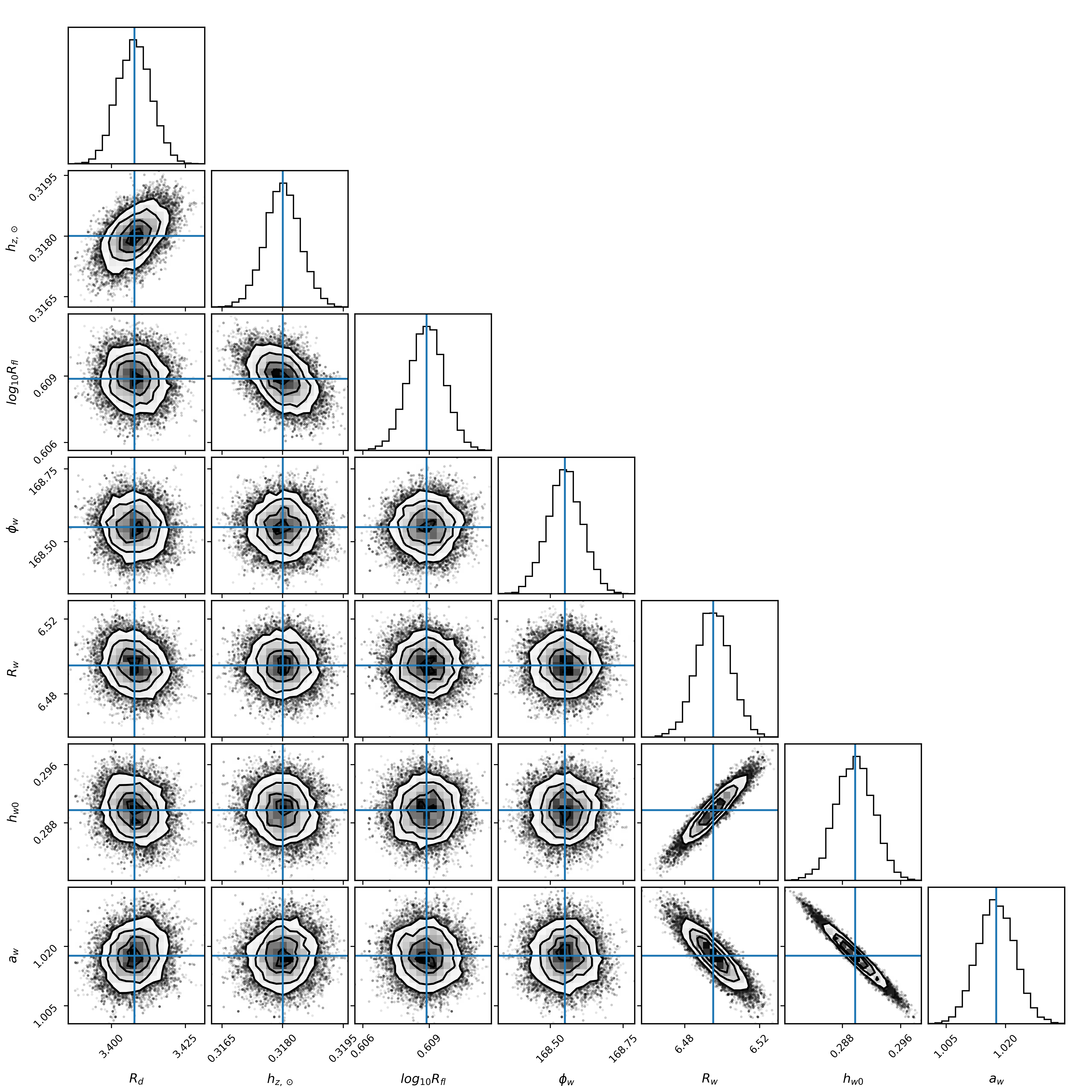}
\caption{Mock case II: Corner plots from the mcmc fitting for the three parameters describing mock case I, with the best-fitting values indicated with blue lines. The true values in this example are:  \rd{}$=3.3$ kpc, \hzsun{}$=0.3$ kpc, $log_{10}$ \rflare{}$=0.6$, \phiwarp{}$=170^\circ$, \rwarp{}$=6.5$ kpc, \awarp{}$=1.0$, \hwarp{}$=0.3$ kpc. } 
\label{fig:mock_case2_fitmcmc}
\end{figure*}

We show here the corner plots of the MCMC fit (1000 iterations) carried out on the two mock cases described in Sect. \ref{sec:mock_results}. In case I we demonstrate the fit on an exponential disc with scale height \rd{}, scale length \hzsun{}, and a flare scale length \rflare{} (Fig. \ref{fig:mock_case1_fitmcmc}). In case II, we generate a mock galaxy identical to that in case I, except that it is also forced to be warped, and is described by four additional parameters (\phiwarp{},\rwarp{},\hwarp{},\awarp{}) as shown in Fig. \ref{fig:mock_case2_fitmcmc}.

\section{Corner plot: Single disc fit}
\begin{figure}[h!]
\resizebox{9cm}{9.cm}
{\includegraphics{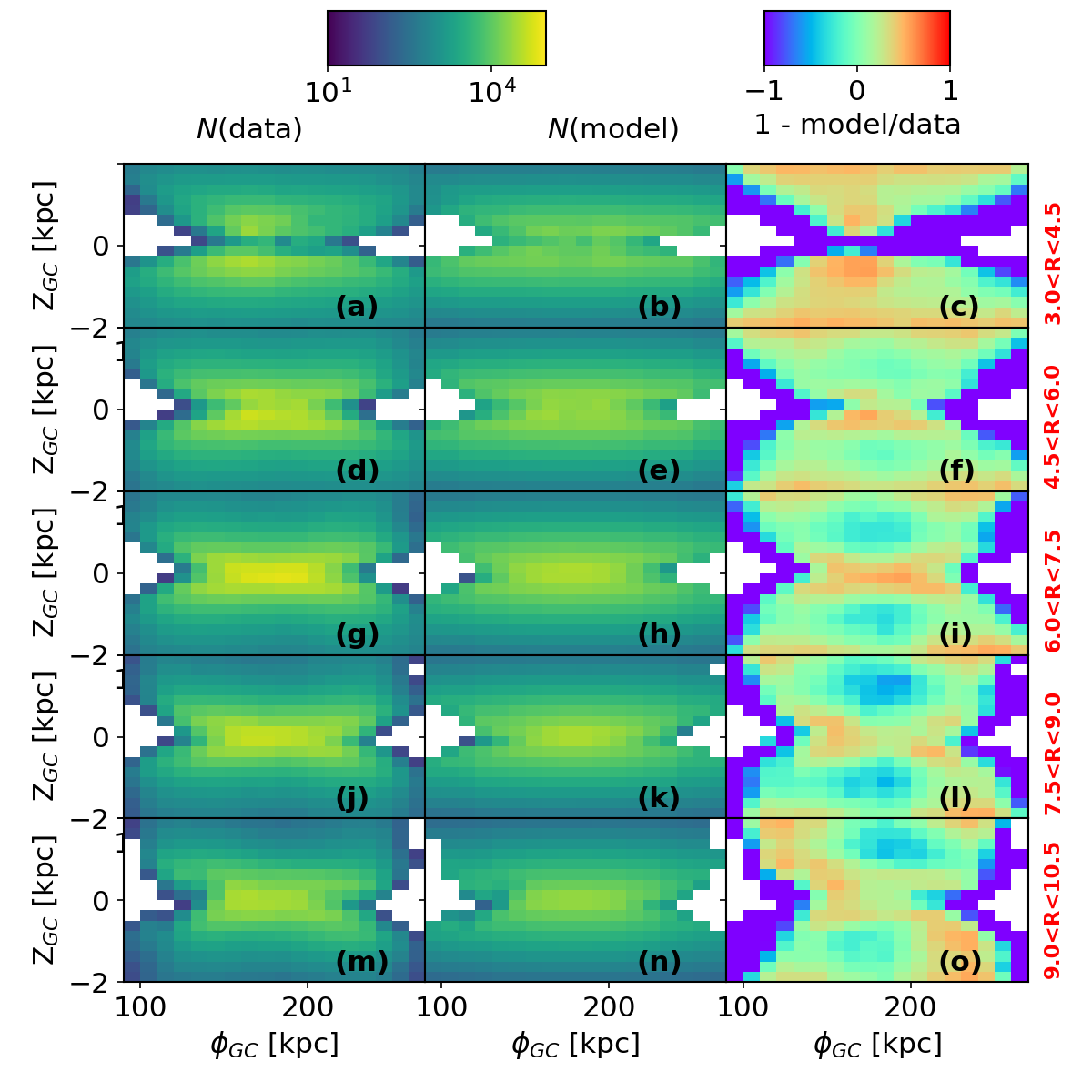}}
\resizebox{9cm}{9.cm}
{\includegraphics{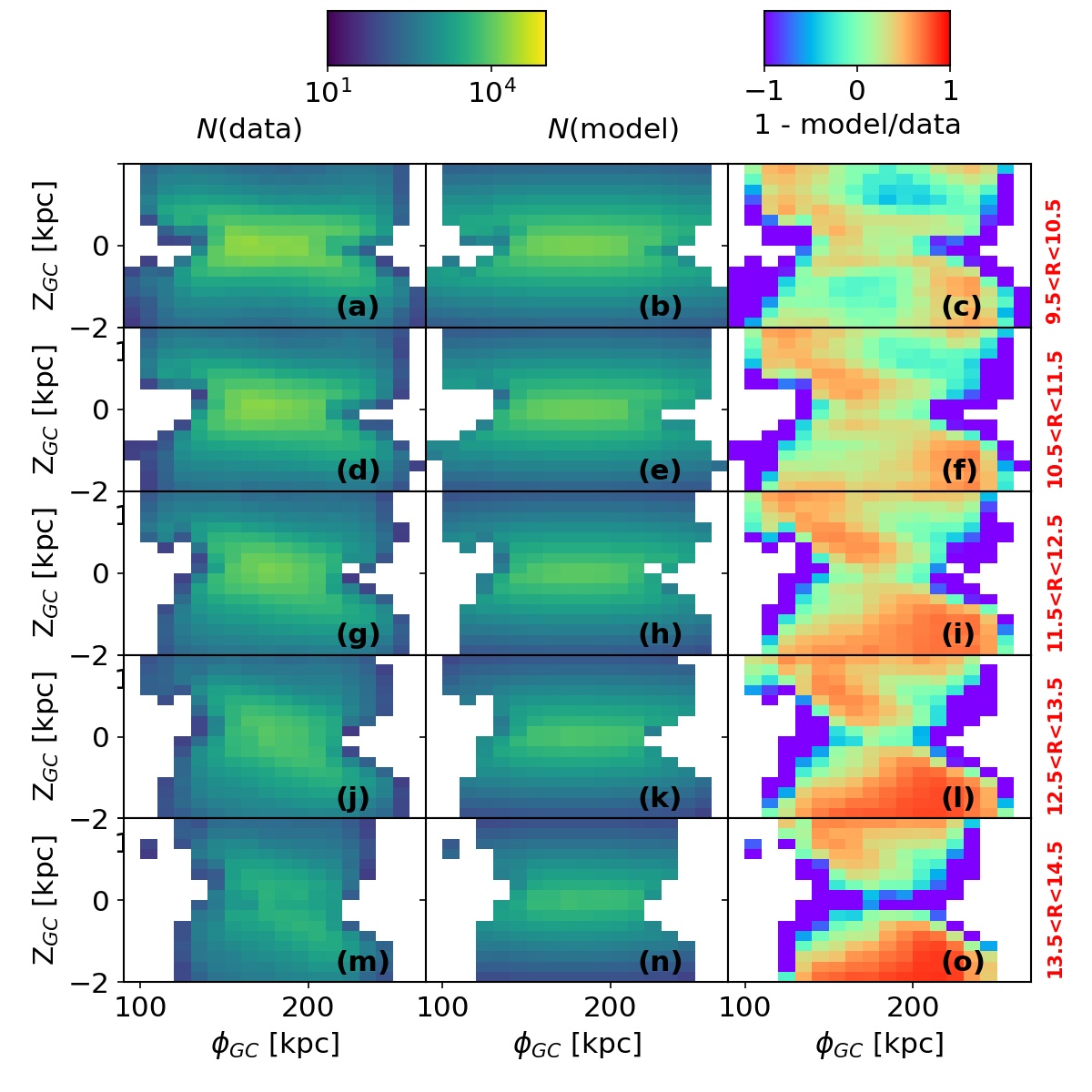}}
\caption{Residuals between the model and data in the $\phi$,\zgal{} projection. The setup is the same as Fig. \ref{fig:residuals_phiz_inner} for the inner disc and Fig. \ref{fig:residuals_phiz_outer} for the outer disc, but it is for Model 1, which allows for only a single disc component (\autoref{tab:bestfitdatamodel}).} \label{fig:residuals_phiz_singdisc}
\end{figure}
\begin{figure}
\includegraphics[width=1.\columnwidth]{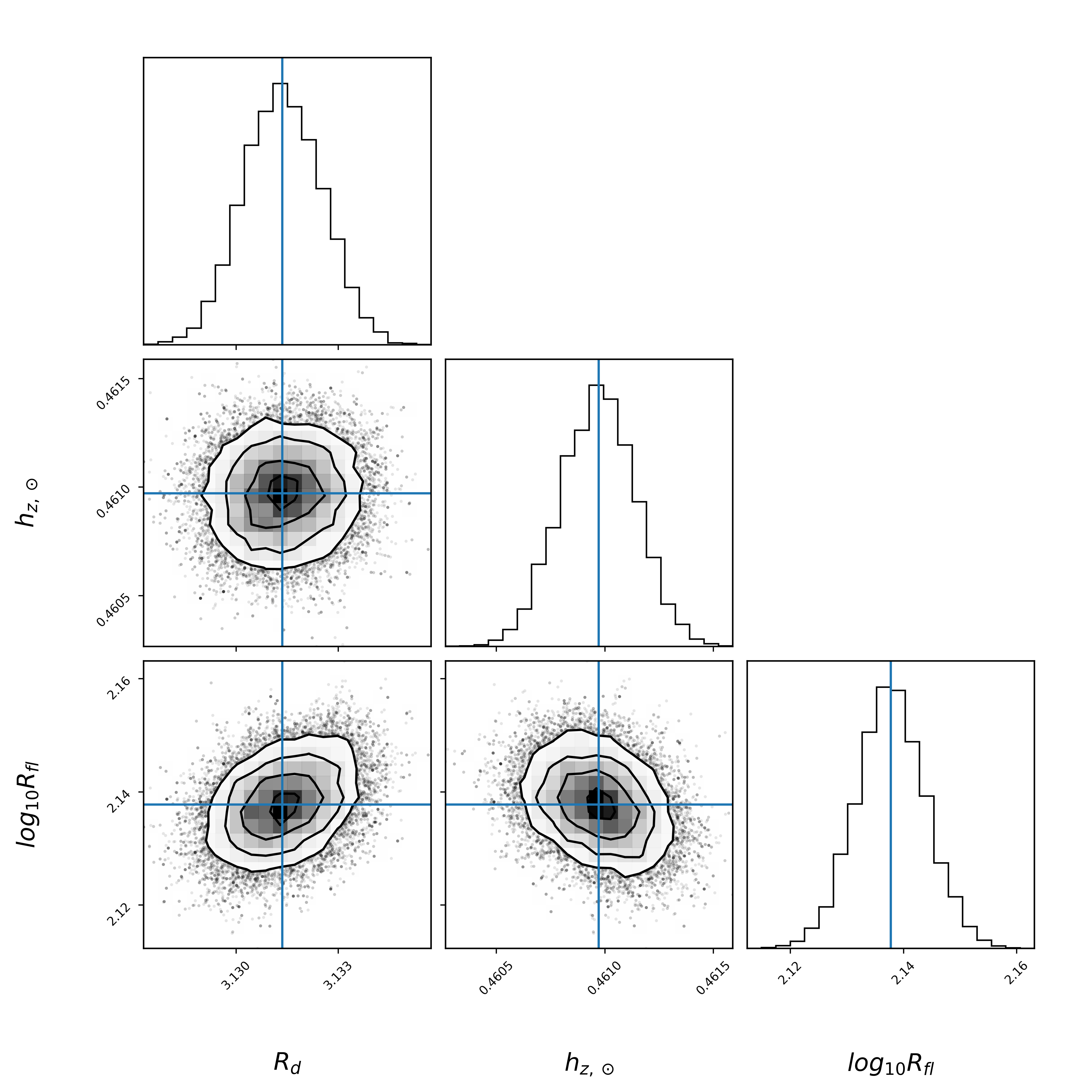}
\caption{Posterior probability distributions for Model 1(\autoref{tab:bestfitdatamodel}) applied to \gaiawise{[RC]}. This model allows for only one disc component.} \label{fig:datafitmcmc_single}
\end{figure}

Figure \ref{fig:datafitmcmc_single} shows the corner plot for the fit to the \gaiawise{}[RC] dataset, for the case assuming only one disc component. Here we fit for the scale length, scale height and flare parameter only (see Model 1 in \autoref{tab:bestfitdatamodel}). The residuals for this fit are shown in Fig. \ref{fig:residuals_phiz_singdisc} in $\phi$-\zgal{} projection for the inner (3$<$\rgal{}$<$10.5 kpc) and the outer disc regions (9.5$<$\rgal{}$<$14.5 kpc).  Compared to the residuals for the double disc case shown in Fig. \ref{fig:residuals_phiz_inner}, \& Fig. \ref{fig:residuals_phiz_outer}, this model does not fit the disc well across all \rgal{}, and is especially problematic in the outer disc region.

\section{Corner plot: Double disc fit}
\begin{figure*}
\includegraphics[width=2.\columnwidth]{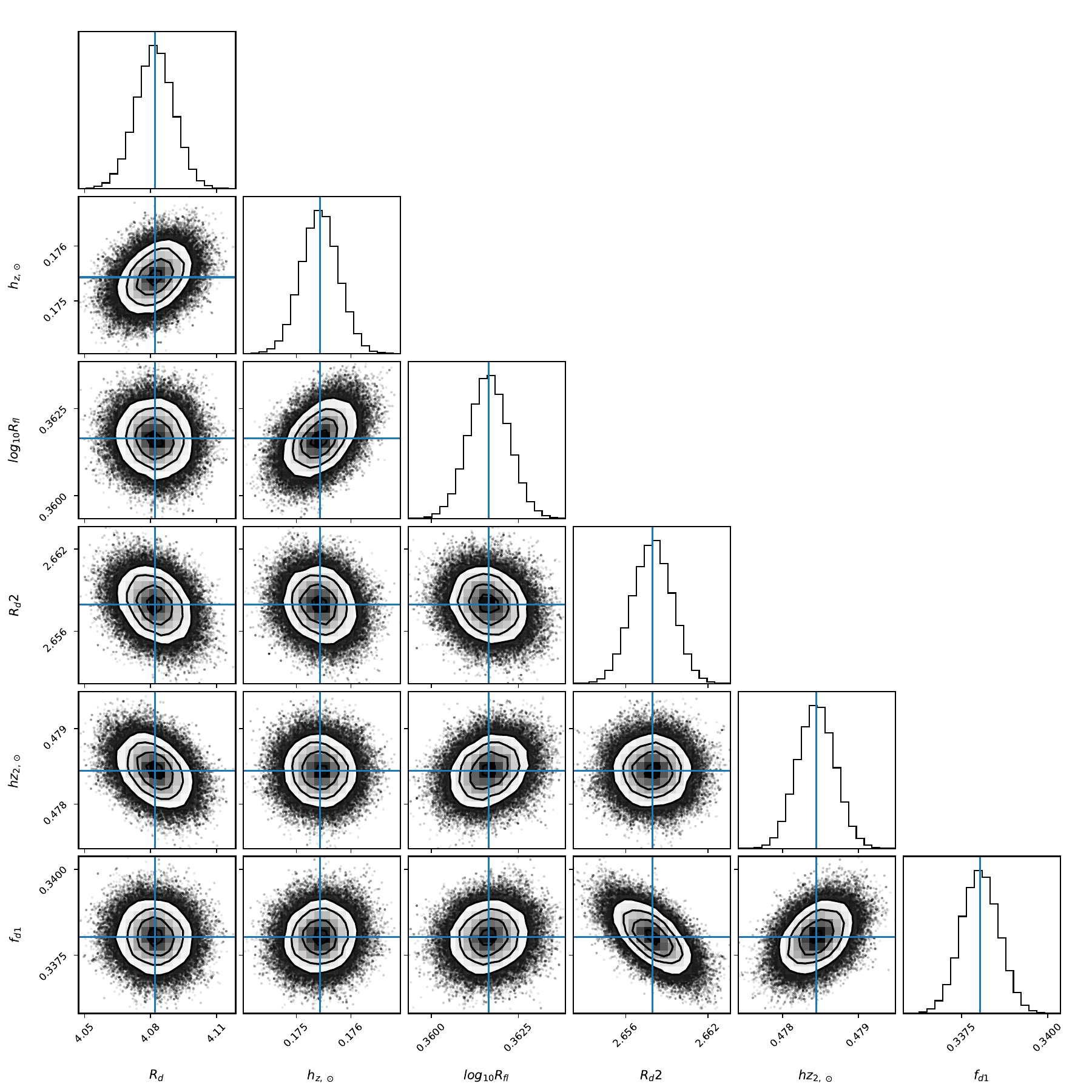}
\caption{Posterior probability distributions for Model 2 (\autoref{tab:bestfitdatamodel}) applied to \gaiawise{[RC]}. This model allows for two discs.} \label{fig:datafitmcmc}
\end{figure*}

Fig. \ref{fig:datafitmcmc} shows the corner plot for fit to the \gaiawise{}[RC] dataset, assuming a double disc model. Here we fit for the scale length, scale height of two independent exponential discs, with one of these allowed to flare (see Model 2 in \autoref{tab:bestfitdatamodel}). We do not fit for the warp in this model.

\section{Warp fitting}
\begin{figure*}
\includegraphics[width=.5\columnwidth]{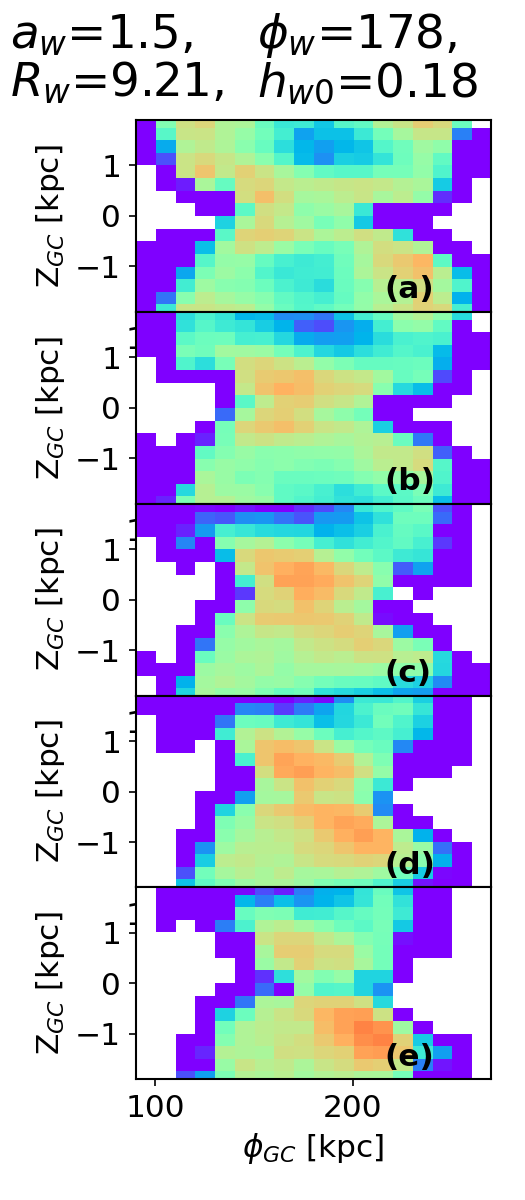}
\includegraphics[width=.5\columnwidth]{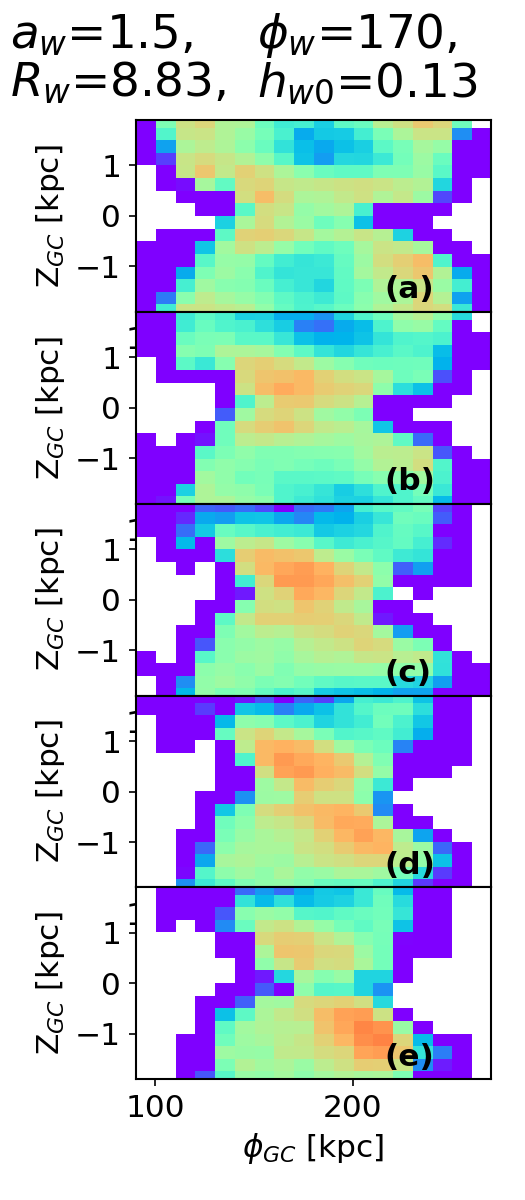}
\includegraphics[width=.5\columnwidth]{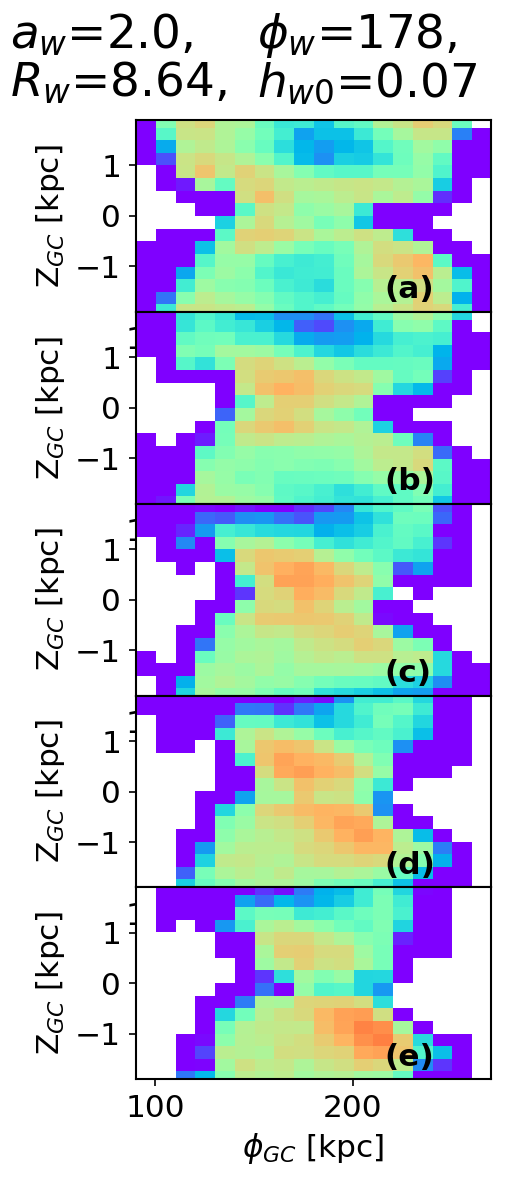}
\includegraphics[width=.5\columnwidth]{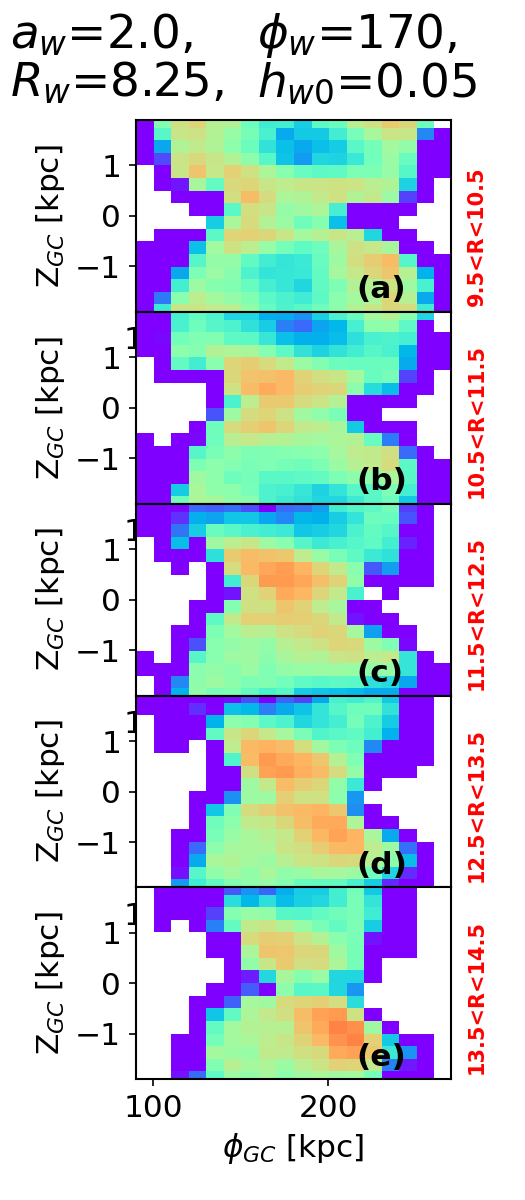}
\caption{Relative residual (1-model/data) for \gaiawise{[RC]}, corresponding to the four cases exploring warp parameters shown in Fig. \ref{fig:warp_fit_test}. For each case, we show the residuals in successive \rgal{} annuli ($9.5$ kpc$<$\rgal{}$<14.5$ kpc). The corner plots corresponding to each case are shown in Fig. \ref{fig:warp_fit_test}.} \label{fig:warp_fit_res_test}
\end{figure*}

\end{appendix}
\end{document}